\documentclass[11pt]{article}
\usepackage{amsmath,amssymb,graphicx,bbm,bm,mathtools}
\usepackage[utf8]{inputenc}
\numberwithin{equation}{section}
\usepackage{color}
\usepackage[dvipsnames]{xcolor}

\usepackage[linktocpage=true]{hyperref}
\hypersetup{
    colorlinks=true,
    linkcolor=blue,
    citecolor=blue,
    pdfpagemode=FullScreen,
    }

\edef\restoreparindent{\parindent=\the\parindent\relax}
\usepackage{parskip}
\restoreparindent

\usepackage{tikz}
\usetikzlibrary{arrows,arrows.meta,intersections, calc,positioning,decorations.pathreplacing,decorations.pathmorphing,shapes}
\usetikzlibrary{patterns}
\usetikzlibrary{decorations.markings}
\usetikzlibrary{knots}

\def\tr{{\rm tr}}

\let\Re\relax
\DeclareMathOperator{\Re}{Re}
\let\Im\relax 
\DeclareMathOperator{\Im}{Im}

\DeclareMathOperator{\Det}{Det}
\DeclareMathOperator{\arcsinh}{arcsinh}

\begin{document}

\begin{titlepage}

\renewcommand{\thefootnote}{\fnsymbol{footnote}}
\begin{flushright}
\begin{tabular}{l}
YITP-24-50\\
\today 
\end{tabular}
\end{flushright}

\vfill
\begin{center}

\noindent{\Large \textbf{The semi-classical saddles in three-dimensional gravity}}

\smallskip

\noindent{\Large \textbf{via holography and mini-superspace approach}}

\vspace{1.5cm}

\noindent{Heng-Yu Chen,$^{a,b}$\footnote{heng.yu.chen@phys.ntu.edu.tw} Yasuaki Hikida,$^b$\footnote{yhikida@yukawa.kyoto-u.ac.jp} Yusuke Taki$^b$\footnote{yusuke.taki@yukawa.kyoto-u.ac.jp} and Takahiro Uetoko$^c$\footnote{uetoko-t@t.kagawa-nct.ac.jp}}

\bigskip

\vskip .6 truecm

\centerline{\it $^a$Department of Physics, National Taiwan University, Taipei 10617, Taiwan }

\medskip

\centerline{\it $^b$Center for Gravitational Physics and Quantum Information, Yukawa Institute for Theoretical Physics, }
\centerline{\it Kyoto University, Kitashirakawa Oiwakecho, Sakyo-ku, Kyoto 606-8502, Japan}

\medskip

\centerline{\it $^c$Department of General Education, National Institute of Technology, Kagawa College,} 
\centerline{\it Chokushicho 355, Takamatsu, Kagawa 761-8058, Japan}
\end{center}

\vfill
\vskip 0.5 truecm
\begin{abstract}

We determine the complex geometries dual to the semi-classical saddles in three-dimensional gravity with positive or negative cosmological constant. We examine the semi-classical saddles in Liouville field theory and interpret them in terms of gravity theory. For this, we describe the gravity theory by Chern-Simons theory and classify the possible saddles based on the homotopy group argument. We further realize the semi-classical saddles using the mini-superspace model of quantum gravity and explicitly determine the integral contour. In the case of positive cosmological constant, we recovered the geometry used for no-boundary proposal of Hartle and Hawking. In the case of negative cosmological constant, the geometry can be identified with Euclidean anti-de Sitter space attached with imaginary radius spheres. The geometry should be unphysical and several arguments on this issue are provided. Partial results were already presented in our earlier letter,
and more detailed derivations and explanations on the results are given along with additional results. In particular, we reproduce the classical Liouville action from the Chern-Simons formulation of dual gravity theory.

\end{abstract}
\vfill
\vskip 0.5 truecm

\setcounter{footnote}{0}
\renewcommand{\thefootnote}{\arabic{footnote}}
\end{titlepage}

\newpage

\hrule
\tableofcontents

\bigskip
\hrule
\bigskip

\section{Introduction}

It remains one of the biggest problems to quantize gravity theory.
We may formulate quantum gravity in terms of path integral over metric fields. In order to make the path integral well-defined, it is sometimes useful to consider complex valued metric fields. 
One of the canonical examples is a black hole with Euclidean spacetime, whose periodicity is proportional to the inverse of temperature \cite{Hartle:1976tp}. 
Another is given by no-boundary proposal of Hartle and Hawking \cite{Hartle:1983ai}, which claims that our universe starts from nothing. After the integration over metric fields, the path integral may be expressed as a sum of semi-classical saddle points dressed by perturbative corrections in the Newton constant $G$. In this sense, the choice of path integral contour is equivalent to that of semi-classical saddles. In fact, the path integral over complex fields have too many saddles than the original ones and it is in general a quite difficult task to choose a correct set of saddle points. 
In this paper, we investigate this issue by making use of a simple example of three-dimensional gravity theory with negative or positive cosmological constant.

It is expected that holographic duality would provide useful tools to examine quantum gravity \cite{Susskind:1994vu}. It was proposed that gravity theory on $(d+1)$-dimensional anti-de Sitter (AdS$_{d+1}$) space (multiplied by an internal manifold) is dual to a $d$-dimensional conformal field theory (CFT$_d$) \cite{Maldacena:1997re}. 
The partition function of gravity theory on AdS$_{d+1}$ is evaluated as
\begin{align} \label{eq:AdSpf}
  \mathcal{Z}_\text{AdS} [\chi_j^{(0)}] = \int \prod_j [\mathcal{D} \chi_j]  e^{i 
 S_\text{AdS} [\chi_j] } \, ,
\end{align}
where $S_\text{AdS}[\chi_j]$ is a bulk action of bulk fields $\chi_j$ including gravity sector with negative cosmological constant.
The path integral is over all $\chi_j$ satisfying the boundary condition $\chi_j = \chi_j^{(0)}$ at the AdS boundary. 
The gravity partition function has a relation    \cite{Gubser:1998bc,Witten:1998qj}
\begin{align}
 \mathcal{Z}_\text{AdS} [\chi_j^{(0)}] = Z_\text{CFT} [\chi_j^{(0)}] \, ,
\end{align}
where $Z_\text{CFT} [\chi_j^{(0)}]$ is the CFT$_d$ partition function with sources $\chi_j^{(0)}$ for CFT operators $\mathcal{O}_j$. Here $\mathcal{O}_j$ are dual to the bulk fields $\chi_j$. In the semi-classical limit, the path integral over $\chi_j$ may be localized at several saddle points $\chi_{j(n)}$, and the partition function may be decomposed into summing over their individual contributions:
\begin{align} \label{AdSexpansion0}
 \mathcal{Z}_\text{AdS}[\chi_j^{(0)}] = \sum_n \mathcal{Z}_n [\chi_j^{(0)}] \, , \quad \mathcal{Z}_n [\chi_j^{(0)}] \sim e^{i S_\text{AdS}[\chi_{j(n)}]} \, .
\end{align}
We shall determine the set of saddle points in the gravity path integral with the help of dual CFT.

We are interested in a semi-classical gravity theory on AdS$_3$, which can be described by a CFT$_2$ with large central charge. 
It is generically expected that generic properties of CFT$_2$ with large central charge can be described by Liouville field theory.
Alternatively, we may regard the Liouville field theory as a CFT$_2$ dual to a concrete gravity theory on AdS$_3$. In order to explain it, let us start by introduding the Gaberdiel-Gopakumar duality \cite{Gaberdiel:2010pz}. The gravity side is given by 3d Prokushkin-Vasiliev theory \cite{Prokushkin:1998bq}. The theory includes higher spin gauge fields of spin $s=2,3,\ldots$, which can be described by a Chern-Simons gauge theory based on a higher spin algebra called as $\mathfrak{hs}[\lambda]$. The higher spin algebra can be truncated into $\mathfrak{sl}(N)$ at $\lambda = N$. The theory also includes two complex scalar fields with mass $\ell^2_\text{AdS} m^2 = - 1 + \lambda^2$ with $\ell_\text{AdS}$ as the AdS radius.
The dual CFT$_2$ is given by $W_N$ minimal model described by a coset
\begin{align} \label{eq:coset}
 \frac{SU(N)_k \times SU(N)_1}{SU(N)_{k+1}}
\end{align}
with the central charge
\begin{align} \label{eq:cosetcentral}
 c = (N -1) \left(1 - \frac{N (N+1)}{(N+k)(N+k+1)}\right)  \, .
\end{align}
We take the 't Hooft limit, where $N,k \to \infty$ but the ratio $\lambda = N/(N + k)$ kept finite. 

We actually use a different version of higher spin gravity obtained by applying the triality relation \cite{Castro:2011iw,Gaberdiel:2012ku,Perlmutter:2012ds}.
The version is effectively obtained by setting $\lambda = N$ in the original one. This means that the gravity theory includes $SL(N)$ Chern-Simons theory including gauge fields of spin $s=2,3,\ldots, N$. In this paper, we focus on the simplest case with $N=2$, where the higher spin gauge theory reduces to pure gravity theory. The theory includes a complex massive scalar field and the other complex massive scalar field is replaced by conical excess geometry. This implies that the gravity theory is non-unitary, which can also be seen from the dual CFT. In order to obtain the CFT$_2$ describing classical gravity, we need to take a large central charge limit as the central charge is inversely proportional to the gravitational coupling:
\cite{Brown:1986nw}
\begin{align} \label{eq:BH}
 c = \frac{3 \ell_\text{AdS}}{2 G} \, .
\end{align}
The expression \eqref{eq:cosetcentral} implies a negative level 
\begin{align} \label{eq:cosetlevel}
k = - 1 - N + \frac{N (N^2-1)}{c} + \mathcal{O} (c^{-2}) \, .
\end{align}
Since a unitary coset model \eqref{eq:coset} is realized by a positive integer $k$, the above equation means that the dual CFT is non-unitary. It was recently shown in \cite{Creutzig:2021ykz} that the correlation functions of the coset theory with non-integer level $k$ can be evaluated in terms of Liouville field theory. Thus we can use Liouville field theory as a definition of the coset theory. We may regard the Liouville field theory as a CFT$_2$ dual to three-dimensional gravity in this sense. In order to go beyond its semi-classical limit, we need to include the effects of matter sector. In other words, the Liouville field theory with large central charge can capture the semi-classical features of pure gravity.%
\footnote{It is known that three-dimensional gravity on AdS$_3$ has a Liouville theory description, see, e.g., \cite{Verlinde:1989ua,Cotler:2019nbi,Verlinde:2024zrh}. However, it is also known that Liouville field theory cannot be holographic dual to three-dimensional gravity. For instance, the effective central charge of Liouville field theory is $c_\text{eff} =1$, which is far from the large central charge limit. Our claim is that the classical limit of three-dimensional gravity is dual to the large central charge limit of the coset \eqref{eq:coset}. Liouville field theory is simply a tool to compute its correlation functions. The relationship between these two Liouville descriptions is not clear for us and is worth investigating.}

We also study de Sitter (dS) gravity in a similar manner.
Currently, there are several versions of holography for dS gravity, but we shall use the original one proposed in \cite{Maldacena:2002vr}, see also \cite{Strominger:2001pn,Witten:2001kn}.
In the gravity theory on dS$_{d+1}$, it is convenient to use the Hartle-Hawking wave functional of universe defined by \cite{Hartle:1983ai}
\begin{align} \label{eq:dSwf}
  \Psi_\text{dS} [\chi_j^{(0)}] = \int \prod_j [\mathcal{D} \chi_j]  e^{i 
 S_\text{dS} [\chi_j] } \, .
\end{align}
Here $S_\text{dS}[\chi_j]$ is a bulk action of bulk fields $\chi_j$ including gravity sector with positive cosmological constant.
The path integral is over all $\chi_j$ satisfying the boundary condition $\chi_j = \chi_j^{(0)}$ at the future infinity. 
It was conjectured in \cite{Maldacena:2002vr} that the wave functional of universe can be computed  as
\begin{align}
\Psi_\text{dS} [\chi_j^{(0)}] = Z_\text{CFT} [\chi_j^{(0)}] \, .
\end{align}
Here $Z_\text{CFT} [\chi_j^{(0)}]$ is the partition function of CFT$_d$ with sources $\chi_j^{(0)}$ for CFT operators $\mathcal{O}_j$ dual to the bulk fields $\chi_j$.
The bulk physical quantity can be evaluated by utilizing the wave functional of universe.
For instance, the gravity partition function is evaluated as
\begin{align} \label{eq:wftopf}
\mathcal{Z}_ \text{dS} = \int \prod_j [\mathcal{D} \chi_j^{(0)}]  \left |\Psi_\text{dS} [\chi_j^{(0)}] \right |^2 \, .
\end{align}
As in the AdS case, the path integral over $\chi_j$ may be localized at saddle points $\chi_{j(n)}$ in the semi-classical limit as
\begin{align} \label{dSexpansion0}
 \Psi_\text{dS}[\chi_j^{(0)}] = \sum_n \Psi_n [\chi_j^{(0)}] \, , \quad \Psi_n [\chi_j^{(0)}] \sim e^{i S_\text{dS}[\chi_{j(n)}]} \, ,
\end{align}
and we shall determine the set of saddle points in the gravity path integral with the help of dual CFT.

The Liouville field theory has a parameter $b$ and the parameter (or more precisely speaking $b^2$) plays the role of the Planck constant. This means that the limit of $b^2 \to 0$ corresponds to the semi-classical limit of Liouville field theory.
The central charge is related to the parameter $b$ as
\begin{align}
Q = b + b^{-1} \, , \quad c = 1 + 6 Q^2  = 1 + 6 (b + b^{-1})^2 \, . \label{eq:central}
\end{align}
Thus the small $b^2$ limit corresponds to  large $c$ limit of Liouville field theory. 
In the AdS case, the central charge is related to gravitational parameters as in \eqref{eq:BH}, which implies that the central charge $c$ is real (or $\text{Re}\, b^{-2} > 0$). On the other hand, in the dS case, the central charge of dual CFT is known to be pure imaginary as \cite{Strominger:2001pn} (see also \cite{Ouyang:2011fs})
\begin{align} \label{eq:BHdS}
c  =  - i   c^{(g)} =  - i \frac{3 \ell_\text{dS}}{2G}
\end{align}
with real $c^{(g)}$. The dual CFT is proposed in \cite{Hikida:2021ese,Hikida:2022ltr,Chen:2022ozy,Chen:2022xse} to be given by a coset model \eqref{eq:coset} with a complex level
\begin{align} \label{eq:cosetleveldS}
k = - 1 - N + i \frac{N (N^2-1)}{c^{(g)}} + \mathcal{O} (c^{-2}) \, .
\end{align}
Its correlation functions can be computed again by the Liouville field theory but with imaginary central charge. As explained in \cite{Hikida:2022ltr,Chen:2023prz,Chen:2023sry}, a careful treatment of sub-leading order shows that $\text{Re} \, b^{-2} < 0$ in this case.

There are several exact results available for the Liouville field theory \cite{Dorn:1994xn,Zamolodchikov:1995aa}. From the behaviors of exact results at the $b^2 \to 0$ limit, we can read off the semi-classical expressions in Liouville field theory \cite{Zamolodchikov:1995aa,Harlow:2011ny}.
The equation of motion  is given by
\begin{align} \label{eq:eom}
\partial \bar \partial \varphi (z , \bar z )  = 2 \lambda e^{\varphi (z , \bar z )} -2 \pi \sum_{j=1}^m \eta_j \delta^2 (z -z_j) \, , 
\end{align}
where the summation over the delta functions corresponds to the insertions of vertex operators.
The complex coordinates of world-sheet are given by $(z , \bar z)$.
We are now interested in the case where the classical Liouville field $\varphi (z , \bar z)$ takes a complex value. 
From the equation of motion \eqref{eq:eom}, we can see that 
\begin{align} \label{varphin}
\varphi_{(n)}(z , \bar z) = \varphi_{(0)} (z , \bar z)+ 2 \pi i n
\end{align}
is a solution as well once $\varphi_{(0)}(z , \bar z)$ is a solution.
This implies that the semi-classical expressions of the correlation functions can realized by a sum over the semi-classical saddles w.r.t.\,$b^2$ labeled by an integer $n$. We can see that, for the two- and three-point functions, the semi-classical expressions are obtained by the sum of $n = -1,0$ for $\text{Re}\, b^{-2} < 0$ and $n=0,1,2,\ldots$ for $\text{Re}\, b^{-2} > 0$, see \cite{Harlow:2011ny}.

We would like to interpret the Liouville saddles in terms of geometry in dual three-dimensional gravity theory.
The case with $\text{Re}\, b^{-2} < 0$ was studied in our previous works
\cite{Hikida:2021ese,Hikida:2022ltr,Chen:2022ozy,Chen:2022xse}, where the Liouville field theory is dual to a gravity theory with positive cosmological constant. We concluded that complex saddles of quantum gravity on dS$_3$ are the geometries used for the no-boundary proposal of Hartle and Hawking \cite{Hartle:1983ai}. We provided evidence by computing the Gibbons-Hawking entropy \cite{Bekenstein:1973ur,Hawking:1975vcx,Gibbons:1976ue,Gibbons:1977mu} from the both sides of the duality and finding agreement. The result is consistent with the argument based on the criteria of allowable complex geometry in \cite{Louko:1995jw,Kontsevich:2021dmb,Witten:2021nzp}.

In this paper, we extend the previous works in several ways. As mentioned above, the exact results of Liouville field theory suggest that there are two saddles for the case with positive cosmological constant. On the other hand, the exact results of the Liouville field theory for $\text{Re}\, b^{-2} > 0$, which is dual to a gravity theory with negative cosmological constant, suggest that there are infinitely many saddles labeled by non-negative integers $n=0,1,2,\dots$. One of our main results is to provide geometrical interpretation of the semi-classical saddle points in dual gravity theory. 
It was argued there that the semi-classical saddles in Liouville field theory can be realized by the non-trivial topological gauge configuration of Chern-Simons formulation \cite{Harlow:2011ny}. Based on the argument, we classify the possible semi-classical saddles of quantum gravity in the Chern-Simons formulation \cite{Achucarro:1986uwr,Witten:1988hc} from the homotopy group of three-sphere, i.e., $\pi_3 (\mathbb{S}^3)$.
Another main result is to provide further supports for our claim on the geometrical interpretation both for the positive and negative cosmological constants. 
For this, we utilize mini-superspace approach to quantum gravity, see \cite{Feldbrugge:2017kzv,Caputa:2018asc,Donnelly:2019pie,DiTucci:2020weq} and a review \cite{Lehners:2023yrj} for recent developments. From the mini-superspace models, we indeed realize the saddle points suggested by dual CFT description in a path-integral formulation of quantum gravity. In case of positive cosmological constant, we obtain results consistent with the previous ones in \cite{Chen:2023prz,Chen:2023sry}. In the case of negative cosmological constant, we identify the geometry corresponding to saddle points of gravity theory as the Euclidean AdS$_3$ but attached with three-spheres with pure imaginary radius. The geometry should be unphysical, which implies the violation of the criteria of allowable complex geometry in \cite{Louko:1995jw,Kontsevich:2021dmb,Witten:2021nzp}.
We provide several arguments on this issue.

The results were partially presented in our previous letter \cite{Chen:2024vpa} and they were derived explicitly and explained in more details along with additional results in this paper.
For instance, in order to obtain insights on the relation between dual theories, we derive the classical action of Liouville field theory directly from the dual gravity theory. This was already done in \cite{Chang:2016ftb} (see also \cite{Chandra:2022bqq,Abajian:2023bqv}) for the case with negative cosmological constant by making use of the Einstein-Hilbert action for the gravity theory. We reproduce the result in the Chern-Simons formulation of gravity theory and extend it to the case with positive cosmological constant.

The rest of this paper is organized as follows.
In the next section, we begin by reviewing the previous works on the semi-classical saddles in Liouville field theory at the large central charge \cite{Harlow:2011ny}. We classify the possible semi-classical saddles of quantum gravity from the homotopy group argument and determine the set of saddles to take from the exact results of dual CFT.
In order to provide further evidence, we examine mini-superspace models of quantum gravity in section \ref{sec:minisuperspace}.
We explicitly determine the integral contours of the mini-superspace models, which reproduce the CFT results. 
In section \ref{sec:claads}, we compute the classical action of bulk gravity theory on backgrounds with insertions of conical defects and reproduce the classical action of Liouville field theory. Section \ref{sec:discussion} is devoted to conclusion and discussion. In particular, we address the issue on the relation to the criteria of allowable complex geometry discussed in \cite{Louko:1995jw,Kontsevich:2021dmb,Witten:2021nzp}. In appendix \ref{app:Liouville3pt}, the analysis in section \ref{sec:2ptresurgence} for two-point functions in Liouville field theory is extended to three-point functions. In appendix \ref{app:regularization}, some subtlety associated with regularization of infinite sum over Liouville saddle points is addressed.
In appendix \ref{app:geometry}, the geometrical interpretation of complex metric realized as saddle points in mini-superspace models is discussed.

\section{Semi-classical saddles of Liouville theory and \texorpdfstring{(A)dS$_3$}{(A)dS3} gravity}
 \label{sec:2ptresurgence}

In our previous works \cite{Chen:2023prz,Chen:2023sry}, we have identified the set of semi-classical saddles in dS$_3$ gravity from the exact results of Liouville field theory. 
The semi-classical limit of gravity theory is known to be dual to the limit of large central charge in Liouville field theory. This limit of the CFT was examined in \cite{Harlow:2011ny} to a large extent, and the results were utilized in \cite{Chen:2023prz,Chen:2023sry}. In the next subsection, we start by reviewing them. In subsection \ref{sec:holography}, we interpret the results of large central charge limit of Liouville field theory in the language of dual gravity theory. In subsection \ref{sec:complexsaddles}, we determine the geometries corresponding to the semi-classical gravity saddles obtained in the previous subsections.

\subsection{Semi-classical limit of Liouville field theory}

In this subsection, we examine the semi-classical saddle points of Liouville field theory.
The action of Liouville field theory is given by
\begin{align} \label{eq:Laction}
S_\text{L}[\phi] = \frac{1}{\pi} \int d ^2 z  \left[ \partial \phi \bar \partial \phi + \frac{Q}{4} \sqrt{\hat g} \mathcal{\hat R} \phi + \pi \mu e^{2 b \phi} \right] \, ,
\end{align}
where $\hat g_{ij}$ is the reference metric of world-sheet, $\hat g = \det \hat g_{ij}$, and $\mathcal{\hat R}$ is the curvature w.r.t.\ $\hat g_{ij}$. We set the reference metric to be the flat one as $\hat g_{ij} = \delta_{ij}$. We use the notation, where $z = \sigma_1 + i \sigma_2$ and $d^2 z = d^2 \sigma = d \sigma_1 d \sigma_2$.
The background charge $Q$ and the central charge $c$ are related to the parameter $b$ as \eqref{eq:central}.
The physical metric is given by $g_{ij} = e^{\frac{2}{Q }\phi} \hat g_{ij}$.
We consider correlation functions of the form:
\begin{align}
 \left \langle \prod_{j=1}^m V_{\alpha_j} (z_j , \bar z_j)  \right \rangle 
= \int \mathcal{D} \phi e^{- S_\text{L}[\phi]} \prod_{j=1}^m V_{\alpha_j} (z_j , \bar z_j) \, .
\end{align}
The vertex operators are given by
\begin{align}
V_\alpha (z , \bar z) = e^{2 \alpha \phi (z , \bar z)}
\end{align}
with conformal weights
\begin{align} \label{eq:weight}
h = \bar h = \alpha (Q - \alpha) \, .
\end{align}
Note that the conformal weights are invariant under the reflection $\alpha \to Q - \alpha$.

We are interested in the large central charge limit. From the expression of central charge in \eqref{eq:central}, the limit can be realized by $b \to 0$ (or $b \to \infty$). We consider the correlation functions in the limit $b \to 0$ while keeping the combination
$\eta = \alpha b$ fixed.
The operator whose conformal weight scales this way is referred as heavy. 
We also keep $\varphi = 2 b \phi$ and $\lambda = \pi \mu b^2$ to be finite in this limit. Near the insertion point of vertex operator, $z = z_j$, the heavy operator behaves as $\varphi (z, \bar z) \sim - 4 \eta_j \ln |z - z_j|$, thus the physical metric is ($z - z_j = r e^{i \theta}$)
\begin{align}
ds^2 \sim \frac{1}{r^{4 \eta_j}} (dr^2 + r^2 d \theta^2) \, .
\end{align}
In order to have a physical interpretation as a conical defect with angle $2 \pi (1 - 2 \eta_j)$, we set
\begin{align} \label{eq:seiberg}
 0 < \eta_j < \frac{1}{2} \, .
\end{align}
This is known as the Seiberg bound \cite{Seiberg:1990eb}.
We will focus on Liouville correlation functions further satisfying
\begin{align} \label{eq:sumeta}
   \sum_{j=1}^m \eta_j < 1 \, .
\end{align}
In this case, the effective metric describes two-dimensional manifold with $m$ conical defects and positive constant curvature.

When the inserted operators are heavy, it is convenient to modify the action such as to include them as a part of action. The modified action becomes 
\begin{align} \label{eq:modLaction}
\begin{aligned}
\tilde S_\text{L}[\varphi] & = \frac{1}{4\pi} \int_{D - \cup_j d_j} d ^2 z  \left[ \partial \varphi \bar \partial \varphi  + 4 \lambda e^{\varphi} \right] - \sum_{j} \left[\eta_j \varphi_j + 2 \eta_j^2 \ln \epsilon_j  \right] + \varphi_R + 2 \ln R\, .
\end{aligned}
\end{align}
Here we regularized the action by working on the world-sheet of disk topology $D$ with radius $R$ in addition with holes $d_i$ for the region $|z - z_j| \leq \epsilon_j$.
We further defined
\begin{align} \label{eq:boundarLf}
  \varphi_j = \frac{1}{2 \pi} \oint_{\partial d_j} \varphi d \theta_j \, , \quad 
 \varphi_R = \frac{1}{2 \pi} \oint_{\partial D} \varphi d \theta \, , 
 \end{align}
where $\theta_j$ and $\theta$  are the angular coordinates for the boundaries of $d_j$ and $D$, respectively.
The equation of motion is \eqref{eq:eom}
in the limit of $R \to \infty$ and $\epsilon_j \to 0$.
Once $\varphi_{(0)}$ is a solution to the equation of motion, then we find that 
$
\varphi_{(n)} = \varphi_{(0)} + 2 \pi i n
$
with integer $n$ is also its solution. Thus we can conclude that there are saddle points labeled by an integer $n$. 
This implies that a correlation function can be written as a sum over contributions from saddles labeled by $n$ dressed by perturbative corrections in $b^2$ as
\begin{align}  \label{eq:expansion}
 \left   \langle \prod_{j=1}^m V_{\alpha_j} (z_j , \bar z _j )  \right \rangle  = \sum_{n} I_n 
\, , \quad 
I_n = e^{- b^{-2}\tilde S_L [\varphi_{(n)}]} \sum_{q=0}^\infty a_q  \left(b^{2}\right)^q  
\end{align}
with
$
 a_0 = 1
$.
The factor $\exp ( -b^{-2} \tilde S_L [\varphi_{(n)}])$ describes a non-perturbative contribution from the saddle labeled by $n$ and the other part corresponds to the perturbative corrections. Here we assumed that $a_q$ do not depend on the label $n$ for the saddle. The perturbative corrections are generically asymptotic series, and the sum of them leads to divergence. A non-perturbative contribution from another saddle can cancel the divergence and in total the expression is unique. In other words, once we have the information on perturbative corrections at higher order, then we can obtain which saddles we should take.
Below, we shall read off the non-perturbative saddles from the exact expressions by taking the heavy limit of correlation functions.

As a simple example, we study the two-point function of heavy operator. The exact expression is known as \cite{Dorn:1994xn,Zamolodchikov:1995aa}
\begin{align}
 \left \langle V_{\alpha} (z_1)   V_{\alpha} (z_2) \right \rangle
= |z_{12}|^{-4 \alpha (Q - \alpha) } \frac{2 \pi}{b^2} [\pi \mu \gamma(b^2)]^{(Q - 2 \alpha)/b} \gamma \left( \frac{2 \alpha}{b} - 1 - \frac{1}{b^2}\right) \gamma (2 b \alpha - b^2) \delta (0) \, .  \label{eq:exact2pt}
\end{align}
Here $z_{ij} = z_i - z_j$ and $\gamma (x) = \Gamma(x)/\Gamma (1 -x)$.
We can read off the non-perturbative saddles by expanding the expression in $b^2$ as in \eqref{eq:expansion}.
For the two-point function, the solutions to the equation of motion \eqref{eq:eom} and the modified classical action \eqref{eq:modLaction} were obtained in \cite{Harlow:2011ny}. In particular, the modified classical action is given by
\begin{align}
\begin{aligned}
 \tilde S_\text{L} [\varphi_{(n)}] &= 2 \pi i (n + 1/2) (1 - 2 \eta) + (2 \eta -1) \ln \lambda  \\
 & \quad + 4 \eta (1 - \eta) \ln |z_{12}| + 2 [(1 - 2 \eta) \ln (1 - 2 \eta) - (1 -2 \eta)] \, .
 \end{aligned}
\end{align}
The small $b$ behavior of the exact expression \eqref{eq:exact2pt} can be read off from the Stirling's formula of the Gamma function. We find that
\begin{align}
 \left \langle V_{\alpha} (z_1)   V_{\alpha} (z_2) \right \rangle
\sim  \left( e^{- \frac{\pi i }{b^2}(1 - 2 \eta)} - e^{\frac{\pi i }{b^2}(1 -2 \eta)} \right) |z_{12}|^{- \frac{4  }{b^2}\eta (1 - \eta) } 
e^{ - \frac{2}{b^2} [ (1 - 2 \eta) \ln (1 - 2 \eta) - (1 - 2 \eta) ] }
\end{align}
for $\text{Re} \ b^{-2} > 0$ and
\begin{align}\label{Reb2l0}
 \left \langle V_{\alpha} (z_1)   V_{\alpha} (z_2) \right \rangle
&\sim \frac{1}{ e^{- \frac{\pi i }{b^2}(1 - 2 \eta)} - e^{\frac{\pi i }{b^2}(1 -2 \eta)} }|z_{12}|^{- \frac{4  }{b^2} \eta (1 - \eta)} 
e^{  - \frac{2}{b^2} [ (1 - 2 \eta) \ln (1 - 2 \eta) - (1 - 2 \eta) ] }
\end{align}
for $\text{Re} \ b^{-2} < 0$.

\subsection{Gravity interpretation via holography}
\label{sec:holography}

Applying the holography, we interpret the expressions of the correlation functions of heavy operators in the semi-classical limit of Liouville field theory in terms of dual gravity theory. 
The dual theory includes three-dimensional gravity with cosmological constant, whose action is
\begin{align} \label{eq:action}
I = I_\text{EH} + I_{\text{GH}} + I_\text{CT} \, .
\end{align}
The Einstein-Hilbert action is given by
\begin{align} \label{eq:IEH}
 I_\text{EH} = - \frac{1}{16 \pi G} \int d^3 x \sqrt{g} (R - 2 \Lambda) \, .
\end{align}
Here $g_{\mu \nu}$ is the metric, $g = \text{det}\, g_{\mu \nu}$ and $R$ is the Ricci scalar w.r.t.\ the metric. We denote the Newton constant by $G$ and the cosmological constant by $\Lambda$. It is given by $\Lambda = \ell_\text{dS}^{-2}$ and $\Lambda = - \ell_\text{AdS}^{-2}$ for the cases of positive and negative cosmological constants, respectively.
The Gibbons-Hawking boundary term is
\begin{align} \label{eq:IGH}
I_\text{GH}  = \frac{1}{8 \pi G} \int d^2 x\sqrt{h} K \, ,
\end{align}
where $h_{ij}$ is the boundary metric, $h = \text{det} \, h_{ij}$ and $K$ is the extrinsic curvature.
We regularize the IR divergence near the boundary by adding a counter term,%
\footnote{The counter term in the case with $\Lambda = \ell^{-2}_\text{dS}$ can be found in \cite{Balasubramanian:2001nb}. They are dealing with Lorentzian geometry, while we start from Euclidean geometry and take an analytic continuation later. This provides the factor $i$ difference in the counter term.}
\begin{align} \label{eq:ICT}
 I_\text{CT} = \frac{1}{8 \pi G} \int d^2 x \sqrt{h}  \sqrt{ - \Lambda} \, . 
\end{align}
In order to understand the results of Liouville field theory in terms of dual gravity theory, it is convenient to move to the first order formulation, where the action can be written as \cite{Achucarro:1986uwr,Witten:1988hc}
\begin{align} \label{eq:CSac0}
&I_\text{EH} = - k I_\text{CS} [A] + k I_\text{CS} [\tilde A] 
\end{align}
up to a boundary term, see \eqref{eq:CSaction0} below.
Here $I_\text{CS}[A]$ is the action of Chern-Simons gauge theory,
\begin{align} \label{eq:CSac}
&I_\text{CS} [A] = \frac{1}{4 \pi} \int \text{tr} \left[ A \wedge dA + \frac{2}{3} A \wedge A \wedge A \right] \, .
\end{align}
The Chern-Simons level is related to the gravitational parameters as
\begin{align}
k = - i \kappa = - i \frac{\ell_\text{dS}}{4G} \, , \quad k = \frac{\ell_\text{AdS}}{4G}
\end{align}
for the cases of positive  and negative cosmological constant, respectively.

We start from the case with positive cosmological constant. The central charge of Liouville field theory is related to parameters of gravity theory on dS$_3$ as in \eqref{eq:BHdS}.
We are interested in the case with heavy scalar operators dual to very massive scalar fields. A scalar field backreacts on the dS$_3$ background, which creates a conical defect. 
The wave functional of the corresponding universe is thus related to a CFT two-point function as
\begin{align} \label{eq:wf2corr}
 \Psi_\text{dS} =  \left \langle V_{\alpha} (z_1 , \bar z_1) V_{\alpha} (z_2 , \bar z_2)  \right \rangle \, ,
\end{align}
which can be decomposed as in \eqref{dSexpansion0}.
We rewrite the contribution from a saddle as 
\begin{align} \label{eq:eachwf}
\Psi_n  =
\exp \left( S^{(n)}_\text{GH}/2 + i \mathcal{I}^{(n)}\right) 
\end{align}
with real $S_\text{GH}^{(n)},\mathcal{I}^{(n)}$.
Here $S_\text{GH}^{(n)}$ is regarded as the contribution to the Gibbons-Hawking entropy \cite{Bekenstein:1973ur,Hawking:1975vcx,Gibbons:1976ue,Gibbons:1977mu} from the saddle labeled by $n$.
From \eqref{eq:central}, we obtain 
\begin{align} \label{eq:reb2}
b^{-2} = - i \frac{c^{(g)}}{6} - \frac{13}{6} + \mathcal{O} ((c^{(g)})^{-1}) \, .
\end{align}
Positive real $c^{(g)}$ implies $\Re b^{-2} < 0$. 
Applying the relation \eqref{eq:wf2corr}, we obtain the semi-classical expression for the wave functional of universe 
into two saddles:
\begin{align}\label{dSexpansion}
 \Psi_\text{dS} =\sum_{n=-1,0}  \Psi_n
\end{align} 
with
\begin{align}
 \Psi_n \sim e^{\frac{\ell_\text{dS}}{4G}(2 n + 1) \pi  (1 - 2 \eta)}  |z_{12}|^{ i \frac{\ell_\text{dS}}{G}\eta (1 - \eta)  } 
e^{ i \frac{\ell_\text{dS}}{2G} [ (1 - 2 \eta) \ln (1 - 2 \eta) - (1 - 2 \eta) ] } \, .
\end{align}
Here the sum is over two saddles with $n = -1,0$. We can read off the contribution to the Gibbons-Hawking entropy from each saddle with $n=-1$ or $n=0$ as
\begin{align} \label{eq:SGH}
 S_\text{GH}^{(n)} = \frac{(2 n +1) \pi \ell_\text{dS} (1 - 2 \eta)}{2G} \, .
\end{align}
The pure phase contribution is
\begin{align} \label{eq:In}
 \mathcal{I}_n \, (\equiv \mathcal{I} )
 =  \frac{\ell_\text{dS}}{G}\eta (1 - \eta) \ln |z_{12}|
 + \frac{\ell_\text{dS}}{2G} [ (1 - 2 \eta) \ln (1 - 2 \eta) - (1 - 2 \eta) ]  \, ,
\end{align}
which is independent on the label $n$ for the saddle point.

We then move to the case with negative cosmological constant.
It has been known for a long time that the symmetry of bulk isometry enhances to Virasoro symmetry near the boundary of AdS$_3$ and its central charge was computed as in \eqref{eq:BH}.
The large $c$ limit is given by taking $b \to 0$ with real $b$, see \eqref{eq:central}.
As before, we are interested in the case with heavy scalar operators, whose dual scalars create conical defects when $0 < \eta_j < 1/2$. The gravity partition function on the corresponding geometry can be computed from the two-point function of dual CFT as
\begin{align}
 \mathcal{Z}_\text{AdS} =  \left \langle V_{\alpha} (z_1 , \bar z_1)  V_{\alpha} (z_2 , \bar z_2)  \right \rangle \, .
\end{align}
We expand the partition function as in \eqref{AdSexpansion0}.
Since the central charge $c$ is now real and positive, we have 
\begin{align}
\text{Re} \, b^{-2} \sim \text{Re} \, \frac{c}{6}  > 0 \, .
\end{align}
For real $b^2$ and $\eta$, we have to make some regularization to perform the series expansion on the phase factor in \eqref{Reb2l0}. 
For our purpose, it is convenient to introduce a regulator by replacing $\ell_\text{AdS}$ by $\ell_\text{AdS} + i \epsilon$ with $\epsilon > 0$.%
\footnote{We may regularize the expression with the opposite sign $\ell_{\text{AdS}}-i\epsilon$. In the case, the series expansion becomes the sum of negative $n=-1,-2,\ldots$. A different regularization is discussed in Appendix \ref{app:regularization}.}
This implies that
\begin{align}
    \text{Im} \, b^{-2} \sim \text{Im} \, \frac{c}{6} \sim \text{Im}\frac{\ell_\text{AdS} + i \epsilon}{4 G} > 0 \, .
\end{align}
With the regulator, we can rewrite
\begin{align}\label{series}
\frac{1}{ e^{- \frac{\pi i }{b^2}(1 - 2 \eta)} - e^{\frac{\pi i }{b^2}(1 -2 \eta)} } \sim  e^{\pi i  (1 - 2 \eta)\frac{\ell_\text{AdS}}{4G}}\sum_{n=0}^\infty e^{  n \pi i  (1 - 2 \eta)\frac{\ell_\text{AdS}}{2G}} \, .
\end{align}
The partition function can be decomposed as
\begin{align}\label{AdSexpansion}
 \mathcal{Z}_\text{AdS} = \sum_{n=0}^\infty \mathcal{Z}_n
\end{align} 
with
\begin{align}
 \mathcal{Z}_n \sim e^{\frac{\ell_\text{AdS}}{4G}(2 n + 1) \pi i (1 - 2 \eta) }  |z_{12}|^{- \frac{\ell_\text{AdS}}{2G}\eta (1 - \eta)   } 
e^{ - \frac{\ell_\text{AdS}}{2G}[ (1 - 2 \eta) \ln (1 - 2 \eta) - (1 - 2 \eta) ]  } \, .
\end{align}
The expression suggests that there are infinitely many saddles labeled by $n=0,1,2,\ldots$. The dependence on the label $n$ appears in the pure phase factor as
\begin{align} \label{eq:eachpf}
\mathcal{Z}_n \sim \Theta_n \mathcal{Z}_0
\end{align}
with
\begin{align} \label{eq:phase}
\Theta_n  = e^{\frac{\ell_\text{AdS}}{2G} n  \pi i (1 - 2 \eta)}
\end{align}
and
\begin{align} \label{eq:Z0}
\mathcal{Z}_0 \sim e^{\frac{\ell_\text{AdS}}{4G}\pi i (1 - 2 \eta)} |z_{12}|^{- \frac{\ell_\text{AdS}}{G}\eta (1 - \eta)  } 
e^{ - \frac{\ell_\text{AdS}}{2G} [ (1 - 2 \eta) \ln (1 - 2 \eta) - (1 - 2 \eta) ] } \, .
\end{align}

\subsection{Semi-classical saddles in three-dimensional gravity}\label{sec:complexsaddles}

In the previous subsection, we interpret the semi-classical expressions of Liouville field theory in terms of gravity theory. In particular, we found that the semi-classical saddles of gravity theory depends on a parameter $n \in \mathbb{Z}$. In the case of positive cosmological constant, the label takes the value of $n=-1,0$ and the dependence appears in contributions to the Gibbons-Hawking entropy.  In the case of negative cosmological constant, the label takes the value of $n=0,1,2,\ldots$ in the current way of regularization, and the dependence appears in relative phases among the different contributions to the partition function. In this subsection, we derive the dependence on the integer parameter $n$ purely in terms of gravity theory.
For simplicity we set $\eta = 0$ for a while. Namely, we consider the wave functional of universe or the gravity partition function without any insertion of conical defects. As discussed in section \ref{sec:claads}, it is not difficult to generalize the analysis to the case with $\eta \neq 0$ or higher point functions once we understand the case without any insertions.

We start by reviewing the result of \cite{Chen:2023prz,Chen:2023sry}, where gravity theory on dS$_3$ was investigated.
We are interested in the geometry of the type used for the no-boundary proposal by Hartle and Hawking \cite{Hartle:1983ai}. Following \cite{Witten:2021nzp}, we consider a complexified metric of three-sphere with the ansatz,
\begin{align} \label{eq:ansatzdS}
 ds^2 = \ell_\text{dS}^2 \left[ \left(\frac{d\theta (u)}{du}\right)^2 d u^2 + \cos ^2 \theta  (u) d \Sigma^2 \right]\, .
\end{align}
Since we are working  without any insertions of conical defects, we set $d \Sigma^2$ as the metric of $\mathbb{S}^2$. Note that we only complexify one direction, which is represented by a holomorphic function $\theta (u)$. The metric of $\mathbb{S}^3$ is obtained by setting $\theta =u$, while the metric of dS$_3$ is given by $\theta = i u$. For no-boundary proposal, we require that the universe starts from nothing and approaches to dS$_3$. We may set the range of the parameter $u$ as $0 \leq u < \infty$. At $u=0$, we require that the universe vanishes, which can be realized with $\theta = (n + 1/2) \pi$. For $u \to \infty$, the universe becomes de Sitter space by setting $\theta = i u$. We can glue the two geometries.
We may set 
\begin{align}
\theta = \left(n + \frac{1}{2} \right) \pi (1 - u) \quad (0 \leq u \leq 1) \, , \quad \theta = i (u - 1 )  \quad ( u  > 1 ) \, . 
\end{align}
In case of pure gravity, the wave functional of universe \eqref{eq:dSwf} is given by
\begin{align} \label{eq:wf}
 \Psi[h ] = \int \mathcal{D} g \exp (- I_\text{EH} - I _\text{GH} - I_{\text{CT}}) \, .
\end{align}
We set $h$ as the metric of $\mathbb{S}^2$ from the ansatz \eqref{eq:ansatzdS}. As discussed above,
the wave functional can be put of the form 
\begin{align}
\Psi[h] \sim \sum_n \exp (S_\text{GH}^{(n)}/2 + i \mathcal{I}^{(n)}) \, .
\end{align}
The sum is over the family of geometries corresponding to saddle points.
The term $S^{(n)}_\text{GH}$ comes from the Euclidean geometry realized for $0 \leq u \leq 1$ and the phase factor $\exp (i \mathcal{I}^{(n)})$ from the Lorentzian region realized for $u > 1$. The contribution to $S_\text{GH}^{(n)}$ from the geometry labeled by $n$ can be evaluated as
\begin{align} \label{eq:GH}
S_\text{GH}^{(n)} = \frac{(2 n + 1) \pi \ell_\text{dS}}{2 G} \, ,
\end{align}
which is the same as \eqref{eq:SGH}. Thus,
the CFT result is reproduced when the geometry is summed over $n = -1,0$.

We can make similar arguments for the AdS$_3$ case as well. We consider an ansatz for the metric of complexified Euclidean AdS$_3$, 
\begin{align} \label{eq:ansatzAdS}
 ds^2 = \ell_\text{AdS}^2 \left[ \left(\frac{d\theta (u)}{du}\right)^2 d u^2 + \sinh ^2 \theta  (u) d \Sigma^2 \right]\, .
\end{align}
We work without any insertions of conical defects and set $d \Sigma^2$ as the metric of $\mathbb{S}^2$. As before, we only complexify one direction represented by  a holomorphic function $\theta (u)$. We choose the range of parameter $u$ as $0 \leq u < \infty$.
We assign the boundary condition of asymptotic Euclidean AdS$_3$, which means that $\theta = u$ for $u \to \infty$. Further, we  assign the condition where the geometry truncates at $u=0$. We thus require that $\theta = n \pi i$ at $u =0$. We may glue the two geometries by setting 
\begin{align}
\theta =  n \pi i (1 - u) \quad (0 \leq u \leq 1) \, , \quad \theta = (u - 1 ) \quad  ( u > 1) \, . 
\end{align}
The geometry is given by the usual Euclidean AdS$_3$ for $u > 1$.
However, the geometry becomes multiply wrapped $\mathbb{S}^3$ with an imaginary radius $i \ell_\text{AdS}$ or with three Lorentzian time directions for $0 \leq u \leq 1$. The two geometries may be connected at a point and the resultant geometry should be unphysical for non-zero $n$. In case of pure gravity, the partition function \eqref{eq:AdSpf} is written as
\begin{align} \label{eq:pf}
 \mathcal{Z}_\text{AdS}[h] = \int \mathcal{D} g \exp (- I_\text{EH} - I _\text{GH} - I_{\text{CT}}) \, .
\end{align}
As in \eqref{AdSexpansion0}, we can expand the gravity partition function by the contributions from the saddles labeled by an integer parameter $n$.
The region $0 \leq u \leq 1$ leads to the continuation of $n$ three-spheres with radius $i \ell_\text{AdS}$. The phase factor contribution from the saddle with $n$ to the gravity partition function is
\begin{align}
 e^{\frac{n \pi i \ell_\text{AdS}}{2 G}} \, .
\end{align}
The CFT result can be reproduced when the geometry is summed over $n = 0,1,\ldots$ up to an overall factor.

We reinterpret the above results with the homotopy argument in terms of Chern-Simons formulation of gravity theory.
We first recall the analysis of \cite{Chen:2023prz,Chen:2023sry} for the dS$_3$ case.
Let us start from Lorentzian dS$_3$, which may be described by a hyper-surface in a flat space-time as
\begin{align}
 - X_0^2 + X_1^2 + X_2 ^2 + X_3^2 = \ell_\text{dS}^2 \, . \label{eq:hyds}
\end{align}
We glue this geometry to Euclidean dS$_3$ (or three-sphere) at $X_0 = 0$, which may be given by
\begin{align}
 {\tilde X}_0^2 + X_1^2 + X_2 ^2 + X_3^2 = \ell_\text{dS}^2 \, .
\end{align}
As in \eqref{eq:wftopf}, we examined the square of absolute value of wave functional to evaluate gravity partition function.
Combining two half-spheres, we obtain one full-sphere.
The gravity theory on the full three-dimensional sphere can be described by $SU(2) \times SU(2)$ Chern-Simons theory. When the gauge configuration corresponds to the $\mathbb{S}^3$ geometry, then it has a non-trivial bulk winding number (or Chern-Simons invariant). Applying a large gauge transformation, we can construct a different gauge configuration with a different bulk winding number.
In the case with an integer Chern-Simons level $m \in \mathbb{Z}$, the large gauge transformation is a symmetry of the theory as it does not change the value of the path integral $e^{i m I_\text{CS}}$. However, in case with non-integer Chern-Simons level, the large gauge transformation is not a symmetry of the theory anymore, and we should determine which set of configuration we should choose. Fortunately, in the dS$_3$ case, only the physical configuration with bulk winding number one (or its opposite orientation with bulk winding number minus one) appears in the semi-classical expression of wave functional of universe. Therefore, the wave functional of universe can be obtained by summing over possible physical geometries.
See \cite{Witten:2021nzp} for related arguments.

Let us apply a similar argument to the AdS$_3$ case.
We start from an Euclidean AdS$_3$ and analytically continue the Euclidean geometry to an Lorentzian one. We describe the Euclidean AdS$_3$ by a hyper-surface on a flat space-time as
\begin{align}
 \tilde{X}_0^2 + X_1^2 + X_2 ^2 - X_3^2 = -  \ell_\text{AdS}^2 \, . \label{eq:hyeads}
\end{align}
The hyper-surface can be obtained by replacing $\ell_\text{dS}$ in \eqref{eq:hyds} with $i\ell_\text{AdS}$.
As in the dS$_3$ case, it is natural to continue the Euclidean AdS$_3$ into a Lorentzian AdS$_3$ as
\begin{align}
 - X_0^2 + X_1^2 + X_2 ^2 - X_3^2 = -  \ell_\text{AdS}^2 \, .
\end{align}
We may glue the two geometries at $i \tilde X_0  = X_0 = 0$. However, the gauge configuration corresponding to the geometry has only trivial bulk winding number. Therefore, there is no corresponding large gauge transformation. Moreover, the volume of the Lorentzian geometry is infinity, and it is difficult to explain the phase factor depending on the integer parameter $n$ labeling the saddle points.
We may perform a different continuation through $X_3 = i \tilde{X}_3$ as 
\begin{align}
 \tilde{X}_0^2 + X_1^2 + X_2 ^2 + \tilde{X}_3^2 = - \ell_\text{AdS}^2  \, .
\end{align}
The geometry is given by a three-sphere with imaginary radius $i \ell_\text{AdS}$ as appeared above. For \eqref{eq:hyeads} with real coordinate, we should set $|X_3| \geq \ell_\text{AdS}$. Thus the two geometries cannot be glued at $X_3 = i \tilde{X}_3 = 0$. Instead, we may glue at $X_3 = i \tilde{X}_3 = \ell_\text{AdS}$, which is realized by $\tilde X_0 = X_1 = X_2 = 0$. In the Chern-Simons formulation of gravity theory, the imaginary radius does not matter as it can be realized by an imaginary level, see, e.g., appendix A of \cite{Chen:2023sry}. The Chern-Simons configuration corresponding to the geometry indeed has a non-trivial winding number (i.e.\ $\pi_3 (\mathbb{S}^3) = \mathbb{Z}$) and this type of configuration can be generated by a large gauge transformation. 
The correspondence between the label of Liouville saddle and the topological number of Chern-Simons description was already pointed out in \cite{Harlow:2011ny}.

\section{Mini-superspace approach to semi-classical gravity}
\label{sec:minisuperspace}

In the previous section, we determined the set of saddle points in three-dimensional gravity theory from dual CFT results. It is desired to see how such saddles are selected in the path integral for the gravity theory or equivalently which contours of path integral we should take. It is however difficult task to do full gravity analysis. In this section, we shall instead take mini-superspace approach to three-dimensional gravity with positive or negative cosmological constant.

\subsection{Wave functional of dS spacetime}
\label{sec:dSmini}

We first consider the positive cosmological constant case with action
\begin{align}
    I=-\frac{1}{16\pi G}\int d^3x\sqrt{g}\left(R-\frac{2}{\ell_{\text{dS}}^2}\right)+I_{\text{bdy}}\,.
\end{align}
Note that the boundary term $I_{\text{bdy}}$ here is not restricted to the ordinary Gibbons-Hawking term \eqref{eq:IGH}. The data of boundary conditions and boundary terms will be discussed later.
Even though we are interested in the Lorentzian dS$_3$ geometry, we adopt the Euclidean notation. Furthermore, the counter term 
\begin{equation}
    I_{\text{CT}}= \frac{1}{8\pi G} \int d^2x \sqrt{h}\sqrt{-\frac{1}{\ell_{\rm dS}^2}}
\end{equation}
at the outer boundary $\tau=1$ will be needed to remove the divergence. 
In the mini-superspace approach, we often take an isotropic ansatz of metric
\begin{align}\label{minidS}
    ds^2=\ell_{\text{dS}}^2\left[N(\tau)^2 d\tau^2+a(\tau)^2d\Omega_2\right]\,,
\end{align}
in the normalized range $0\le\tau\le1$. 

The quantity of our interest is the wave functional of universe:
\begin{align}
    \Psi=\int\mathcal{D}g_{\mu\nu}\, e^{-I-I_{\text{CT}}}\,.
\end{align}
The domain of the path integral should be taken as the set of metrics in which the diffeomorphism equivalent metrics are identified. It has been shown in \cite{Halliwell:1988wc} that after the gauge fixing in the path integral with the mini-superspace ansatz \eqref{minidS}, the resulting path integral takes the form
\begin{align}\label{WFoU}
    \Psi=\int_{\mathcal{C}}dN\int \mathcal{D}a\, e^{-I[a;N]-I_{\text{CT}}}\,
\end{align}
where the action is\footnote{Here we have an ambiguity of choosing a branch of the square root. Throughout this paper, we take it as satisfying $\sqrt{N^2}=N$. }
\begin{align}\label{dSminiaction}
    I[a;N]=-\frac{\ell_{\text{dS}}}{2G}\int_0^1d\tau \,N\left(\frac{1}{N^2}\left(\frac{da}{d\tau}\right)^2-a^2+1\right)+ (\text{boundary contributions})\,,
\end{align}
i.e.\ the $\tau$-dependence of $N(\tau)$ is removed and the resulting form of $\Psi$ includes the ordinary integral for $N$. 

Here the remaining data to be input for the expression \eqref{WFoU} are the boundary conditions of $a(\tau)$ at the outer boundary $\tau=1$ and the inner boundary $\tau=0$, and the integration contour $\mathcal{C}$ of $N$. On the outer boundary, we only consider the Dirichlet boundary condition $a(1)=a_1$ with a positive real constant $a_1>1$. The inner boundary is more non-trivial: In the dS$_4$ case, the Dirichlet boundary condition is considered in e.g.\ \cite{Halliwell:1988ik} and the Neumann boundary condition is proposed in \cite{Feldbrugge:2017kzv}. Hence we will consider both the Dirichlet-Dirichlet and Neumann-Dirichlet cases as the boundary conditions at the inner boundary and the outer boundary, respectively. 

The second information, the integration contour $\mathcal{C}$, has also been discussed for a long time. When we take a contour asymptotically close to the positive imaginary axis, which seems to give an asymptotically dS spacetime, the integral \eqref{WFoU} is not well-defined since the integrand does not converge along the contour. One way to define a well-defined integral is to extend the contour to a path in the complex plane $\mathbb{C}$ and apply the Picard-Lefschetz theory to find an appropriate contour. The Picard-Lefschetz theory is usually applied to an analytic continuation of an integral when a parameter involved in the integrand is extended to a complex value. In such situations, of course, the original integral before the continuation is well-defined. 
{A peculiar point in the gravitational path integral is that there is no canonical way to choose $\mathcal{C}$ due to the lack of knowledge about quantum gravity, apart from the physical interpretations of the saddle point values of $\Psi$. Our novel point is that the contour can be determined such that the saddle point values match with the CFT results via dS holography.}

\subsubsection{Dirichlet-Dirichlet}
First we consider the case where the Dirichlet boundary conditions are imposed both on $\tau=0,1$. In this case, we do not need any additional boundary contributions to \eqref{dSminiaction}. Therefore we analyze the action:
\begin{align}
    I[a;N]=-\frac{\ell_{\text{dS}}}{2G}\int_0^1d\tau \,N\left(\frac{1}{N^2}\left(\frac{da}{d\tau}\right)^2-a^2+1\right)\,.
\end{align}
The equation of motion with respect to $a$ is given by
\begin{align}
    \frac{d^2a}{d\tau^2}+N^2a=0\,.
\end{align}
A solution to this equation with the boundary conditions $a(0)=0,a(1)=a_1$ imposed is
\begin{align}\label{dSabar}
    \bar{a}^{(N)}(\tau)=\frac{a_1}{\sin N}\sin\left(N\tau\right)\,,
\end{align}
which also leads to
\begin{align}
    I[\bar{a}^{(N)};N]=-\frac{\ell_{\text{dS}}}{2G}\left(N+a_1^2\cot N\right)\,.
\end{align}

Next we evaluate the perturbative contribution around the saddle $\bar{a}^{(N)}(r)$. Expanding $a(r)$ as $a(r)=\bar{a}^{(N)}(r) +A(r)$ with $A(0)=A(1)=0$, we find
\begin{align}
\begin{aligned}
    \Psi&=\int_{\mathcal{C}}dN\,e^{-I[\bar{a}^{(N)};N]-I_{\text{CT}}}\int\mathcal{D}A\,\exp\left[\frac{\ell_{\text{dS}}}{2GN}\int_0^1d\tau\, A(\tau)\left(-\frac{d^2}{d\tau^2}-N^2\right)A(\tau)\right]\\
    &=\int_{\mathcal{C}}dN\,e^{-I[\bar{a}^{(N)};N]-I_{\text{CT}}}\Det\left[\frac{\ell_{\text{dS}}}{2\pi GN}\left(\frac{d^2}{d\tau^2}+N^2\right)\right]^{-\frac{1}{2}} \ .
    \end{aligned}
\end{align}
We would like to calculate the functional determinant. This determinant looks similar to the one-loop contribution of free scalar field, while the mass and the overall constant take complex values. See also \cite{Denef:2009kn} for more detailed discussion of such an analytic continuation. We solve the eigenvalue equation
\begin{align}
    \frac{\ell_{\text{dS}}}{2\pi GN}\left(A''+N^2A\right)=\lambda A\,,
\end{align}
under the boundary conditions $A(0)=A(1)=0$. 
Then we find the spectrum 
\begin{align}
    \lambda_m=\frac{N\ell_{\text{dS}}}{2\pi G}-\frac{m^2\pi\ell_{\text{dS}}}{2 GN} \quad (m=0,1,\ldots)\,.
\end{align}
The functional determinant is formally expressed in terms of the eigenvalues as  
\begin{align}
    \Det\left[\frac{\ell_{\text{dS}}}{2\pi GN}\left(\frac{d^2}{dr^2}+N^2\right)\right]=\prod_{m=0}^{\infty}\lambda_m=\frac{N\ell_{\text{dS}}}{2\pi G}\prod_{m=1}^\infty\left(\frac{N\ell_{\text{dS}}}{2\pi G}-\frac{m^2\pi\ell_{\text{dS}}}{2GN}\right) \ .
\end{align}
To evaluate the infinite product, which needs to be regularized due to the divergence, it is useful to consider the ratio
\begin{align}\begin{aligned}
    \frac{\prod_{m=1}^\infty\left(\frac{N\ell_{\text{dS}}}{2\pi G}-\frac{m^2\pi\ell_{\text{dS}}}{2 GN}\right)}{\prod_{m=1}^\infty\left(-\frac{m^2\pi\ell_{\text{dS}}}{2 GN}\right)}=\prod_{m=1}^\infty\left(1-\frac{N^2}{\pi^2m^2}\right)=\frac{1}{N}\sin N
\end{aligned}\end{align}
where we have used a formula $\sinh z=z\prod_{m=1}^\infty(1+\frac{z^2}{m^2\pi^2})$. To evaluate the denominator, we use the zeta function regularization
\begin{align}
    \prod_{m=1}^\infty\left(-\frac{m^2\pi\ell_{\text{dS}}}{2 GN}\right)=e^{-\zeta_\lambda'(0)},
\end{align}
where 
\begin{align}\label{zetalambda}
    \zeta_\lambda(s):=\sum_{m=1}^\infty\left(-\frac{m^2\pi\ell_{\text{dS}}}{2GN}\right)^{-s}=\left(-\frac{\pi\ell_{\text{dS}}}{2GN}\right)^{-s}\zeta(2s)
\end{align}
with the Riemann zeta function $\zeta(s)$. By using the identity \eqref{zetalambda}, we have
\begin{align}
    \zeta_\lambda'(0)=\log\left(-\frac{2GN}{\pi\ell_{\text{dS}}}\right) \zeta(0)+2\zeta'(0)=-\log\left(-2\sqrt{\frac{2\pi GN}{\ell_{\text{dS}}}}\right)\,.
\end{align}
Summarizing the above calculations, we obtain the regularized functional determinant
\begin{align}\label{DetdS}
    \Det\left[\frac{\ell_{\text{dS}}}{2\pi GN}\left(\frac{d^2}{dr^2}+N^2\right)\right]= - 2\sqrt{\frac{\ell N}{2\pi G}}\sin N\,.
\end{align}
Ignoring the irrelevant overall constant, we arrive at the expression of the partition function
\begin{align}\label{DDwithfluc}
    \Psi=\int_{\mathcal{C}}dN\,\left(\frac{1}{\sqrt{N}\sin N}\right)^{\frac{1}{2}}e^{-I[\bar{a}^{(N)};N]-I_{\text{CT}}} \ .
\end{align}
Though the pre-factor in front of the exponential will not give any effects on the saddle points, it may contain information about singularities or branch cuts. In the current case, the pre-factor does not have any poles, but $N=\pi\mathbb{Z}$ are the branch points coming from $\sqrt{\sin N}$.

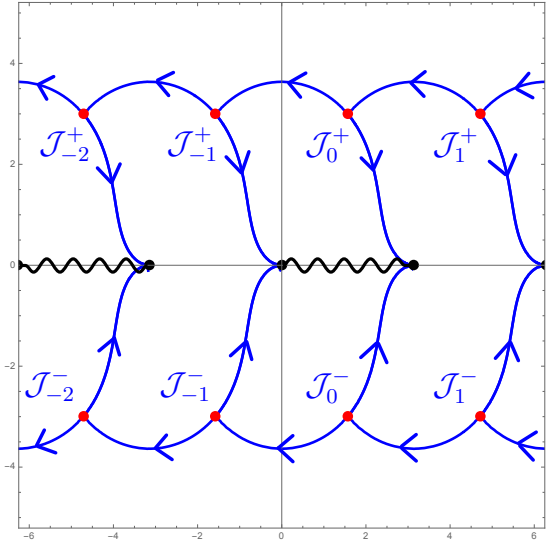
\begin{figure}
    \centering
    \begin{tikzpicture}[xshift=-1cm]
        \node at (0,0) {\includegraphics[width=0.45\linewidth]{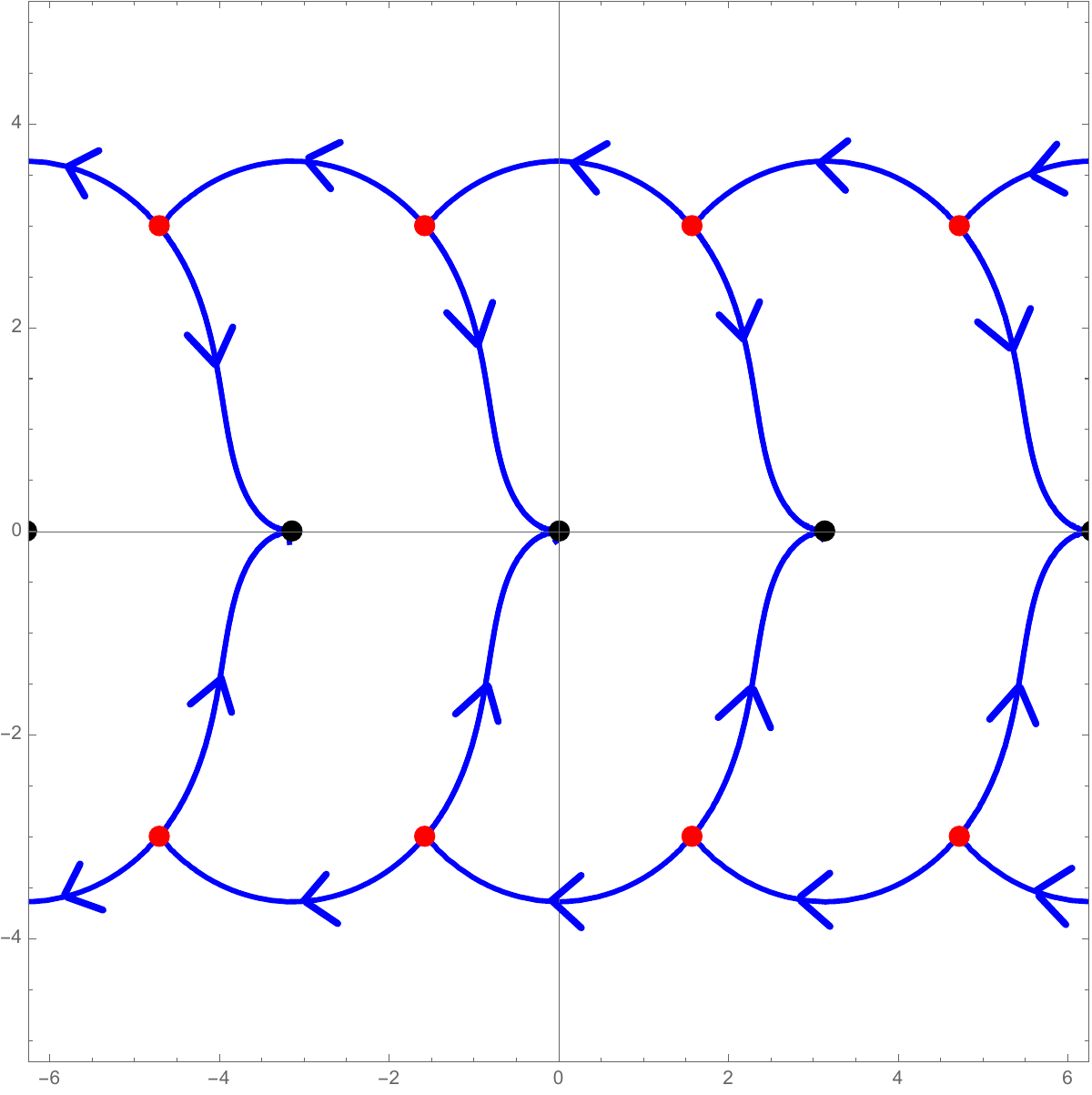}};
        \begin{scope}
            \node[blue] at (0.7,1.7) {\large $\mathcal{J}^+_{0}$};
            \node[blue] at (2.4,1.7) {\large $\mathcal{J}^+_{1}$};
            \node[blue] at (-1.1,1.7) {\large $\mathcal{J}^+_{-1}$};
            \node[blue] at (-2.8,1.7) {\large $\mathcal{J}^+_{-2}$};
        \end{scope}
        \begin{scope}
            \node[blue] at (0.7,-1.5) {\large $\mathcal{J}^-_{0}$};
            \node[blue] at (2.4,-1.5) {\large $\mathcal{J}^-_{1}$};
            \node[blue] at (-1.2,-1.5) {\large $\mathcal{J}^-_{-1}$};
            \node[blue] at (-3,-1.5) {\large $\mathcal{J}^-_{-2}$};
        \end{scope}
        \begin{scope}
            \draw[very thick,decorate,decoration=snake] (0.1,0.1)--(1.9,0.1);
            \draw[very thick,decorate,decoration=snake] (-1.7,0.1)--(-3.4,0.1);
        \end{scope}
    \end{tikzpicture}
    \caption{The steepest descent paths for the dS$_3$ case with Dirichlet-Dirichlet boundary conditions. The red dots denote the saddle points and the wavy lines correspond to branch cuts.}
    \label{fig:dS_DD}
\end{figure}

Now we are at the stage to perform the saddle point approximation. 
From the equation $\partial I[\bar{a}^{(N)};N]/\partial N=0$, the saddle points should satisfy 
\begin{align}\label{saddleeqDD}
    \sin N=\pm a_1\,.
\end{align}
Since we are restricting to the region $a_1>1$, the saddle points are listed as
\begin{align}
    N^+_m&=\left(m+\frac{1}{2}\right)\pi+i\log\left(a_1+\sqrt{a_1^2-1}\right)\,,\\
    N^-_m&=\left(m+\frac{1}{2}\right)\pi-i\log\left(a_1+\sqrt{a_1^2-1}\right)\,,
\end{align}
and the on-shell action on each saddle point is 
\begin{align}
    I[\bar{a}^{(N^+_m)};N^+_m]&=-\frac{\ell_{\text{dS}}}{2G}\left(m\pi-ia_1\sqrt{a_1^2+1}+i\log\left(a_1+\sqrt{a_1^2-1}\right)\right)\,, \\
    I[\bar{a}^{(N^-_m)};N^-_m]&=-\frac{\ell_{\text{dS}}}{2G}\left(m\pi+ia_1\sqrt{a_1^2+1}-i\log\left(a_1+\sqrt{a_1^2-1}\right)\right)\,.
\end{align}
The saddle points and the steepest descents are depicted in figure \ref{fig:dS_DD}. We denote $\mathcal{J}^+_m$ and $\mathcal{J}^-_m$ as the steepest descents associated with $N^+_m$ and $N^-_m$, respectively.
The orientations of steepest descents are defined to be positive in the positive directions of the real and imaginary axes.
We can see that the steepest descent paths in figure \ref{fig:dS_DD} show a curious behavior: A steepest descent path that emanates from a saddle point can end at another saddle point. This is actually a critical behavior in the sense that the theory exactly lies on the Stokes line. Precisely speaking, the saddle point approximation is not well-defined in such a situation (see e.g. \cite{Witten:2010cx,Lehners:2023yrj}). This is actually a specific feature in three dimensions, which cannot be observed in four dimensions \cite{Honda:2024aro}. As we can find from \eqref{DDwithfluc}, there are branch points at $N=\pi \mathbb{Z}$ coming from $\sqrt{\sin N}$. We choose the branch cuts as connecting $N=2n\mathbb{Z}$ and $N=(2n+1)\mathbb{Z}$ as depicted in figure \ref{fig:dS_DD}. 

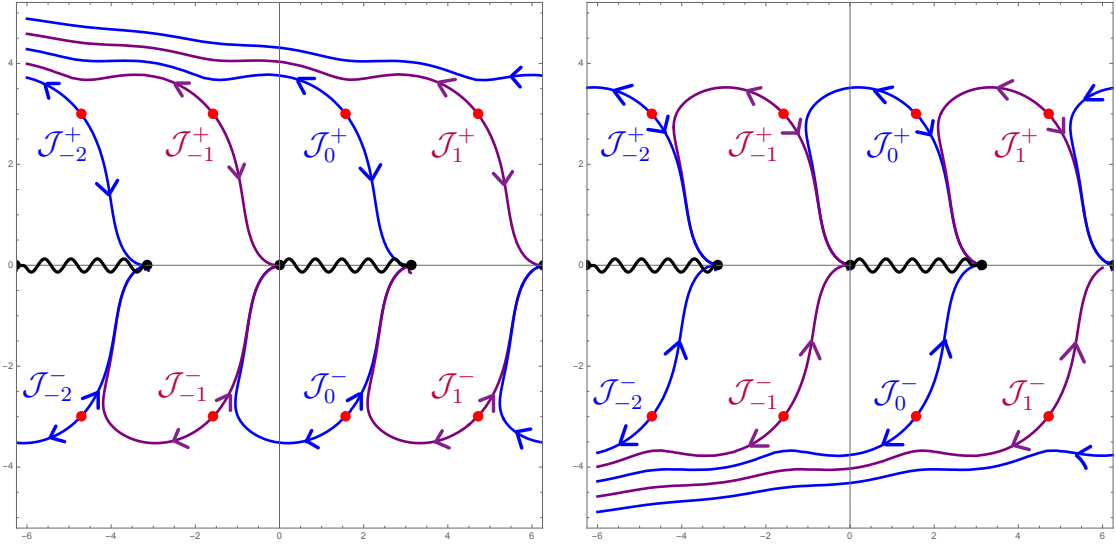
\begin{figure}
    \centering
    \begin{tikzpicture}[xshift=-1cm]
        \node at (0,0) {\includegraphics[width=0.45\linewidth]{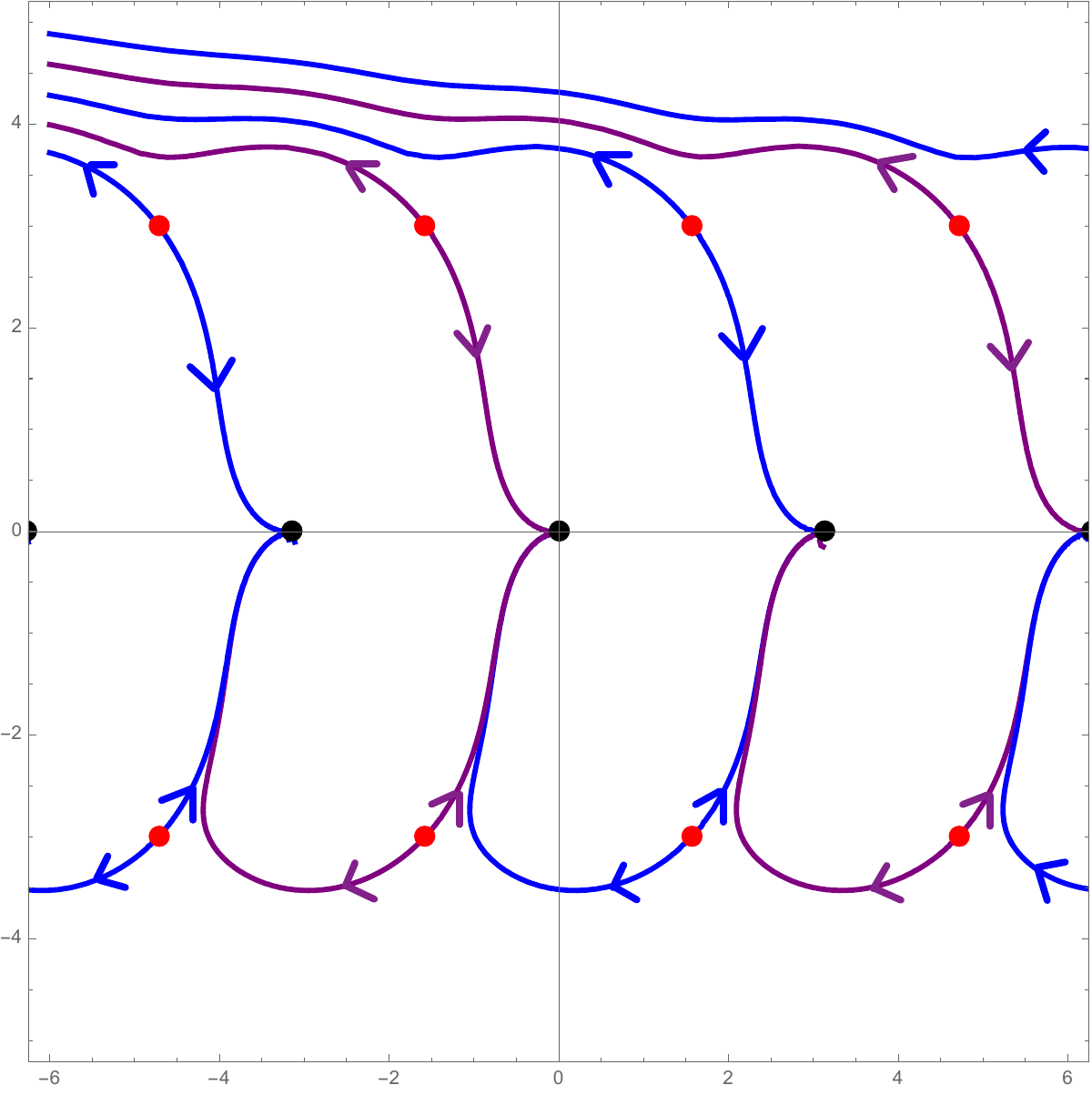}};
        \begin{scope}
            \node[blue] at (0.7,1.7) {\large $\mathcal{J}^+_{0}$};
            \node[purple] at (2.4,1.7) {\large $\mathcal{J}^+_{1}$};
            \node[purple] at (-1.1,1.7) {\large $\mathcal{J}^+_{-1}$};
            \node[blue] at (-2.8,1.7) {\large $\mathcal{J}^+_{-2}$};
        \end{scope}
        \begin{scope}
            \node[blue] at (0.7,-1.5) {\large $\mathcal{J}^-_{0}$};
            \node[purple] at (2.4,-1.5) {\large $\mathcal{J}^-_{1}$};
            \node[purple] at (-1.2,-1.5) {\large $\mathcal{J}^-_{-1}$};
            \node[blue] at (-3,-1.5) {\large $\mathcal{J}^-_{-2}$};
        \end{scope}
        \begin{scope}
            \draw[very thick,decorate,decoration=snake] (0.1,0.1)--(1.9,0.1);
            \draw[very thick,decorate,decoration=snake] (-1.7,0.1)--(-3.4,0.1);
        \end{scope}
    \end{tikzpicture}
    \begin{tikzpicture}
        \node at (0,0) {\includegraphics[width=0.45\linewidth]{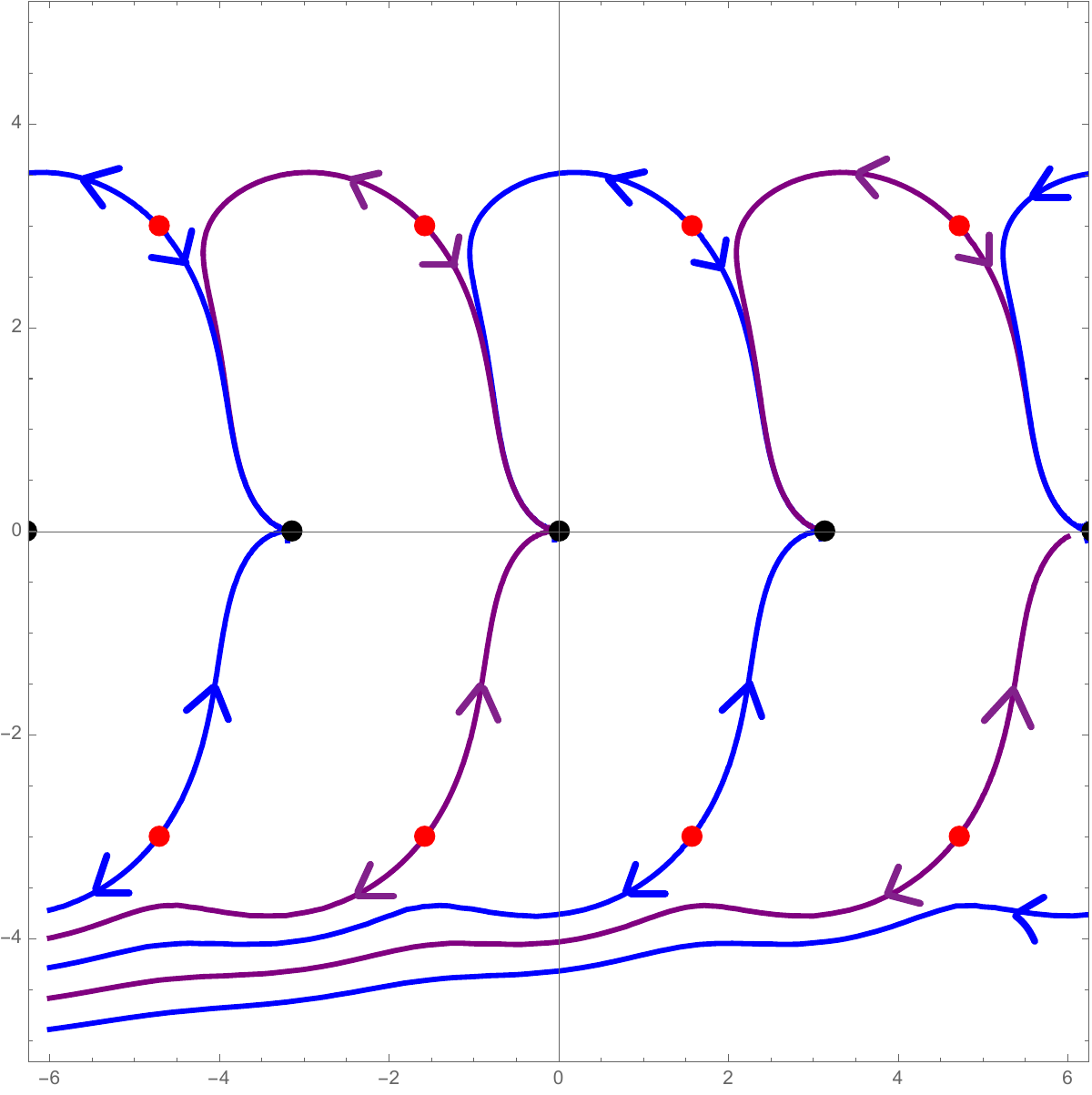}};
        \begin{scope}
            \node[blue] at (0.6,1.7) {\large $\mathcal{J}^+_{0}$};
            \node[purple] at (2.3,1.7) {\large $\mathcal{J}^+_{1}$};
            \node[purple] at (-1.2,1.7) {\large $\mathcal{J}^+_{-1}$};
            \node[blue] at (-2.9,1.7) {\large $\mathcal{J}^+_{-2}$};
        \end{scope}
        \begin{scope}
            \node[blue] at (0.7,-1.6) {\large $\mathcal{J}^-_{0}$};
            \node[purple] at (2.4,-1.6) {\large $\mathcal{J}^-_{1}$};
            \node[purple] at (-1.2,-1.6) {\large $\mathcal{J}^-_{-1}$};
            \node[blue] at (-3,-1.6) {\large $\mathcal{J}^-_{-2}$};
        \end{scope}
        \begin{scope}
            \draw[very thick,decorate,decoration=snake] (0.1,0.1)--(1.9,0.1);
            \draw[very thick,decorate,decoration=snake] (-1.7,0.1)--(-3.4,0.1);
        \end{scope}
    \end{tikzpicture}
    \caption{The left panel shows the plot when $\ell_{\text{dS}}\to \ell_{\text{dS}} + i \epsilon$. The right panel shows the plot when $\ell_{\text{dS}}\to \ell_{\text{dS}} - i \epsilon$. }
    \label{fig:lshift}
\end{figure}

To avoid this problem, we try to slightly shift the parameter $\ell_{\text{dS}} $ in the imaginary direction $\ell_{\text{dS}}\to\ell_{\text{dS}}\pm i\epsilon$, which are depicted in figure \ref{fig:lshift}. The configurations of steepest descent paths highly depend on the sign of the shift. 
When we only focus on the upper half plane, the right panel in figure \ref{fig:lshift} is in the same phase as the AdS$_3$ case as we will see in the next section, while the left panel is very different. 
The transition between these phases causes the Stokes phenomenon. In fact, this Stokes phenomenon looks very similar to that for Gamma function, see e.g. Appendix C of \cite{Harlow:2011ny}. 

Now we have a question: Which direction of the shift, $\ell_{\text{dS}}\to\ell_{\text{dS}} \pm i\epsilon$, is physically appropriate for the dS$_3$ case? One answer is given by the dual CFT point of view via dS/CFT correspondence, where the central charge of the dual CFT is\footnote{Note that our formula for the central charge is different from \cite{Chen:2023prz,Chen:2023sry}, where the Newton constant is already normalized so that the central charge involves all quantum corrections. The notation used here is rather ordinary one, same as the original Brown-Henneaux formula \cite{Brown:1986nw}. }
\begin{align}
    c=-i\frac{3\ell_{\text{dS}}}{2G}+13+\mathcal{O}(G)\,.
\end{align}
In the gravity interpretation, the constant term ``$13$'' comes from the one-loop contributions. This contribution can also be realized as a shift $\ell_{\text{dS}}\to \ell_{\text{dS}}+ i\epsilon$, which corresponds to the left panel of figure \ref{fig:lshift} (Note that we adopt the notation $\ell_{\text{AdS}}=-i\ell_{\text{dS}}$). 

Finally let us discuss the contour $\mathcal{C}$ for the $N$ integral. Our goal is to find a contour that reproduces the CFT calculation \eqref{dSexpansion}. 
Each saddle point $N^{\pm}_m$ gives the leading contribution to the wave functional of universe 
\begin{align}
    \Psi^{\pm}_m\sim e^{\frac{(2m+1)\ell_{\text{dS}}\pi}{4G}}(2a_1)^{\mp i \frac{\ell_{\text{dS}}}{2G}\pm \frac{\epsilon}{2 G}} \ .
\end{align}
Due to $\epsilon$ in the exponent, $\Psi^-_m$ from the saddles in the lower half plane are suppressed in the large $a_1$ approximation.
Naively to preserve future time direction, one would choose the integration contour as $i\mathbb{R}_+$. Though the integral along this contour diverges, it can be deformed to coincide $\mathcal{J}^+_{-1}$. However, this contour does not seem to be physical because $\Psi$ has only the suppressing saddle. 
To match with the CFT result \eqref{dSexpansion}, we have to pick up two saddles $\mathcal{J}^+_0,\mathcal{J}^+_{-1}$ from the upper half plane. Therefore we propose the contour that comes from the positive imaginary infinity, rounds the branch cut lying between $N=0$ and $N=\pi$, and goes back to the positive imaginary infinity. This contour can be deformed to
\begin{align}
    -\mathcal{J}^+_{-1}+\mathcal{J}^-_0+\mathcal{J}^+_0\,,
\end{align}
where the signatures take $-1$ if the orientations of the steepest descent and the contour are opposite. Since the contribution from $\mathcal{J}^-_0$ is suppressed as explained above, the semi-classical approximation of the wave functional of universe is obtained as
\begin{align}
    \Psi\sim\left(e^{\frac{\pi\ell_{\text{dS}}}{4G}}-e^{-\frac{\pi\ell_{\text{dS}}}{4G}}\right)(2a_1)^{i\frac{\ell_{\text{dS}}}{2G}} \ .
\end{align}
Identifying $c^{(g)}=3\ell_{\text{dS}}/2G$, this actually reproduces \eqref{dSexpansion} with $\eta=0$.

It might be difficult to obtain geometrical interpretation of the complex metric realised by a saddle point. We provide supporting arguments in appendix \ref{app:geometry}. Here we simply show that the complex geometry realized as a saddle can be put into the form of the one discussed in subsection \ref{sec:complexsaddles}.
As in \cite{Lehners:2021mah} we may consider a coordinate system which interpolate between the two types of metrics as
\begin{align}\label{Ttau}
T(\tau) =  - \left(n + \frac{1}{2} \right) \pi (1 - \tau)^q + i \ln (2 a_1 ) \tau^q \, .
\end{align}
Due to the Cauchy's theorem, the integral does not depends on the path integral contour over the metrics.
We can see that the time coordinate reduces to the one realized by a saddle with $q = 1$ up to a constant shift and approaches the one used for the no-boundary proposal of Hartle and Hawking \cite{Hawking:1975vcx} for $q \to \infty$.

The behavior of mini-superspace analysis for three-dimensional gravity is highly different from the four dimensional case summarized in \cite{Lehners:2023yrj}. The most distinct point is the infinite number of saddle points. The four dimensional model has finite number of saddle points that can be regarded as saddle points of Airy functions. As described in \cite{Witten:2021nzp}, such a behavior of the number of saddle points depend on whether the spacetime dimension is even or odd. When we analytically continue one coordinate of $D$-dimensional sphere $S^D$ to the no-boundary geometries, there exist infinite saddle points for odd $D$. For even $D$, precisely speaking, the infinite saddles degenerate in the sense that the on-shell actions take the same value among those saddles. This observation is consistent with our analysis. Therefore the extension to higher dimensions are expected to have similar structure to our analysis in odd-dimensional mini-superspace model.

\subsubsection{Neumann-Dirichlet}
So far we have only considered the Dirichlet boundary condition at the initial boundary $\tau=0$. This is the most straightforward boundary condition in order to construct the no-boundary saddle points. 
Nevertheless, it is also possible to impose Neumann boundary condition at $\tau=0$. 
Although the initial value of the scaling factor $a(0)$ then does not always take zero in the path integral, we might adjust the boundary condition so that there exists a saddle point we want. 
Indeed, such attempts have been tried in \cite{DiTucci:2019dji,DiTucci:2019bui}, where the authors construct certain no-boundary saddle points from the other boundary conditions in four dimensions. 
Our purpose to match the mini-superspace calculations with the CFT results has already been accomplished from the Dirichlet-Dirichlet setup in the previous section. 
In this subsection, however, we will try another boundary condition; a Neumann boundary condition $a'(0)=N$. The value of $a'(0)$ has been chosen so that the theory has the physically relevant saddle $\mathcal{J}^+_0$ as in Dirichlet case. The Neumann-Dirichlet condition is, of course, not the only possibility of other boundary conditions. There are various choices, e.g. Ribon boundary condition, as studied in detail in \cite{DiTucci:2019bui}. We leave the study of such other boundary conditions as a future work. 

When we impose a Neumann boundary condition, the boundary term at $\tau=0$ should be modified. 
In terms of $a$ and $N$, the action becomes 
\begin{align}
    I[a;N]=-\frac{\ell_{\text{dS}}}{2G}\int_0^1d\tau\, N\left(\frac{1}{N^2}\left(\frac{da}{d\tau}\right)^2-a^2+1\right)-\frac{\ell_{\text{dS}}}{GN}\left.a\frac{da}{d\tau}\right|_{\tau=0}\,.
\end{align}
A solution to the equation of motion 
\begin{align}
    \frac{d^2a}{d\tau^2}+N^2a=0
\end{align}
with boundary conditions $\left.\frac{da}{d\tau}\right|_{\tau=0}=N$ and $a(1)=a_1$ is 
\begin{align}
    \bar{a}^{(N)}(\tau)=\frac{a_1}{\cos N}\cos(N\tau)+\frac{\sin( N(\tau-1))}{\cos N}\,.
\end{align}
The on-shell action is then 
\begin{align}
    I[\bar{a}^{(N)};N]=-\frac{\ell_{\text{dS}}}{2G}\left[N+\frac{2a_1}{\cos N}-(1+a_1^2)\tan N\right]\,.
\end{align}
{The derivative with respect to $N$ is 
\begin{align}\label{derivI}
    \frac{\partial I[\bar{a}^{(N)};N]}{\partial N}=\frac{\ell_{\text{dS}}}{2G}\frac{(\sin N-a_1)^2}{\cos^2 N}\,.
\end{align}
Differing from the previous case, it turns out that the solutions to $\sin N=a_1$ are the second order saddle points rather than the first ones. Namely, the saddle points have the vanishing second order derivative as well:
\begin{align}
     \frac{\partial^2 I[\bar{a}^{(N)};N]}{\partial N^2} = 0\, .
\end{align}
In terms of the analysis of steepest ascend/descend paths, we consider the fluctuations around each of these saddle points. This requires us to expand to the third order derivatives, which leads to three steepest ascending/descending directions respectively. Such paths are depicted in figure \ref{fig:dS_ND}. In this case, the saddle point that corresponds to $n=-1$ in \eqref{dSexpansion} does not exist. 
Therefore the current boundary condition does not reproduce the CFT result. 
\begin{figure}
    \centering
    \begin{tikzpicture}[xshift=-1cm]
        \node at (0,0) {\includegraphics[width=0.45\linewidth]{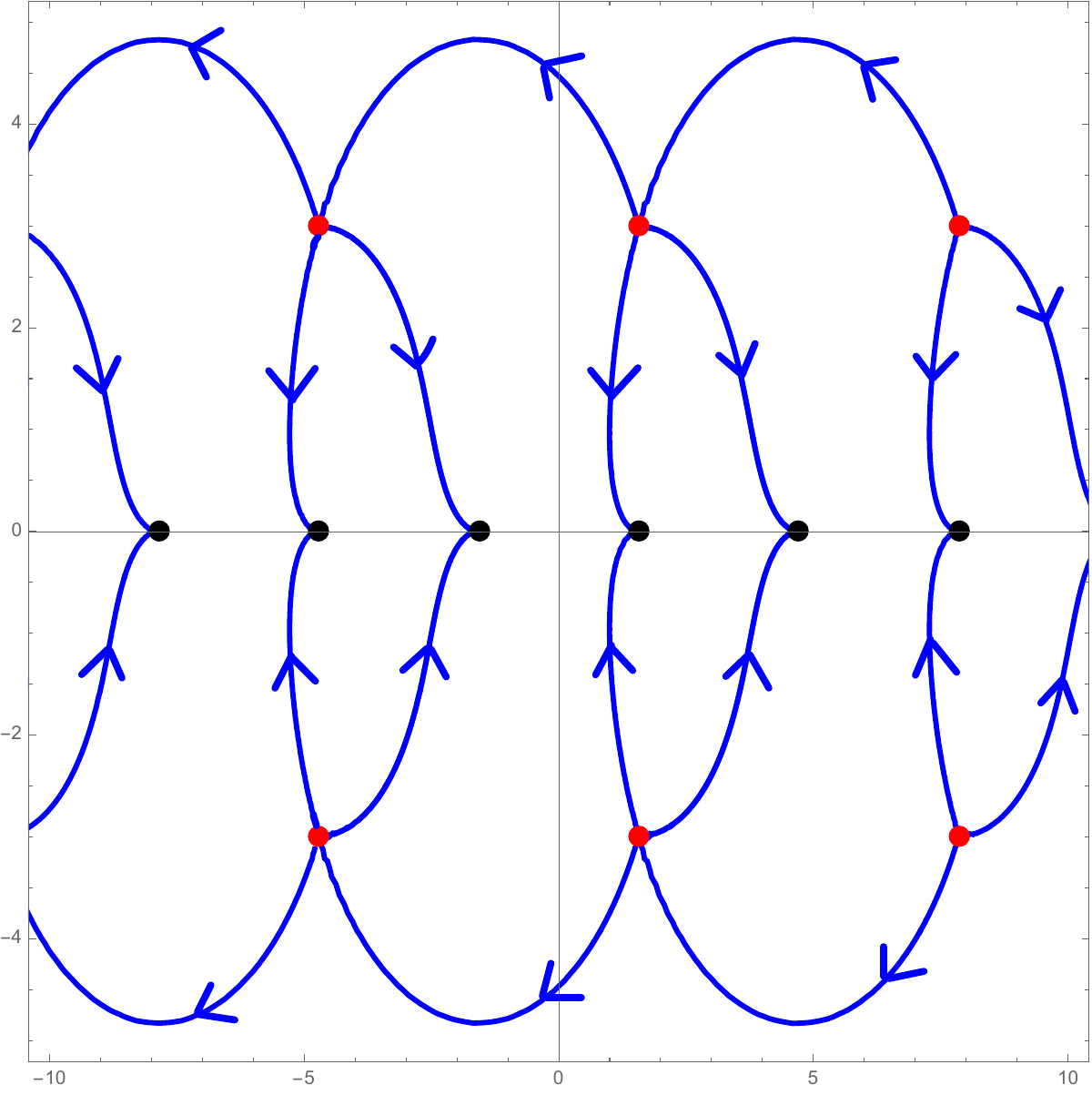}};
    \end{tikzpicture}
    \caption{The steepest descent paths for the dS$_3$ case with Neumann-Dirichlet boundary conditions. The red dots denote the saddle points. }
    \label{fig:dS_ND}
\end{figure}

\subsection{Partition function of AdS gravity}
\label{sec:AdSmini}

Next let us move on to the case with negative cosmological constant $\Lambda=-1/\ell_{\text{AdS}}^2$, where the action is given by
\begin{align}\label{EHaction}
    I=-\frac{1}{16\pi G}\int d^3x\sqrt{g}\left(R+\frac{2}{\ell_{\text{AdS}}^2}\right)+I_{\text{bdy}}\,.
\end{align}
We assume the almost same mini-superspace ansatz as in the dS$_3$ case:
\begin{align}\label{minisuperspace}
    ds^2=\ell_{\text{AdS}}^2\left[N(r)^2dr^2+a(r)^2d\Omega_2^2\right]\,,
\end{align}
where $d\Omega_2^2$ denotes the metric of the unit two-sphere $\mathbb{S}^2$, and $r$ is the radial coordinate normalized as taking the values in the range $0\le r \le 1$. This again has two boundaries at $r=0$ and $r=1$. We will need the counter term at $r=1$ \cite{Balasubramanian:1999re}
\begin{align}
    I_{\text{CT}}=\frac{\ell_{\text{AdS}}}{8\pi G}\int_{r=1}d^2x\sqrt{\gamma}=\frac{\ell_{\text{AdS}}}{2G}a(1)^2
\end{align}
in order to compensate the infinite volume divergence.

We are interested in the following gravitational partition function
\begin{align}
    \mathcal{Z}=\int \mathcal{D}g_{\mu\nu} \,e^{-I-I_{\text{CT}}}\,,
\end{align}
which again reduces to 
\begin{align}\label{pfmini}
    \mathcal{Z}=\int_{\mathcal{C}}dN\int \mathcal{D}a\,e^{-I[a;N]-I_{\text{CT}}}
\end{align}
after the gauge fixing. The action takes the following form
\begin{align}\label{actionmini}
    I[a;N]=-\frac{\ell_{\text{AdS}}}{2G}\int_0^1dr\, N\left(\frac{1}{N^2}\left(\frac{da}{dr}\right)^2+a^2+1\right)+ (\text{boundary contributions})\,.
\end{align}

In the Dirichlet-Dirichlet case, we do not need to add any additional boundary terms to the expression \eqref{actionmini}, therefore we have
\begin{align}
    I[a;N]=-\frac{\ell_{\text{AdS}}}{2G}\int_0^1dr\, N\left(\frac{1}{N^2}\left(\frac{da}{dr}\right)^2+a^2+1\right)\,.
\end{align}
The equation of motion for $a(r)$ is given by 
\begin{align}
    a''-N^2a=0 \ .
\end{align}
We require that the manifold truncates at $r =0$. We thus impose the following Dirichlet boundary conditions:
\begin{align}
    a(0)=0 \ ,\qquad a(1)=a_1\,.
\end{align}
Under these boundary conditions, a solution $\bar{a}(r)$ to the EOM is given by 
\begin{align}\label{DDsolution}
    \bar{a}^{(N)}(r)=\frac{a_1}{\sinh N}\sinh (Nr)\,.
\end{align}
Inserting this solution to the action, we have 
\begin{align}
    I[\bar{a}^{(N)};N]=-\frac{\ell_{\text{AdS}}}{2G}\left[N+a_1^2\coth N\right]\,.
\end{align}

Next we evaluate the perturbative contribution around the saddle $\bar{a}^{(N)}(r)$. Expanding $a(r)$ as $a(r)=\bar{a}^{(N)}(r) +A(r)$ with boundary conditions $A(0)=A(1)=0$, we find
\begin{align}
\begin{aligned}
    \mathcal{Z}&=\int_{\mathcal{C}}dN\,e^{-I[\bar{a}^{(N)};N]-I_{\text{CT}}}\int\mathcal{D}A\,\exp\left[\frac{\ell_{\text{AdS}}}{2GN}\int_0^1drA(r)\left(-\frac{d^2}{dr^2}+N^2\right)A(r)\right]\\
    &=\int_{\mathcal{C}}dN\,e^{-I[\bar{a}^{(N)};N]-I_{\text{CT}}}\Det\left[\frac{\ell_{\text{AdS}}}{2\pi GN}\left(\frac{d^2}{dr^2}-N^2\right)\right]^{-\frac{1}{2}} \ .
\end{aligned}
\end{align}
The calculation of the functional determinant is almost same as the positive cosmological constant case. 
The result is identical to the analytic continuation $\ell_{\text{dS}}\to i\ell_{\text{AdS}},\ N\to iN$ of \eqref{DetdS}, that is,
\begin{align}
    \Det\left[\frac{\ell_{\text{AdS}}}{2\pi GN}\left(\frac{d^2}{dr^2}-N^2\right)\right]=  2\sqrt{\frac{\ell N}{2\pi G}}\sinh N\,
\end{align}
up to the overall sign.
Ignoring the irrelevant overall constant, we obtain the expression of the partition function
\begin{align}
    \mathcal{Z}=\int_{\mathcal{C}}dN\,\left(\frac{1}{\sqrt{N}\sinh N}\right)^{\frac{1}{2}}e^{-I[\bar{a}^{(N)};N]-I_{\text{CT}}} \, .
\end{align}
We can see from the pre-factor in the integral that $N=\pi i\,\mathbb{Z}$ are branch points.

Let us discuss the saddle point approximation. The stationary points satisfying $\partial I[\bar{a}^{(N)};N]/\partial N=0$ are listed as
\begin{align}
    N^+_m&= \arcsinh a_1+\pi i m  \quad ( m\in\mathbb{Z} )\,,\\
    N^-_m&=-\arcsinh a_1+\pi i m \quad ( m\in\mathbb{Z} ) \,,
\end{align}
and the on-shell action for each saddle point is 
\begin{align}
    I[\bar{a}^{(N^+_m)};N^+_m]&=-\frac{\ell_{\text{AdS}}}{2G}\left(im\pi+a_1\sqrt{a_1^2+1}+\arcsinh a_1\right) \ , \\
    I[\bar{a}^{(N^-_m)};N^-_m]&=-\frac{\ell_{\text{AdS}}}{2G}\left(im\pi-a_1\sqrt{a_1^2+1}-\arcsinh a_1\right) \ .
\end{align}
The configurations of the saddle points and the steepest descent paths are shown in figure \ref{fig:DDthimbles_modified}. 
In a way similar to the positive cosmological constant case, some steepest descents flow into other saddle points. 
Therefore the present theory lies on the Stokes ray, which makes the integral ill-defined.  
Furthermore, there is another problem in figure \ref{fig:DDthimbles_modified}: Each plus thimble $\mathcal{J}^+_m$ crosses a branch cut. Since each branch point is surrounded by a thimble $\mathcal{J}^+_m$, the branch cut emanating from the branch point necessarily crosses with $\mathcal{J}^+_m$. Although this does not lead to any ill-definedness of the integral, this would make the analysis highly complicated. 
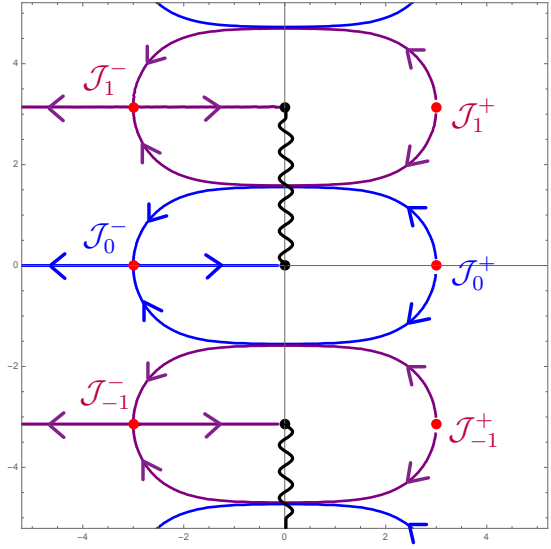
\begin{figure}
    \centering
    \begin{tikzpicture}[xshift=-1cm]
        \node at (0,0) {\includegraphics[width=0.45\linewidth]{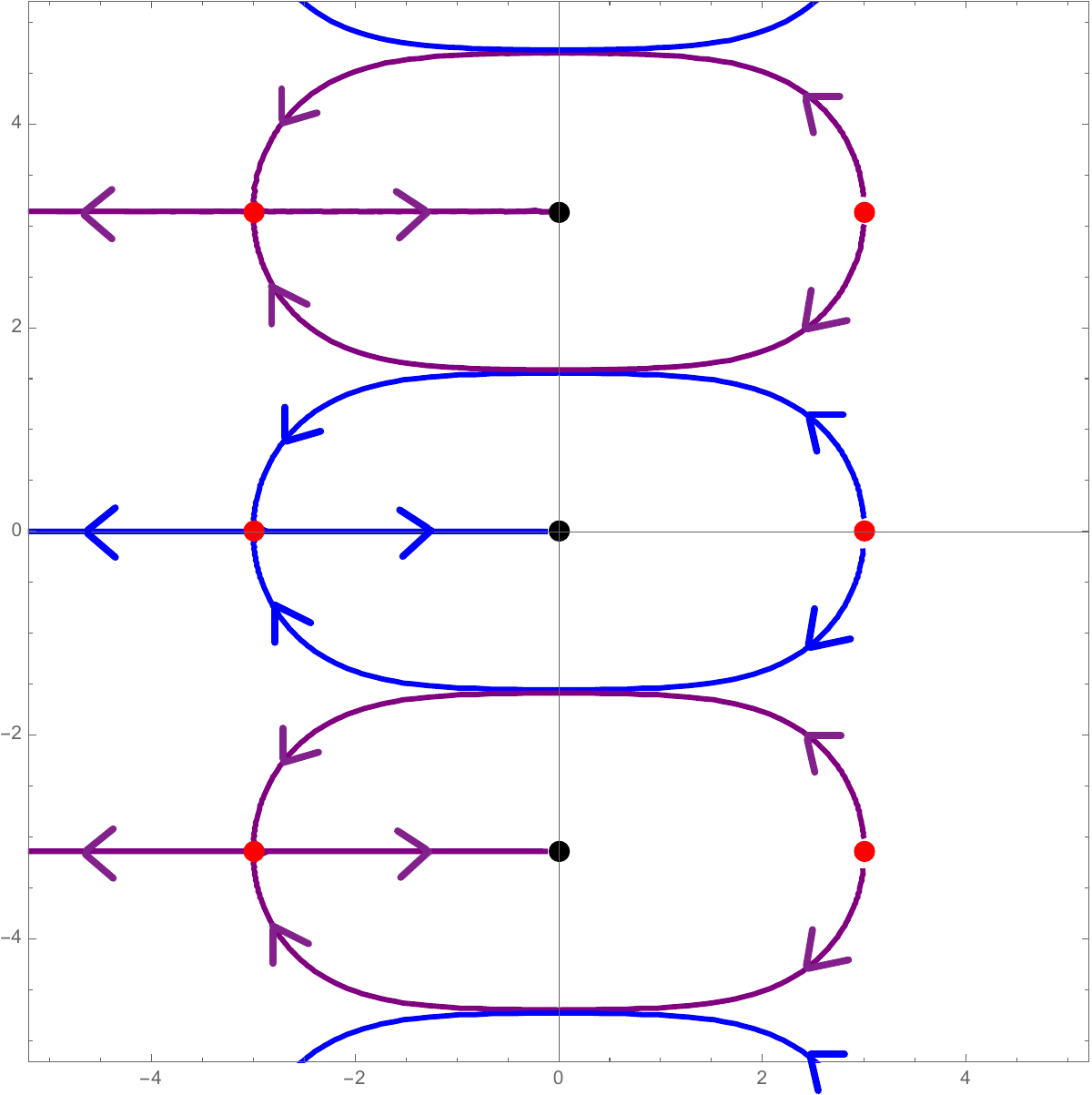}};
        \begin{scope}
            \node[blue] at (2.6,0) {\large $\mathcal{J}^+_0$};
            \node[purple] at (2.6,2.1) {\large $\mathcal{J}^+_1$};
            \node[purple] at (2.6,-2.1) {\large $\mathcal{J}^+_{-1}$};
        \end{scope}
        \begin{scope}
            \node[blue] at (-2.3,0.5) {\large $\mathcal{J}^-_0$};
            \node[purple] at (-2.3,2.6) {\large $\mathcal{J}^-_1$};
            \node[purple] at (-2.3,-1.6) {\large $\mathcal{J}^-_{-1}$};
        \end{scope}
        \begin{scope}
            \draw[very thick,decorate,decoration=snake] (0.1,0.1)--(0.1,2.2);
            \draw[very thick,decorate,decoration=snake] (0.1,-2)--(0.1,-3.4);
        \end{scope}
    \end{tikzpicture}
    \caption{The steepest descent paths for the AdS$_3$ case with Dirichlet-Dirichlet boundary conditions. The red dots denote the saddle points and the wavy lines correspond to branch cuts.}
    \label{fig:DDthimbles_modified}
\end{figure}

Again, we slightly shift the AdS radius in the imaginary direction as $\ell_{\text{AdS}}\to\ell_{\text{AdS}}\pm i\epsilon$. The steepest descent paths after the regularization are depicted in figure \ref{fig:DDthimbles_shift}. After the regularization, the theory leaves the Stokes ray. Furthermore, we can take the branch cut from each branch point so as not to cross any steepest descent paths. In contrast to the positive cosmological constant case, the signature in front of $i\epsilon$ does not change the final result, as we will see explicitly. 
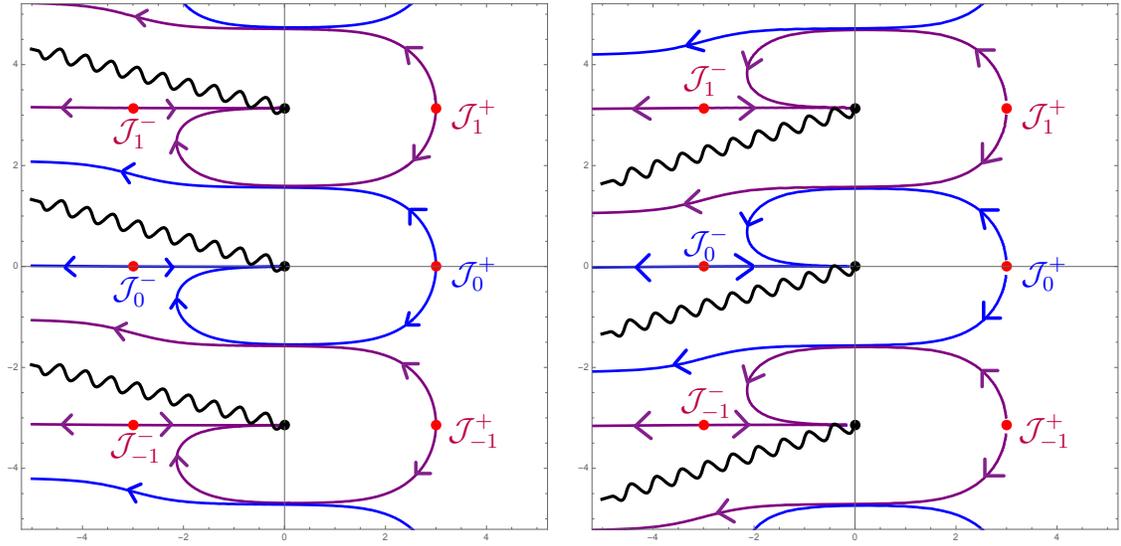
\begin{figure}
    \centering
    \begin{tikzpicture}[xshift=-1cm]
        \node at (0,0) {\includegraphics[width=0.45\linewidth]{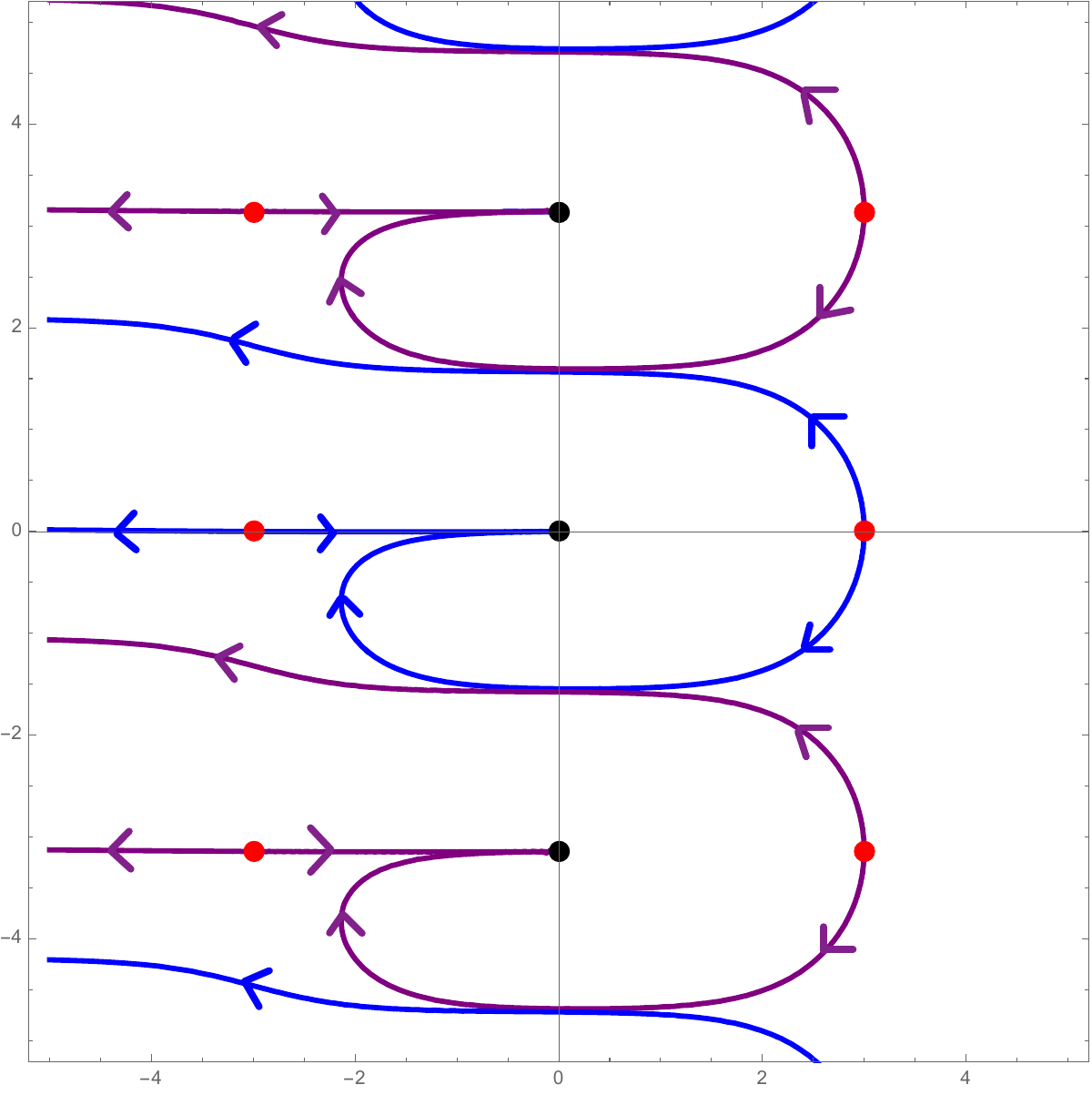}};
        \begin{scope}
            \node[blue] at (2.6,0) {\large $\mathcal{J}^+_0$};
            \node[purple] at (2.6,2.1) {\large $\mathcal{J}^+_1$};
            \node[purple] at (2.6,-2.1) {\large $\mathcal{J}^+_{-1}$};
        \end{scope}
        \begin{scope}
            \node[blue] at (-1.9,-0.2) {\large $\mathcal{J}^-_0$};
            \node[purple] at (-1.9,1.9) {\large $\mathcal{J}^-_1$};
            \node[purple] at (-1.9,-2.3) {\large $\mathcal{J}^-_{-1}$};
        \end{scope}
        \begin{scope}
            \draw[very thick,decorate,decoration=snake] (0.1,0.1)--(-3.3,1);
            \draw[very thick,decorate,decoration=snake] (0.1,-2)--(-3.3,-1.2);
            \draw[very thick,decorate,decoration=snake] (0.1,2.2)--(-3.3,3);
        \end{scope}
    \end{tikzpicture}
    \begin{tikzpicture}
        \node at (0,0) {\includegraphics[width=0.45\linewidth]{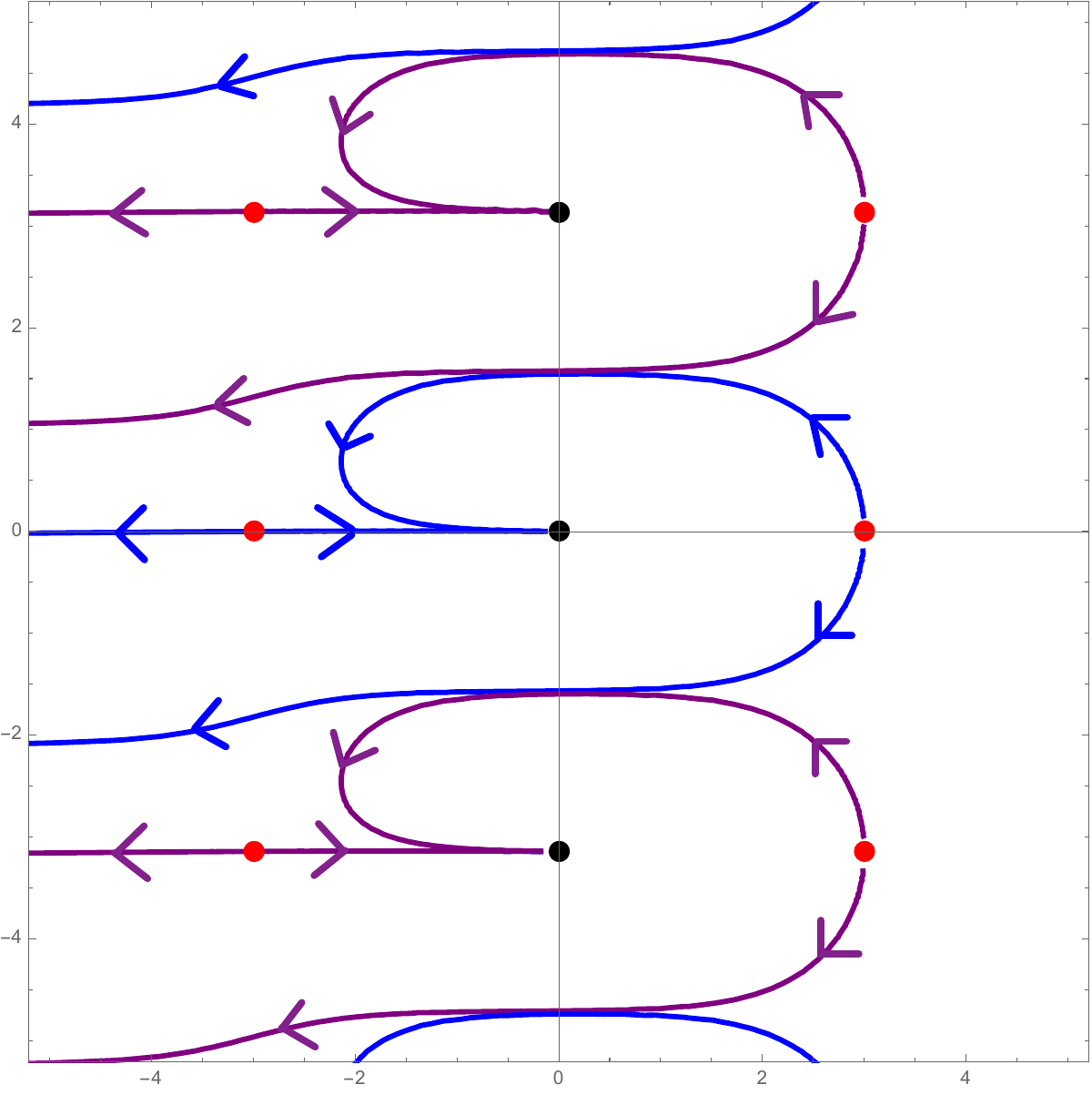}};
        \begin{scope}
            \node[blue] at (2.6,0) {\large $\mathcal{J}^+_0$};
            \node[purple] at (2.6,2.1) {\large $\mathcal{J}^+_1$};
            \node[purple] at (2.6,-2.1) {\large $\mathcal{J}^+_{-1}$};
        \end{scope}
        \begin{scope}
            \node[blue] at (-1.9,0.4) {\large $\mathcal{J}^-_0$};
            \node[purple] at (-1.9,2.6) {\large $\mathcal{J}^-_1$};
            \node[purple] at (-1.9,-1.7) {\large $\mathcal{J}^-_{-1}$};
        \end{scope}
        \begin{scope}
            \draw[very thick,decorate,decoration=snake] (0.1,0.1)--(-3.3,-0.8);
            \draw[very thick,decorate,decoration=snake] (0.1,-2)--(-3.3,-3);
            \draw[very thick,decorate,decoration=snake] (0.1,2.2)--(-3.3,1.2);
        \end{scope}
    \end{tikzpicture}
    \caption{
    The left panel shows the plot when $\ell_{\text{AdS}}\to \ell_{\text{AdS}} + i \epsilon$. The right panel shows the plot when $\ell_{\text{AdS}}\to \ell_{\text{AdS}} - i \epsilon$.}
    \label{fig:DDthimbles_shift}
\end{figure}

The leading contributions from the saddle points $N^+_m,N^-_m$ are 
\begin{align}
    \mathcal{Z}^\pm_m\sim e^{\frac{im\pi\ell_{\text{AdS}}}{2G}}(2a_1)^{\pm \frac{\ell_{\text{AdS}}}{2G}}
\end{align}
in the large $a_1$ regime. We can easily see that the contributions from the minus saddles $N^-_m$ are suppressed. A naive contour is the positive real axis $\mathcal{C}=\mathbb{R}_+$. 
Since the path integral along this contour diverges, we have to deform the contour to be expressed as a summation of some steepest descents. 
The proper deformation depends on the regularization of $\ell_{\text{AdS}}$. 
For $\ell_{\text{AdS}}+i\epsilon$, we deform the contour upwards such that it can be expressed as
\begin{align}
    \mathcal{C}\to\sum_{m=0}^{\infty}\mathcal{J}^+_m-\sum_{m=1}^{\infty}\mathcal{J}^-_m\,,
\end{align}
along which the semi-classical limit of the partition function is
\begin{align}\label{Zupwards}
    \mathcal{Z}\sim \sum_{m=0}^{\infty}e^{\frac{i\pi m(\ell_{\text{AdS}}+i\epsilon)}{2G}}(2a_1)^{\frac{\ell_{\text{AdS}}+i\epsilon}{2G}}=-\frac{2ie^{-\frac{i\pi \ell_{\text{AdS}}}{4G}}}{\sin\left(\frac{\ell_{\text{AdS}}\pi}{4G}\right)}(2a_1)^{\frac{\ell_{\text{AdS}}}{2G}}\,.
\end{align}
Thanks to the $i\epsilon$ regularization, the series is indeed convergent. On the other hand, for $\ell_{\text{AdS}}-i\epsilon$, we deform the contour downwards:
\begin{align}
    \mathcal{C}\to -\sum_{m=1}^{\infty} \mathcal{J}^+_{-m}+\sum_{m=0}^\infty \mathcal{J}^-_{-m}\,.
\end{align}
Along this contour, the partition function approximates 
\begin{align}
    \mathcal{Z}\sim -\sum_{m=-\infty}^{0}e^{\frac{i\pi m(\ell_{\text{AdS}}-i\epsilon)}{2G}}(2a_1)^{\frac{\ell_{\text{AdS}}-i\epsilon}{2G}}=-\frac{2ie^{\frac{i\pi \ell_{\text{AdS}}}{4G}}}{\sin\left(\frac{\ell_{\text{AdS}}\pi}{4G}\right)}(2a_1)^{\frac{\ell_{\text{AdS}}}{2G}}\,.
\end{align}
Up to the overall phase factor, the leading contribution is identical to \eqref{Zupwards}. Therefore the different regularizations $\ell_{\text{AdS}}\pm i\epsilon$ pick up different saddles, but finally lead to the almost same result. Furthermore our results are consistent with the CFT result \eqref{AdSexpansion}. We then conclude that the naive contour $\mathcal{C}=\mathbb{R}_+$ is the appropriate one consistent in our perspective via AdS/CFT. 

We can assign the Neumann boundary condition at $r=0$. The analysis is similar to the dS$_3$ case, and the CFT results are not expected to be reproduced. Therefore, we do not analyze this case in details.
The metric for the geometry realized by a saddle is complex and a geometrical interpretation is discussed in appendix \ref{app:geometry}.
We can also introduce a new radial coordinate:
\begin{align}
R(r) = - n \pi i (1 - r)^q + \ln (2 a_1) r^q \, ,
\end{align}
which interpolates between the complex geometry realized by a saddle and the one discussed in subsection \ref{sec:complexsaddles}.
Namely, the radial coordinate is the one used for a saddle point for $q=1$ and approaches to the one discussed below \eqref{eq:ansatzAdS} for $q \to \infty$.

\section{Classical Liouville action from dual gravity}
\label{sec:claads}

In section \ref{sec:2ptresurgence}, we studied the semi-classical limit of two-point functions in Liouville field theory and interpret them in terms of gravity theory. For the case with positive cosmological constant, CFT correlator is related to the wave functional of universe, which can be decomposed into the sum of contributions from saddle points. Each contribution can be decomposed into \eqref{eq:eachwf} with \eqref{eq:SGH} and \eqref{eq:In}. The contribution to \eqref{eq:SGH} depends on the label of saddle points $n$, and it was largely examined so far. Similarly, 
for the case with negative cosmological constant, CFT correlator is related to the gravity partition function with the decomposition \eqref{eq:eachpf} and the $n$-dependent contribution \eqref{eq:phase} was examined to a large extent. In this section, we focus on the contribution independent of $n$. For the case with negative cosmological constant, it corresponds to evaluating classical bulk action on AdS$_3$ with the insertion of conical defects satisfying $\sum_j \eta_j < 1$. It was already done in \cite{Chang:2016ftb} in terms of Einstein-Hilbert action, see also \cite{Chandra:2022bqq,Abajian:2023bqv}. In the next subsection, we reproduce the result in terms of Chern-Simons formulation as it is important to work with the formulation for understanding of $n$-dependent contributions. 
In subsection \ref{sec:dStoLiouville}, we extend the analysis to the case with positive cosmological constant by analytically continuing the bulk parameters.

Before working on the saddle independent contributions, we would like to comment on how the saddle dependent contributions change when the number of conical defects is increased.
In the Liouville field theory, we have exact expressions for two-point functions as in \eqref{eq:exact2pt} and three-point functions as in \eqref{eq:exact3pt} with \eqref{eq:DOZZ}. For these cases, we can examine the semi-classical limit of these expressions, see \cite{Zamolodchikov:1995aa,Harlow:2011ny}. In this section, we consider $m$-point function with $0 < \eta_j < 1$ and $\sum_j \eta_j < 1$. In that case, each insertion creates a conical defect with defect angle $2 \pi (1 - 2 \eta_j)$ and the surface with $m$ such conical defects and positive scalar curvature can be constructed. In the previous analysis, we have worked with two-sphere as the space where the dual CFT lives. In the current case, we need to replace the two-sphere by the manifold with $m$ conical defects.
In particular, we can compute the area of the two-dimensional manifold described by the physical metric as
\begin{align} \label{eq:physicalarea}
 \int d^2 z e^{\varphi} = 4 \pi (1 - \sum_j \eta_j ) \, .
 \end{align}
This implies that the phase factor \eqref{eq:phase} in the gravity partition function is now replaced as
\begin{align}  \label{eq:phasem}
\Theta_n  = e^{\frac{\ell_\text{AdS}} {2G} n   \pi i (1 - \sum_j \eta_j)} \, ,
\end{align}
which reduces to \eqref{eq:phase} for $m=2$  and \eqref{eq:phase3} for $m=3$.
In subsection \ref{sec:dStoLiouville}, we extend the analysis to the case with positive cosmological constant. The same argument holds for the case as well. Namely, the Gibbons-Hawking term in the wave functional of universe \eqref{eq:eachwf} should be given by 
\begin{align} \label{eq:SGHm}
 S_\text{GH}^{(n)} = \frac{(2 n +1) \pi \ell_\text{dS} (1 - \sum_j \eta_j)}{2G} \, ,
\end{align}
which reduces to \eqref{eq:SGH} for $m=2$ and \eqref{eq:GH3} for $m=3$.
In this way, we can extend the saddle dependent contributions to the case with $m$ insertions of conical defects. 

\subsection{Classical gravity action with negative cosmological constant}

In this subsection, we reproduce the analysis of \cite{Chang:2016ftb} (see also \cite{Chandra:2022bqq,Abajian:2023bqv}) in terms of  the Chern-Simons formulation of gravity theory. The bulk action may be given by \eqref{eq:action}, which consists of bulk and boundary actions. We first examine the bulk contribution from the Einstein-Hilbert action \eqref{eq:IEH} and then move to the boundary contributions from the Gibbons-Hawking term \eqref{eq:IGH} and the counter term \eqref{eq:ICT}.

\subsubsection{Bulk action}

We start by evaluating the bulk action in the Chern-Simons formulation at the semi-classical limit. The action is
\begin{align} \label{eq:CSaction0}
\begin{aligned}
I_\text{EH} &= - \frac{1}{16 \pi G} \int_{\mathcal {M}} d^3 x \sqrt{g} (R + 2 \ell_\text{AdS}^{-2}) \\
&= k I_\text{CS} [A] - k I_\text{CS} [\tilde A] - \frac{k}{4 \pi} \int_{\partial \mathcal{M}} \text{tr} [ A \wedge \tilde A ] \, , 
\end{aligned}
\end{align}
where the Chern-Simons action is given in \eqref{eq:CSac}. Note that we need to add the boundary term in order to reproduce the Einstein-Hilbert action precisely speaking, see, e.g., \cite{Corichi:2015nea,Takayanagi:2020njm}.
In order to express the gauge fields taking values in $\mathfrak{sl}(2)$ Lie algebra, we introduce the generators in $2 \times 2$ matrix form as
\begin{align} \label{eq:sl2generator}
\quad J_{0}=\left(
\begin{array}{cc}
1/2 & 0 \\
0 & -1/2 \\
\end{array} 
\right)~,\quad J_{+1}=\left(
\begin{array}{cc}
0 & 0 \\
1/2 & 0 \\
\end{array} 
\right) ~,\quad J_{-1}=\left(
\begin{array}{cc}
0 & 1/2 \\
0 & 0 \\
\end{array} 
\right) \, .
\end{align}
It turns out to be convenient to define
\begin{align} \label{eq:eta}
\eta_{ab} = 2 \text{tr}(J_a J_b) \, , \quad \epsilon_{abc} = 4 i \text{tr}(J_a J_b J _c) \, ,
\end{align}
where $\eta_{ab}$ is used to raise or lower the indices $a,b,\ldots$.

We would like to consider the case with $m$ insertions of conical defects. For this, we assume the form of the metric as
\begin{align}
\label{eq:metric}
\ell_\text{AdS}^{-2} ds^2 = d \rho^2 + \sinh^2 \rho e^{\varphi(z,\bar z)} d z d \bar z \, ,
\end{align}
where $\varphi (z , \bar z)$ is an arbitrary function of $z,\bar z$.
For instance, the metric of AdS$_3$ is expressed with
\begin{align} \label{eq:spheremetric}
e^{\varphi (z,\bar z)}dz d \bar z = \frac{4 d z d \bar z }{(1 + |z|^2)^2} \, .
\end{align}
We insert the conical defects with deficit angle $2\pi (1 - 2 \eta_j)$ at $z = z_j$, where we assume that $0 < \eta_j < 1/2$ and $\sum_j \eta_j < 1$.
Near the conical defects and the infinity, the function $\varphi (z , \bar z )$ should behave as
\begin{align}
\begin{aligned} \label{eq:varphiass}
&\varphi (z , \bar z  ) \sim - 4 \eta_i \ln |z - z_j| &&  (z \sim z_j) \, , \\
&\varphi (z , \bar z) \sim - 4 \ln |z | && (z \to \infty) \, .
\end{aligned}
\end{align}
As in \cite{Chandra:2022bqq}, we introduce an IR cut off near the AdS boundary at $\rho = \rho_0 (z , \bar z , \epsilon)$, where $\rho_0$ satisfies
\begin{align}\label{eq:boundaries}
\begin{aligned}
& \rho_0 (z , \bar z , \epsilon) = \ln \left( \frac{2 }{ \epsilon } \right) - \frac{ \varphi (z,\bar z) }{2}  \quad (|z - z_i| > \epsilon_i)  \, , \\
& \rho_0 (z , \bar z , \epsilon) = \ln \left( \frac{2 }{ \epsilon } \right ) + 2 \eta_j \ln \epsilon_j - \frac{C_j}{2}  \quad (|z - z_j| < \epsilon_j)  
\end{aligned}
\end{align}
with $C_j$ being the values of $\varphi (z , \bar z )$ of the order $\mathcal{O} ((\epsilon_j)^0)$ at $z \sim z_j$. The induced metric at the boundary $\rho = \rho_0 (z , \bar z , \epsilon)$ is
\begin{align} \label{eq:induced}
\ell_\text{AdS}^{-2} ds^2 = \left( \frac{1}{\epsilon^2} - \frac{e^\varphi}{2}\right) dz d \bar z + \frac14 (\partial \varphi dz + \bar \partial \varphi d \bar z)^2 + \mathcal{O} (\epsilon^2)
\end{align}
away from the conical deficits.

The connections corresponding to the metric \eqref{eq:metric} are given by
\begin{align} 
&A = \frac{1}{2}\left(
\begin{array}{cc}
1 & 0 \\
0 & -1 \\
\end{array} 
\right) d \rho + \frac{1}{4}\left(
\begin{array}{cc}
-\partial \varphi d z + \bar \partial \varphi d \bar z & 2 e^{-\rho+\varphi/2} d \bar z \\
- 2 e^{\rho+\varphi/2} d z & \partial \varphi d z - \bar \partial \varphi d \bar z \\
\end{array} 
\right) \, ,\\
&\tilde A = - \frac{1}{2}\left(
\begin{array}{cc}
1 & 0 \\
0 & -1 \\
\end{array} 
\right) d \rho + \frac{1}{4}\left(
\begin{array}{cc}
-\partial \varphi d z + \bar \partial \varphi d \bar z &  2 e^{\rho+\varphi/2} d \bar z \\
 - 2 e^{-\rho+\varphi/2} d z & \partial \varphi d z - \bar \partial \varphi d \bar z \\
\end{array} 
\right) \,,
\end{align}
or
\begin{align} \label{eq:adsgauge}
\begin{aligned}
&A = - \frac{1}{2} ( \partial \varphi dz - \bar \partial \varphi d \bar z) J_0 + e^{-\rho + \varphi/2} J_{-1} d \bar z - e^{\rho + \varphi/2} J_{+1} d z+ J_0 d \rho \, , \\
&\tilde A = - \frac{1}{2} ( \partial \varphi dz - \bar \partial \varphi d \bar z) J_0  +  e^{\rho + \varphi/2} J_{-1} d \bar z - e^{-\rho + \varphi/2} J_{+1} d z - J_0 d \rho \, .
\end{aligned}
\end{align}
Here we assign a condition for $\varphi (z , \bar z )$ as
\begin{align} \label{eq:eomvarphi}
\partial \bar \partial \varphi(z, \bar z) = - \frac{1}{2}e^{\varphi(z,\bar z)}\, ,
\end{align}
then we can check that the flatness condition $d A + A \wedge A = 0$ is satisfied.
The connection is related the vielbein and spin connection as
\begin{align} 
A = (\omega^a + i \ell_\text{AdS}^{-1}  e^a) J_a \, , \quad \tilde A = (\omega^a - i \ell_\text{AdS}^{-1} e^a ) J_a 
\end{align}
with $a = \pm 1,0$.
The relation to the metric is
\begin{align} \label{eq:gmunu}
g_{\mu \nu} = \frac{\ell_\text{AdS}^{2}}{2} \text{tr} [(A_\mu - \tilde A_\mu) (A_\nu - \tilde A_\nu)] \, .
\end{align}
From \eqref{eq:adsgauge}, we can read off as
\begin{align}
\begin{aligned}
&\ell_\text{AdS}^{-1} e^0 = - i d \rho \, , \quad \ell_\text{AdS}^{-1} e^{+1} =  i \sinh \rho e^{\varphi/2} d z \, , \quad \ell_\text{AdS}^{-1} e^{-1} = i \sinh \rho  e^{\varphi/2} d \bar z \, , \\
&\omega^0 = \frac{\bar \partial \varphi d \bar z -\partial \varphi d z }{2} \, , \quad \omega^{+1} = - \cosh \rho e^{\varphi/2} d z \, , \quad \omega^{-1} =  \cosh \rho e^{\varphi/2} d \bar z \, .
\end{aligned}
\end{align}
We can check that the torsion free condition $d e^a + \omega^a{}_b \wedge e^b = 0$ is satisfied.

We are now prepared to compute the bulk action \eqref{eq:CSaction0}.
Applying \eqref{eq:adsgauge}, we find
\begin{align}
&\text{tr} A \wedge dA = d \rho \wedge dz \wedge d \bar z \left[ \bar \partial \partial \varphi \text{tr} (J_0 J_0) - 2 e^{\varphi} \text{tr} (J_{-1} J_{+1})  \right]  = \left(\frac12 \bar \partial \partial \varphi - \frac12 e^\varphi \right) d \rho \wedge dz \wedge d \bar z\, , \nonumber \\
&\text{tr} A \wedge A \wedge A = d \rho \wedge dz \wedge d \bar z \cdot 6 \cdot e^\varphi \text{tr} (J_0 J_{-1} J_{+1}) =  \frac{3}{4} e^\varphi d \rho \wedge dz \wedge d \bar z\, ,
\end{align}
which lead to 
\begin{align}
\begin{aligned}
\int_\mathcal{M} \text{tr} \left[ A \wedge dA + \frac{2}{3} A \wedge A \wedge A \right]
&= \int dz \wedge d \bar z \int_0^{\rho_0} d \rho  \frac12 \bar \partial \partial \varphi   \\
&=  \int dz \wedge d \bar z \left[ \frac12 \bar \partial \partial\varphi  \left(\ln \left(\frac{2}{\epsilon}\right) - \frac{\varphi}{2} \right)\right] 
\end{aligned}
\end{align}
with $\rho_0 = \ln (2/\epsilon) - \varphi/2$.
We obtain the same expression from the sector of $\tilde A$.
We further find
\begin{align}
\begin{aligned}
\int_{\partial \mathcal{M}}\text{tr} A \wedge \tilde A &= \int_ \mathcal{M} \text{tr} (d A \wedge \tilde A - A \wedge d \tilde A ) \\
& =  \int d\rho \wedge d z \wedge d \bar z \left[- 2 \partial \bar \partial  \varphi \text{tr} (J_0 J_0) - 4 \cosh  (2 \rho ) e^{\varphi} \text{tr} (J_{-1} J_{+1})  \right]  \\
&\simeq  \int d z \wedge d \bar z \left[- \partial \bar \partial \varphi \left( \ln \left(\frac{2}{\epsilon} \right)- \frac{\varphi}{2} \right) - \frac{1}{\epsilon^2}  \right] \, .
\end{aligned}
\end{align}
Combining the above, we obtain 
\begin{align} \label{eq:classicalEH}
\begin{aligned}
 I_\text{EH}  \simeq  \frac{k}{4 \pi} \int  d z \wedge d \bar z \left[2  \partial \bar \partial \varphi \left( \ln \left(\frac{2}{\epsilon} \right)- \frac{\varphi}{2} \right) + \frac{1}{\epsilon^2}  \right]    \, . 
\end{aligned}
 \end{align}

\subsubsection{Adding boundary action}
\label{sec:boundary}

As in \eqref{eq:action}, we need to add Gibbons-Hawking boundary term \eqref{eq:IGH} and the counter term \eqref{eq:ICT}. In the Chern-Simons formulation, they are given by  (see, e.g., \cite{Corichi:2015nea,Takayanagi:2020njm})
\begin{align}
\begin{aligned} \label{eq:IGHCS}
I_\text{GH} + I_\text{CT} &=  -\frac{1}{8 \pi G } \int_{\partial {\mathcal {M}}} d^2 x \sqrt{h}( K - \ell_\text{AdS}^{-1})  \\
 &= \frac{k}{4 \pi \ell_\text{AdS}} \int_{\partial {\mathcal M}} \left[ 2 P_{ab} i e^a \wedge \omega^b + 2 \epsilon_{abc} e^a \wedge n^b d n^c + \ell_\text{AdS}^{-1}i \epsilon_{abc} n^a e^b \wedge e^c \right] \, .
\end{aligned}
\end{align}
We have defined $P_{ab} = \eta_{ab} - n_a n_b$,
where the normal vector is
\begin{align}
n^a = \mathcal{N} e^a_\nu N^\nu \, , \quad \mathcal{N} = (\eta_{ab} e^a_\nu N^\nu e^b_\mu N^\mu)^{-1/2} \, .
\end{align}
Here we have introduced
\begin{align}
 N_\rho = 1 \, , \quad N_z = \frac{1}{2} \partial \varphi \, , \quad N_{\bar z } = \frac{1}{2} \bar \partial \varphi \, ,
\end{align}
or 
\begin{align}
 N^\rho = \ell_\text{AdS}^{-2}  \, , \quad N^{\bar z} = \ell_\text{AdS}^{-2}  \partial \varphi \sinh ^{-2} \rho e^{- \varphi } \, , \quad N^{ z} = \ell_\text{AdS}^{-2}  \bar \partial \varphi \sinh ^{-2} \rho e^{- \varphi }\, .
\end{align}
Thus we have
\begin{align}
 n^0 = - i  \hat{\mathcal{N}} \, , \quad n^{-1} =   i \hat{\mathcal{N}}\partial \varphi \sinh ^{-1} \rho e^{- \varphi/2} \, , \quad n^{+1} =  i  \hat{\mathcal{N}} \bar \partial \varphi \sinh ^{-1} \rho e^{- \varphi/2} 
\end{align}
with
\begin{align}
 \hat{\mathcal{N}} =  (-1 - \partial \varphi \bar \partial \varphi \sinh^{-2} \rho e^{-\varphi})^{-1/2} \, .
\end{align}
Near $\rho \sim \rho_0 = \ln (2/\epsilon) - \varphi/2$, the normal vector becomes
\begin{align}
  n^0 \simeq  - 1 +  \frac{\epsilon^2}{2} \partial \varphi \bar \partial \varphi  \, , \quad  n^{-1} \simeq \epsilon \partial \varphi \, , \quad  n^{+1} \simeq \epsilon \bar \partial \varphi \, .
\end{align}

Let us evaluate the first two terms in \eqref{eq:IGHCS}, which can be rewritten as
\begin{align} \label{eq:classicalGH0}
\begin{aligned}
&    \frac{k}{2 \pi \ell_\text{AdS}} \int_{\partial {\mathcal M}} \left[ P_{ab} i e^a \wedge \omega^b +  \epsilon_{abc} e^a \wedge n^b d n^c\right] \\
 & \quad =  \frac{k}{2 \pi}  \int_{\partial {\mathcal M}}  \text{tr} [A \wedge \tilde A ] - \frac{k }{2 \pi \ell_\text{AdS}} \int_{\partial \mathcal{M} } i e^a \wedge n^b (n_a \omega_b + i \epsilon_{abc} d n^c) \, .
\end{aligned}
\end{align}
We therefore obtained
\begin{align}
\begin{aligned}
& \left .\int i e^{+1} n_{+1} \wedge n^{+1} \omega_{+1} \right|_{\rho = \rho_0 } = \frac{1}{4} \left .\int  i (n^{+1} n^{-1}) e^{+1}  \wedge  \omega^{-1} \right|_{\rho = \rho_0 }\\
& \quad \simeq   \frac{\ell_\text{AdS}}{4} \left . \int dz \wedge d \bar z  i ( \epsilon^2 \bar \partial \varphi \partial \varphi ) \cdot ( \tfrac{i}{4}e^{2 \rho + \varphi} ) \right|_{\rho = \rho_0 } \simeq - \frac{\ell_\text{AdS}}{4} \int d z \wedge d \bar z \bar \partial \varphi \partial \varphi 
\end{aligned}
\end{align}
and the same contribution from $\left .\int  i e^{-1} n_{-1} \wedge n^{-1} \omega_{-1} \right|_{\rho = \rho_0} $.
We also found
\begin{align}
\begin{aligned}
&\left .\int  i e^{-1} n_{-1} \wedge n^{0} \omega_{0} \right|_{\rho = \rho_0 } \simeq \frac12 \left .\int i (n^{+1} n^{0}) e^{-1}  \wedge  \omega^{0} \right|_{\rho = \rho_0}\\
& \quad \simeq   \frac{\ell_\text{AdS}}{2}  \left . \int dz \wedge d \bar z i  ( - \epsilon \bar \partial \varphi ) \cdot \left( \frac{i}{4} \partial \varphi e^{\rho + \varphi/2} \right) \right|_{\rho = \rho_0} \simeq  \frac{\ell_\text{AdS}}{4}  \int d z \wedge d \bar z \bar \partial \varphi \partial \varphi 
\end{aligned}
\end{align}
and the same expression from $\left .\int  i e^{+1} n_{+1} \wedge n^{0} \omega_{0} \right|_{\rho = \rho_0 } $.
Furthermore, we have
\begin{align}
\begin{aligned}
\int_{\partial \mathcal{M}}  i e^{0} n_{0} \wedge n^{0} \omega_{0}  &\simeq - \int_{ \mathcal{M}}  i e^{0} n_{0} \wedge n^{0} d \omega_{0} \simeq -\ell_\text{AdS}\int d \rho \wedge d z \wedge d \bar z \partial \bar \partial \varphi \\& \simeq -\ell_\text{AdS} \int dz \wedge d \bar z \partial \bar \partial \varphi \left( \ln \left( \frac{2}{\epsilon}\right) - \frac{\varphi}{2}  \right) 
\end{aligned}
\end{align}
and
\begin{align}
\begin{aligned}
 \int_{\partial \mathcal{M}} i e^{0} n_{0} \wedge n^{-1} \omega_{-1} &\simeq - \frac12 \int_{\mathcal{M}} 
i e^{0} n^{0} \wedge \left( d n^{-1} \wedge  \omega^{+1} + n^{-1} d \omega^{+1}  \right) \\
&\simeq  \frac{\ell_\text{AdS}}2 \int
dz \wedge d \bar z \left( \bar \partial \partial \varphi + \frac12 \partial \varphi \bar \partial   \varphi \right) \, .
\end{aligned}
\end{align}
The same contribution is obtained from $\int_{\partial \mathcal{M}}  i e^{0} n_{0} \wedge n^{+1} \omega_{+1}  $.
We then obtain
\begin{align}
\begin{aligned}
&\left .\int \epsilon_{+ 0 -} e^{+1} n^{0} \wedge d n^{-1}  \right|_{\rho = \rho_0 } \simeq \frac{i \ell_\text{AdS}}{2} \int dz \wedge d \bar z  \frac{i}{ \epsilon}  \cdot (- \epsilon \bar \partial \partial \varphi) \simeq   \frac{\ell_\text{AdS}}{2} \int dz \wedge d \bar z  \bar \partial \partial \varphi
\end{aligned}
\end{align}
and the same contribution from $\left .\int \epsilon_{- 0 +} e^{-} n^{0} \wedge d n^{+}  \right|_{\rho = \rho_0 } $.
Thus the first term in \eqref{eq:IGHCS} may be written as
\begin{align} \label{eq:classicalGH1}
\begin{aligned}
&    \frac{k}{2 \pi  \ell_\text{AdS}} \int_{\partial {\mathcal M}} \left[ P_{ab} i e^a \wedge \omega^b +  \epsilon_{abc} e^a \wedge n^b d n^c\right] \\
 & \quad =   \frac{k}{2 \pi}  \int_{\partial {\mathcal M}}  \text{tr} [A \wedge \tilde A ] - \frac{k }{2 \pi  \ell_\text{AdS}} \int_{\partial \mathcal{M} }i e^a \wedge n^b (n_a \omega_b + i \epsilon_{abc} d n^c) \\
 & \quad \simeq  - \frac{k}{2 \pi} \int dz \wedge d \bar z \left[ \frac{1}{\epsilon^2}  + \frac12\partial \varphi \bar \partial \varphi  \right] \, .
 \end{aligned} 
\end{align}
The last term in \eqref{eq:IGHCS} is computed as 
\begin{align}
\begin{aligned}
 &\int_{\partial \mathcal{M}} \epsilon_{0+-}[n^0 (e^{+1} \wedge e^{-1} + e^{-1} \wedge e^{+1})  ] \\
 &\quad \simeq  \frac{i}{2} \cdot 2  \ell_\text{AdS}^2  \int dz \wedge d \bar z \left(1  -  \frac{\epsilon^2}{2} \partial \varphi \bar \partial \varphi \right) \cdot \left( - \frac{1}{\epsilon^2}  + \frac{1}{2} e ^\varphi \right) \\
& \quad = i \ell_\text{AdS}^2  \int dz \wedge d \bar z \left[- \frac{1}{\epsilon^2} + \frac12 \partial \varphi \bar \partial \varphi +\frac12 e^\varphi \right] 
\end{aligned}
\end{align}
and
\begin{align}
\begin{aligned}
 \int_{\partial \mathcal{M}}\epsilon_{+ - 0} [n^{+1} (e^{-1} \wedge e^0 + e^0 \wedge e^{-1})  ] &\simeq   i \int_{\mathcal{M}} [d n^{+1} \wedge e^{-1} + n^{+1} d e^{-1}]\wedge e^0 \\ &\simeq   i  \ell_\text{AdS}^2\int  d z \wedge d \bar z \left[- \partial  \bar \partial \varphi - \frac12 \partial \varphi \bar \partial \varphi\right] \, .
\end{aligned}
\end{align}
The same comes from $ \int_{\partial \mathcal{M}}\epsilon_{- + 0} [n^{-1} (e^{+1} \wedge e^0 + e^0 \wedge e^{+1})  ] $.
In summary, we find
\begin{align}\label{eq:classicalGH2}
   \frac{k}{4 \pi \ell_\text{AdS}^2 } \int_{\partial {\mathcal M}} \left[ i\epsilon_{abc} n^a e^b \wedge e^c  \right]  \simeq   \frac{k}{4 \pi} \int d z \wedge d \bar z \left[  \frac{1}{\epsilon^2} - \frac{3 }{2}e^{\varphi} + \frac{1}{2}\partial \varphi \bar \partial \varphi \right] \, .
\end{align}

We now have the expressions of all the bulk and boundary contributions.
Combining the all, we obtain
\begin{align}
\begin{aligned}
&I_\text{EH} + I_\text{GH} + I_\text{CT}\\
& \quad \simeq \frac{k}{4 \pi}  \int d z \wedge d \bar z \left[ - \frac12  \partial \varphi \bar \partial \varphi - \frac12 e^{\varphi} - \varphi \partial \bar \partial \varphi  - e^{\varphi}\left(1 +  \ln \left(\frac{2}{\epsilon}\right)\right) \right] \\
& \quad \simeq \frac{c}{6} \left[ \frac{1}{4 \pi} \int d^2 z \left(  \partial \varphi \bar \partial \varphi - e^{\varphi} - 2 \bar \partial (\varphi \partial \varphi) \right) - 2  (1 - \sum_j \eta_j )\left(1 +  \ln \left(\frac{2}{\epsilon}\right)\right) \right] \, ,
\end{aligned}
\end{align}
where we have used  $\int dz \wedge d \bar z = 2  \int d ^2 z$ and \eqref{eq:physicalarea}.
Applying the behaviors of $\varphi (z , \bar z)$ near $z \sim z_j$ and $z \to \infty$ in \eqref{eq:varphiass}, we can show that  \eqref{eq:boundarLf} are rewritten as
\begin{align}
 \varphi_j = \frac{i}{4 \pi \eta_j} \oint_{\partial d_j} dz \varphi \partial \varphi \, , \quad  \varphi_R = \frac{i}{4 \pi } \oint_{\partial D} dz \varphi \partial \varphi \, .
\end{align}
Therefore, we find
\begin{align}
\begin{aligned}
&I_\text{EH} + I_\text{GH} + I_\text{CT}\simeq\frac{c}{6} \left[ \tilde S_L [\varphi] - 2 \ln R + 2 \sum_j \eta_j^2 \ln \epsilon_j - 2 ( 1 - \sum_j \eta_j )\left(1 + \ln \left( \frac{2}{\epsilon} \right) \right)\right] \, .
\end{aligned}
\end{align}
Here $\tilde S_L[\varphi]$ was defined in \eqref{eq:modLaction} and we set $\lambda = -1/4$.
This means that the semi-classical action of Liouville field theory can be reproduced from the dual gravity theory once the classical Liouville field $\varphi (z , \bar z )$ is identified with the function in the ansatz of metric \eqref{eq:metric}.

\subsection{Classical gravity action with positive cosmological constant}
\label{sec:dStoLiouville}

In the previous subsection, we have evaluated the classical bulk action for gravity theory with negative cosmological constant. In the case, the geometry corresponding to the saddle point with $n=0$ is Euclidean AdS$_3$ with conical defect insertions. Thus, we can simply make use of previous works and the extra contributions provide $n$-dependent phase factor \eqref{eq:phasem} as explained so far.
In this subsection, we extend the analysis to the case with positive cosmological constant. In this case, we should take care of the fact that only the half of Lorentzian dS$_3$ with conical defect insertions is relevant for the term $\mathcal{I}_n = \mathcal{I}$ introduced in \eqref{eq:eachwf}.
The half of 3-sphere connected to the Lorentzian geometry contributes to $S_\text{GH}^{(n)}$ in \eqref{eq:eachwf}, which is given by \eqref{eq:SGHm} in this case.

As above, the gravity action consists of the bulk and boundary contributions. We begin with the bulk contribution, which is described by the Einstein-Hilbert action as
\begin{align} \label{eq:CSactiondS0}
\begin{aligned}
I_\text{EH} &= - \frac{1}{16 \pi G} \int_{\mathcal {M}} d^3x \sqrt{g} (R - 2 \ell_\text{dS}^{-2}) \\
&=-  i \kappa I_\text{CS} [A] + i \kappa  I_\text{CS} [\tilde A] + \frac{i \kappa}{4 \pi} \int_{\partial \mathcal{M}} \text{tr} [ A \wedge \tilde A ]
\end{aligned}
\end{align}
with the Chern-Simons action \eqref{eq:CSac}.
We use the ansatz for the metric 
\begin{align} \label{eq:dsmetric}
   \ell_\text{dS}^{-2} ds^2 = - d \zeta^2 + \cosh^2 \zeta e^{\varphi (z , \bar z)} dz d \bar z
\end{align}
with $\varphi (z, \bar z)$ as a function of $z , \bar z$.
In other words, we perform the analytic continuation from the AdS case as
\begin{align}
 \ell_\text{AdS} \to - i \ell_\text{dS} \, , \quad 
 k \to - i \kappa \, , \quad \rho \to \zeta + \frac{\pi i}{2} \, .
\end{align}
If we set $\varphi (z , \bar z )$ as in \eqref{eq:spheremetric}, then the metric \eqref{eq:dsmetric} describes the Lorentzian dS$_3$.
Generically, the conical defects with deficit angle $2\pi (1 - 2 \eta_i)$ are inserted at $z = z_j$ with satisfying $0 < \eta_j < 1/2$ and $\sum_j \eta_j < 1$.
We introduce an IR cut off near the future infinity at $\zeta = \zeta_0 (z , \bar z , \epsilon)$ with
\begin{align}
\begin{aligned}
& \zeta_0 (z , \bar z , \epsilon) = \ln \left( \frac{2 }{ \epsilon } \right) - \frac{ \varphi (z,\bar z) }{2}  \quad (|z - z_i| > \epsilon_i)  \, , \\
& \zeta_0 (z , \bar z , \epsilon) = \ln \left( \frac{2 }{ \epsilon } \right ) + 2 \eta_j \ln \epsilon_j - \frac{C_j}{2}  \quad (|z - z_j| < \epsilon_j) \, .
\end{aligned}
\end{align}
Here we set $C_j$ as the values of $\varphi (z , \bar z )$ of the order $\mathcal{O} ((\epsilon_i)^0)$ at $z \sim z_j$.  The induced metric at the boundary $\zeta = \zeta_0 (z , \bar z , \epsilon)$ (away from the conical deficits) is
\begin{align} \label{eq:induceddS}
\ell_\text{dS}^{-2} ds^2 = \left( \frac{1}{\epsilon^2} + \frac{e^\varphi}{2}\right) dz d \bar z - \frac14 (\partial \varphi dz + \bar \partial \varphi d \bar z)^2 + \mathcal{O} (\epsilon^2) \, .
\end{align}
The gauge configuration corresponding to the metric \eqref{eq:dsmetric} is given by
\begin{align} \label{eq:dsgauge}
\begin{aligned}
&A = - \frac{1}{2} ( \partial \varphi dz - \bar \partial \varphi d \bar z) J_0 - i e^{-\zeta + \varphi/2} J_{+1} d \bar z - i e^{\zeta + \varphi/2} J_{-1} d z+ J_0 d \zeta \, , \\
&\tilde A = - \frac{1}{2} ( \partial \varphi dz - \bar \partial \varphi d \bar z) J_0  + i  e^{\zeta + \varphi/2} J_{+1} d \bar z + i e^{-\zeta + \varphi/2} J_{-1} d z - J_0 d \zeta \, ,
\end{aligned}
\end{align}
where $\varphi (z , \bar z )$ is assumed to satisfy
\begin{align} \label{eq:eomdS}
 \partial \bar \partial \varphi (z , \bar z ) = - \frac12 e^{\varphi (z , \bar z )} \, .
\end{align}
The vielbein and spin connections are related to the gauge fields by
\begin{align}
A = (\omega^a  -\ell_\text{dS}^{-1} e^a) J_a  \, , \quad \tilde A = (\omega^a +\ell_\text{dS}^{-1} e^a)  J_a
\end{align}
with $a = \pm 1 , 0$, and the relation between gauge fields and metric is
\begin{align} \label{eq:gmunudS}
g_{\mu \nu} = -\frac{\ell_\text{dS}^{2}}{2} \text{tr} [(A_\mu - \tilde A_\mu) (A_\nu - \tilde A_\nu)] \, .
\end{align}

We begin by evaluating the bulk contribution. We obtain
\begin{align}
\begin{aligned}
\int_\mathcal{M} \left[ A \wedge dA + \frac{2}{3} A \wedge A \wedge A \right]
&= \int dz \wedge d \bar z \int_0^{\zeta_0} d \zeta  \frac12 \bar \partial \partial \varphi  \\
&=  \int dz \wedge d \bar z \left[ \frac12 \bar \partial \partial\varphi  \left(\ln \left(\frac{2}{\epsilon} \right) - \frac{\varphi}{2} \right) \right] 
\end{aligned}
\end{align}
and the same expression from the sector of $\tilde A$.
Note that the integration range of $\zeta$ is for $0 \leq \zeta \leq \zeta_0$, where $\zeta = 0$ corresponds to the region where the Lorentzian dS geometry is connected to the Euclidean sphere geometry.
We further find
\begin{align}
\begin{aligned}
\int_{\partial \mathcal{M}}\text{tr} A \wedge \tilde A &=  \int d\zeta \wedge d z \wedge d \bar z \left[- 2 \partial \bar \partial  \varphi \text{tr} (J_0 J_0) + 4 \cosh  (2 \zeta ) e^{\varphi} \text{tr} (J_{+1} J_{-1})  \right]  \\
&\simeq  \int d z \wedge d \bar z \left[ - \partial \bar \partial \varphi \left( \ln \left(\frac{2}{\epsilon} \right)- \frac{\varphi}{2} \right) + \frac{1}{\epsilon^2}  \right] \, .
\end{aligned}
\end{align}
Combining them, we find 
\begin{align} \label{eq:classicalEHds}
\begin{aligned}
 I_\text{EH}  \simeq  
 - \frac{i \kappa}{4 \pi} \int  d z \wedge d \bar z  \left[2 \partial \bar \partial \varphi \left(  \ln \left( \frac{2}{\epsilon}\right) - \frac{\varphi}{2} \right)  +  \frac{1}{\epsilon^2}   \right] \, . 
\end{aligned}
 \end{align}
We then consider the boundary contributions given by the Gibbons-Hawking term \eqref{eq:IGH} and the counter term \eqref{eq:ICT}. In the Chern-Simons formulation, they are given by  
\begin{align}
\begin{aligned} \label{eq:IGHds}
I_\text{GH} + I_\text{CT}&=  -\frac{1}{8 \pi  G} \int_{\partial {\mathcal {M}}} \sqrt{h}( K - i \ell_\text{dS}^{-1})  \\
 &=  \frac{\kappa}{4 \pi \ell_\text{dS}} \int_{\partial {\mathcal M}} \left[ 2 P_{ab} i e^a \wedge \omega^b + 2 \epsilon_{abc} e^a \wedge n^b d n^c - \ell_\text{dS}^{-1}\epsilon_{abc} n^a e^b \wedge e^c \right] \, .
\end{aligned}
\end{align}
The first term in \eqref{eq:IGHds} can written as
\begin{align} \label{eq:classicalGH1ds}
\begin{aligned}
&   \frac{\kappa}{2 \pi  \ell_\text{dS}} \int_{\partial {\mathcal M}} \left[ P_{ab} i e^a \wedge \omega^b +  \epsilon_{abc} e^a \wedge n^b d n^c\right] 
 \simeq  \frac{i \kappa}{2 \pi} \int dz \wedge d \bar z  \left[- \frac{1}{\epsilon^2}   + \frac12 \partial \varphi \bar \partial \varphi   \right] 
 \end{aligned} 
\end{align}
and the last term in \eqref{eq:IGHds} is
\begin{align}\label{eq:classicalGH2ds}
  - \frac{ \kappa}{4 \pi \ell_\text{dS}^2 } \int_{\partial {\mathcal M}} \left[ \epsilon_{abc} n^a e^b \wedge e^c  \right]  \simeq   -\frac{i \kappa}{4 \pi} \int d z \wedge d \bar z \left[ - \frac{1}{\epsilon^2} - \frac{3}{2} e^{\varphi}+  \frac{1}{2}\partial \varphi \bar \partial \varphi \right] \, .
\end{align}
In total, we obtain
\begin{align}
\begin{aligned}
&I_\text{EH} + I_\text{GH} + I_\text{CT}\\
& \quad \simeq - i \frac{c^{(g)}}{6} \left[ \frac{1}{4 \pi} \int d^2 z \left(  \partial \varphi \bar \partial \varphi - e^{\varphi} - 2 \bar \partial (\varphi \partial \varphi) \right) - 2  (1 - \sum_j \eta_j )\left(1 +  \ln \left(\frac{2}{\epsilon}\right)\right) \right] \\
& \quad \simeq - i \frac{c^{(g)}}{6} \left[ \tilde S_L [\varphi] - 2 \ln R + 2 \sum_j \eta_j^2 \ln \epsilon_j - 2 ( 1 - \sum_j \eta_j )\left(1 + \ln \left( \frac{2}{\epsilon} \right) \right)\right] 
\end{aligned}
\end{align}
with $\tilde S_L [\varphi]$ in \eqref{eq:modLaction} and $\lambda = -1/4$. In this way, we have reproduced the classical Liouville action from the dual gravity theory even with positive cosmological constant.

\section{Discussion}
\label{sec:discussion}

In this paper, we determined the geometries of semi-classical saddle points in three-dimensional gravity theory with positive or negative cosmological constant.%
\footnote{In the Chern-Simons formulation, it is not known how to properly take the flat limit of the case with cosmological constant and which CFT is holographic dual to the flat space gravity. It should be important to examine the issue.}
In the case of positive cosmological constant, we confirmed the results obtained in our previous works \cite{Chen:2023prz,Chen:2023sry}, where the geometry is identified with the one used for the no-boundary proposal by Hartle and Hawking \cite{Hartle:1983ai}. In the case of negative cosmological constant, the semi-classical expressions of Liouville correlators suggest that there are infinitely many semi-classical saddles \cite{Harlow:2011ny}. We identified the geometries dual to the semi-classical saddles as Euclidean AdS$_3$ attached with possibly many three-spheres whose radii become purely imaginary. Thus the geometries are unphysical except for the special case with pure AdS$_3$.

This issue might be one of our important findings, so we would like to discuss it further. A closely related question was addressed recently by \cite{Witten:2021nzp} based on the previous works \cite{Louko:1995jw,Kontsevich:2021dmb}.
In order to define a path integral over complex metrics, it is natural to integrate over ``allowable'' complex geometry satisfing some criteria.
The authors considered the free action of $q$-form field strength $F_{j_1 \dots j_q}$ given by
\begin{align}
 I_q = \frac{1}{2 q!} \int_\mathcal{M} d^D x \sqrt{\text{det} g} g^{i_1 j_1} \dots g^{i_q j_q} F_{i_1 \dots i_q} F_{j_1 \dots j_q} \, . 
\end{align}
Here $q$ must run over $q=0,\ldots,D$.
We would like to make the path integral to converge for the allowable geometries.
The condition for the convergence of path integral is
\begin{align} \label{eq:criteria}
    \text{Re} \sqrt{\text{det} g} g^{i_1 j_1} \dots g^{i_q j_q} F_{i_1 \dots i_q} F_{j_1 \dots j_q} > 0 
\end{align}
for real non-zero $q$-form field strength $F_{j_1 \dots j_q}$ with $q=0,\ldots,D$. Diagonalizing the metric as
\begin{align}
    g_{ij} = \lambda_i \delta_{ij} \, , 
\end{align}
the condition \eqref{eq:criteria} can be written as
\begin{align}
    \sum_{j=1}^D |\text{arg} \lambda_j| < \pi \, .
\end{align}
The condition can be satisfied at most one negative sign in the metric, so our metric with three negative signs does not satisfy the criteria.

This fact itself might not be a problem in the current setup in the following sense. We are working with only topological Chern-Simons gravity and possibly scalar fields on the geometry. The action for a real scalar field may be written as
\begin{align}
 I_q = \frac{1}{2 } \int_M d^D x \sqrt{\text{det} g} \left[ g^{i j} \partial_i \phi \partial_j \phi + m^2 \phi^2 \right] \, .
\end{align}
The condition corresponds to the case with $q=0,1$ in \eqref{eq:criteria}.
We are only dealing with Chern-Simons gauge fields, so we do not need to assign the condition with $q=2$, which is the case for Yang-Mills gauge field. In our case, we may be able to realize that 
\begin{align}
 \pi < \sum_{j=1}^3 \text{arg}  \lambda_j < 3 \pi
\end{align}
by properly introducing a regulator. The path integral converges even for geometry with the metric and in particular
the weaker condition can be satisfied even by the three-sphere with purely imaginary radius. Concretely, the contribution to the classical action from imaginary radius three-spheres is%
\begin{align}
  I = - \frac{\ell_\text{AdS} + i \epsilon}{16 \pi G} \int_M d^3 x\sqrt{\text{det}g} (R - \Lambda) = \frac{\ell_\text{AdS} + i \epsilon}{4 \pi } \int_M d^3 x\sqrt{\text{det}g} 
 = \frac{- i \pi n (\ell_\text{AdS} + i \epsilon)}{2 G} \, ,
\end{align}
whose real part is positive for $n > 0$.

Even the decomposition with unphysical geometries itself is not a problem in the sense of the convergence of path integral, one may feel uncomfortable with this fact. We could make several comments on it. If we adopt (an analytic continuation of) Gaberdiel-Gopakumar duality \cite{Gaberdiel:2010pz} mentioned in the introduction, then the unitary of the CFT is violated. The gravity side is indeed unphysical as it includes conical excess solutions \cite{Castro:2011iw,Gaberdiel:2012ku,Perlmutter:2012ds}. Thus, unphysical configurations in gravity theory are not in contradiction with holography.
However, the large central charge of Liouville field theory may be used as an effective theory of unitary CFT dual to a physical gravity theory. For instance, even each contribution to gravity partition function has a non-trivial phase factor as in \eqref{eq:phasem}, the sum over all phase factors leads to a real value up to an overall phase as
\begin{align}
 - i e^{\pi i  (1 - \sum_j \eta_j)\frac{\ell_\text{AdS}} {4G}} \sum_{n=0}^\infty e^{ n  \pi i  (1 - \sum_j \eta_j)\frac{\ell_\text{AdS}} {2G}} = \frac{1}{2 \sin \pi  (1 - \sum_j \eta_j)\frac{\ell_\text{AdS} }{4G} }
\end{align}
It might be possible to change the metric of geometry corresponding to a non-trivial saddle to physical one by a certain gauge transformation. Or it could be possible to have a different decomposition with different saddles whose geometries are physical. The latter possibility was addressed for the semi-classical saddles of four-point blocks of generic CFT$_2$ in \cite{Benjamin:2023uib}.

There are additional open problems we would like to pursue furthermore. Exact expressions are available for two- and three-point functions in Liouville field theory, and we have utilized them. It is an important problem to extend the analysis to higher point functions, where exact expressions have not been obtained yet generically, see \cite{Benjamin:2023uib} for a recent attempt. We also focused on the correlation functions of heavy operators satisfying $0 < \eta_j < 1/2$ and $\sum_j \eta_j <1$. We would like to go beyond these regions of parameters. We may consider the heavy operator with $\eta_j = 1/2 + i \mathbb{R}$, which is dual to a black hole solution. It is also intriguing to work on a CFT base manifold of higher genus topology, such as torus. In the case with  $\sum_j \eta_j > 1$ and negative cosmological constant, the dual geometry might be obtained with the metric of the form
\begin{align}
\label{eq:metricwormhole}
\ell_\text{AdS}^{-2} ds^2 = d \rho^2 + \cosh^2 \rho e^{\varphi(z,\bar z)} d z d \bar z \, ,
\end{align}
where the physical metric $d \Sigma^2 = e^{\varphi(z,\bar z)} d z d \bar z $ describes a two-dimensional manifold with negative constant curvature. The metric describes a Euclidean wormhole geometry \cite{Maldacena:2004rf}, which has two boundaries at $\rho = \pm \infty$. We may apply the current analysis to the case where two CFT live at the different boundaries as examined in \cite{Chandra:2022bqq} or the wormhole geometry truncates at $\rho = 0$ and connects to a close universe as suggested in \cite{Abajian:2023bqv}. 
We examined $SL(2)$ Chern-Simons gauge theory as a description of three-dimensional pure gravity. In the Chern-Simons formulation, it might be not so difficult to extend the analysis to higher spin gravity. We may use known results on $SL(N)$ Toda field theory in \cite{Fateev:2007ab} to examine higher spin gravity described by $SL(N)$ Chern-Simons gauge theory. It might be more interesting to use the 't Hooft limit of the coset model \eqref{eq:coset} while keeping its unitarity to describe Prokushkin-Vasiliev theory within its physical region. 
There are attempts to relate the higher spin holography to stringy one by introducing extended supersymmetry as in \cite{Gaberdiel:2013vva,Gaberdiel:2014cha}
and \cite{Creutzig:2011fe,Creutzig:2013tja,Creutzig:2014ula}.
The current analysis is supposed to be applicable to other dimensions as well. In the case of negative cosmological constant, we have many concrete examples to investigate.
Even in the case of positive cosmological constant, there are several concrete examples available, such as, dS$_4$/CFT$_3$ as in \cite{Anninos:2011ui} and dS$_2$/CFT$_1$ as in \cite{Cotler:2019nbi,Maldacena:2019cbz}.
In particular in the case of four spacetime dimensions with negative cosmological constant, the mini-superspace approach has been applied in \cite{Caputa:2018asc}. After including all the relevant pieces such as Gibbons-Hawking and contact terms, the saddle point of the gravitational path integral surprisingly reproduced the Airy function appearing in the supersymmetric $\mathbb{S}^3$ partition of the dual field theory despite lacking manifest supersymmetry in the gravity computation.
It would be quite interesting to see if similar match also occurs in our three-dimensional mini-superspace computation. In this case it would be natural to compare the result with the $\mathbb{S}^2$ supersymmetric partition function for the appropriate two-dimensional superconformal field theories. If so, it would provide us further support for pursuing higher-dimensional cases and understand such highly non-trivial matches.

\subsection*{Acknowledgements}

We are grateful to Pawel Caputa, Shinji Hirano, Masazumi Honda, Hayato Kanno, Hikaru Kawai, Jun Nishimura and Tadashi Takayanagi for useful discussions.  The work is partially supported by Grant-in-Aid for Transformative Research Areas (A) ``Extreme Universe'' No.\,21H05187. The work of H.\,Y.\,C. is supported in part by Ministry of Science and Technology (MOST) through the grant 110-2112-M-002-006-. H.\, Y.\, C.\, would also like to thank Yukawa Institute for Theoretical Physics, Kyoto University for the hospitality when this work was being finished. The work of Y.\ H. is supported by JSPS Grant-in-Aid for Scientific Research (B) No. 23H01170. Y.\,T. is supported by Grant-in-Aid for JSPS Fellows No.\,22KJ1971. T.\,U. is supported by JSPS Grant-in-Aid for Early-Career Scientists No.\,22K14042.

\appendix

\section{Liouville three-point functions}
\label{app:Liouville3pt}

In section \ref{sec:2ptresurgence}, we have examined the semi-classical limit of Liouville two-point functions and their interpretation in terms of dual gravity theory. In this appendix, we extend the analysis to Liouville three-point function. In the next subsection, we introduce so-called Upsilon function, which will be used to express the three-point function, and study its properties, in particular, its asymptotic expansion. In subsection \ref{app:L3pt}, we study the non-perturbative saddles of Liouville three-point function and its interpretation in terms of gravity theory.

\subsection{Asymptotic series of Upsilon function}
In this appendix, following \cite{Harlow:2011ny}, we summarize some useful properties of $\Upsilon_b(z)$ function including an expression for its asymptotic series expansion in \eqref{upsexp}.
The Upsilon function $\Upsilon_b(z)$ is defined as a unique function that satisfies the recursion relations
\begin{align}
    \Upsilon_b(z+b)&=\gamma(bz)b^{1-2bz}\Upsilon_b(z)\ ,\\
    \Upsilon_b(z+1/b)&=\gamma(z/b)b^{2z/b-1}\Upsilon_b(z)\ .
\end{align}
For $\Re b>0$ and $0<\Re z<\Re Q$, Upsilon function $\Upsilon_b(z)$ has the integral form
\begin{align} \label{eq:upsilon}
    \log\Upsilon_b(z)=\int_{0}^\infty \frac{dt}{t}\left[\left(\frac{Q}{2}-z\right)^2e^{-t}-\frac{\sinh^2\left(\left(\frac{Q}{2}-z\right)\frac{t}{2}\right)}{\sinh\left(\frac{bt}{2}\right)\sinh\left(\frac{t}{2b}\right)}\right].
\end{align}
We shall consider the case with $z=\mathcal{O}(b^{-1})$, appearing in three-point functions with heavy operator insertions. A convenient parameterization is 
\begin{align}
    z=\frac{x}{b}+\frac{b}{2}
\end{align}
with real $x$ in the region $0<x<1/2$. 
From the reflection property $\Upsilon_b(z)=\Upsilon_b(Q-z)$ and the recursion relations, we can extend the result to any real $x\in \mathbb{R}$. By scaling the integral variable $t$ by $b$, we find
\begin{align}
    \log \Upsilon_b\left(\frac{x}{b}+\frac{b}{2}\right)=\int_0^{e^{i\arg b^{-1}}\infty}\frac{dt}{t}\left[\frac{1}{b^2}\left(x-\frac{1}{2}\right)^2e^{-bt}-\frac{\sinh^2\left(\left(x-\frac{1}{2}\right)\frac{t}{2}\right)}{\sinh\left(\frac{t}{2}\right)\sinh\left(\frac{b^2t}{2}\right)}\right] \ .
\end{align}

With the help of 
\begin{align}
    \log z=\int_0^\infty\frac{dt}{t}(e^{-t}-e^{-zt})
\end{align}
and 
\begin{align}
    \frac{1}{\sinh x}=\frac{1}{x}\sum_{n=0}^\infty c_n(2x)^{2n},\qquad c_n=-(1-2^{1-2n})\frac{B_{2n}}{(2n)!}\ ,
\end{align}
we have 
\begin{align}\begin{aligned}
    \log \Upsilon_b\left(\frac{x}{b}+\frac{b}{2}\right)=&-\frac{1}{b^2}\left(x-\frac{1}{2}\right)^2\log b+\frac{1}{b^2}F(x) \\
    &-\sum_{n=1}^\infty c_n b^{4n-2}\int_0^{e^{i\arg b^{-1}}\infty}dt\frac{2t^{2n-2}\sinh^2\left(\left(x-\frac{1}{2}\right)\frac{t}{2}\right)}{\sinh\left(\frac{t}{2}\right)} \ .
\end{aligned}\end{align}
By defining a function $F(x)$ by 
\begin{align}
    F(x)=\int_0^\infty\frac{dt}{t}\left[\left(x-\frac{1}{2}\right)^2-\frac{2}{t}\frac{\sinh^2\left(\left(x-\frac{1}{2}\right)\frac{t}{2}\right)}{\sinh\left(\frac{t}{2}\right)}\right]
\end{align}
and the polygamma function
\begin{align}
    \psi^{(m)}(x):=\frac{d^{m+1}}{dx^{m+1}}\log\Gamma(x)=(-1)^{m+1}\int_0^\infty dt\frac{t^me^{-xt}}{1-e^{-t}}
\end{align}
for $m\ge1$, we obtain the asymptotic series (see also appendix A in \cite{Collier:2018exn})
\begin{align}\begin{aligned}
    \log \Upsilon_b\left(\frac{x}{b}+\frac{b}{2}\right)=&-\frac{1}{b^2}\left(x-\frac{1}{2}\right)^2\log b+\frac{1}{b^2}F(x) \\
    &+\sum_{n=1}^\infty c_n b^{4n-2}\left[\frac{1}{2}\psi^{(2n-2)}(1-x)+\frac{1}{2}\psi^{(2n-2)}(x)-\psi^{(2n-2)}\left(1/2\right)\right] \ .
\end{aligned}\end{align}
By using the formula
\begin{align}
    \psi^{(m)}(z)=(-1)^{m+1} m!\zeta(m+1,z)
\end{align}
with $\zeta(s,z)$ being the Hurwitz zeta function, we can rewrite as
\begin{align}\begin{aligned}
   & \log \Upsilon_b\left(\frac{x}{b}+\frac{b}{2}\right)=-\frac{1}{b^2}\left(x-\frac{1}{2}\right)^2\log b+\frac{1}{b^2}F(x) \\
    & \quad -\sum_{n=1}^\infty (2n-2)! c_n b^{4n-2}\left[\frac{1}{2}\zeta(2n-1,x)+\frac{1}{2}\zeta(2n-1,1-x)-\zeta(2n-1,1/2)\right] \ .
\end{aligned}\end{align}

Let us consider the large $s$ approximation of $\zeta(s,z)$. For $\Re s>1$ and $\Re z>0$, $\zeta(s,z)$ has the integral representation
\begin{align}
    \zeta(s,z)=\frac{1}{\Gamma(s)}\int_0^\infty\frac{t^{s-1}e^{-zt}}{1-e^{-t}}dt=\frac{s^s}{\Gamma(s)}\int_{-\infty}^\infty\frac{e^{s(\phi-ze^\phi )}}{1-e^{-se^\phi}}d\phi \ ,
\end{align}
where the second equality is realized by changing the variable as $x=se^\phi$. The saddle point equation is 
\begin{align}
    1-ze^\phi-\frac{e^{\phi-se^\phi}}{1-e^{-se^\phi}}=0 \ .
\end{align}
In the large $s$ limit, the final term drops out. We then obtain the solutions
\begin{align}
    \phi=-\log z+2\pi iN \ , \quad N\in\mathbb{Z} \ .
\end{align}
Therefore the saddle point approximation for the solution labeled by $N$ will be 
\begin{align}
    \zeta(s,z)\sim \frac{e^{2\pi isN}}{z^s} \ ,
\end{align}
where the terms depending on $e^{-s/z}$ are ignored. In particular, when $s$ is an integer, the $N$-dependence disappears. The important point is that the dependence of $\zeta(s,z)$ in $s$ is just a power of $s$.

By using the above approximation of $\zeta(s,z)$ and the asymptotic formula of the Bernoulli number,
\begin{align}
    B_{2n}\sim (-1)^{n+1}\frac{2(2n)!}{(2\pi)^{2n}} \ ,
\end{align}
we have 
\begin{align}\begin{aligned}\label{upsexp}
   & \log \Upsilon_b\left(\frac{x}{b}+\frac{b}{2}\right)\sim -\frac{1}{b^2}\left(x-\frac{1}{2}\right)^2\log b+\frac{1}{b^2}F(x) \\
    & \quad +\sum_{n=1}^\infty \frac{(-1)^{n+1}(1-2^{1-2n})(2n-2)!}{(2\pi)^{2n}} (b^2)^{2n-1}\left[\left(\frac{1}{x}\right)^{2n-1}+\left(\frac{1}{1-x}\right)^{2n-1}-2^{2n}\right].
\end{aligned}\end{align}
Regarding the perturbative part as the expansion in $b^2$, with only odd-power terms surviving, this is the resurgent series.

\subsection{Non-perturbative saddles of three-point functions}
\label{app:L3pt}

In this subsection, we examine three-point functions of heavy operators in Liouville field theory. The exact expression of three-point function is known as the Dorn-Otto-Zamolodchikov-Zamolodchikov (DOZZ) formula, which is given by \cite{Dorn:1994xn,Zamolodchikov:1995aa}
\begin{align} \label{eq:exact3pt}
 \left \langle V_{\alpha_1} (z_1 , \bar z_1)   V_{\alpha_2} (z_2 , \bar z_2) V_{\alpha_3} (z_3 , \bar z_3) \right \rangle
= \frac{C(\alpha_1 , \alpha_2 , \alpha_3)}{|z_{12}|^{2 (h_1 + h_2 - h_3)} |z_{13}|^{2 (h_1 + h_3 - h_2)} |z_{23}|^{2 (h_2 + h_3 - h_1)}}
\end{align}
with 
\begin{align} \label{eq:DOZZ}
\begin{aligned}
&C (\alpha_1 , \alpha_2 , \alpha_3) = [\pi \mu \gamma (b^2) b^{2 - 2 b^2}]^{(Q - \sum_j \alpha_j)/b} \\
& \quad \times \frac{\Upsilon_0 \Upsilon_b(2 \alpha_1)\Upsilon_b(2 \alpha_2)\Upsilon_b(2 \alpha_2)}{\Upsilon_b(\sum_j \alpha_j - Q)\Upsilon_b(\alpha_1 + \alpha_2 - \alpha_3 - Q)\Upsilon_b(\alpha_2 + \alpha_3 - \alpha_2)\Upsilon_b(\alpha_3 + \alpha_1 - \alpha_3)} \, .
\end{aligned}
\end{align}
The Upsilon function $\Upsilon_b (x)$ is defined in \eqref{eq:upsilon} and $\Upsilon_0 = \left. d \Upsilon_b (x)/dx \right|_{x = 0}$.
As in the case of two-point function, we read off the saddle points for $b \to 0$ from the expansion as in \eqref{eq:expansion}. We set $\eta_j = \alpha_j b$ satisfying
\begin{align}
 0 < \eta_i < \frac{1}{2} \, , \quad \sum_j \eta_j < 1 \, , \quad 0 < \eta_i + \eta_j - \eta_k \quad (i \neq j \neq k) \, .
\end{align}
The geometrical meaning of the conditions may be found in \cite{Harlow:2011ny}. 
In that paper, solutions to the equation of motion \eqref{eq:eom} were obtained and the modified classical action \eqref{eq:modLaction} was evaluated as
\begin{align}
\begin{aligned}
&\tilde S_\text{L} [\varphi_{(n)}] = ( \sum_j \eta_j -1 ) \ln \lambda + (\delta_1 + \delta_2 - \delta_3) \ln |z_{12}|^2+ (\delta_1 + \delta_3 - \delta_2) \ln |z_{13}|^2 \\& \quad + (\delta_2 + \delta_3 - \delta_1) \ln |z_{23}|^2 + F(\eta_1 + \eta_2 - \eta_3) +  F(\eta_1 + \eta_3 - \eta_2) +  F(\eta_2 + \eta_3 - \eta_1) \\& \quad+ F(\sum_j \eta_j ) - F(2 \eta_1) - F(2 \eta_2) - F(2 \eta_3) - F(0) \\& \quad + 2 [ (1 - \sum_j \eta_j) \ln (1 - \sum_j \eta_j) - (1 - \sum_j \eta_j) ] + 2 \pi i (n + 1/2) (1 - \sum_j \eta_j)
\end{aligned}
\end{align}
for each saddles. Here we defined $\delta_j = \eta_j (1 - \eta_j)$ and a function
\begin{align}
F(\eta) = \int_\frac{1}{2}^\eta \ln \gamma (x) dx \, .
\end{align}
The small $b$ behavior of three-point function \eqref{eq:DOZZ} can be read off from the asymptotic behavior of the Upsilon function as
\begin{align}
\begin{aligned}
&C(\alpha_1 , \alpha_2 , \alpha_3)
\sim [ e^{- \frac{\pi i }{b^2}(1 - \sum_j \eta_j)} - e^{\frac{\pi i }{b^2}(1 - \sum_j \eta_j)} ] \\
& \quad \times
e^{  - \frac{1}{b^2} [ (\sum_j \eta_j -1) \ln \lambda + F(\eta_1 + \eta_2 - \eta_3) +  F(\eta_1 + \eta_3 - \eta_2) +  F(\eta_2 + \eta_3 - \eta_1) + F(\sum_j \eta_j ) ] } \\
& \quad \times e^{ - \frac{1}{b^2} [ - F(2 \eta_1) - F(2 \eta_2) - F(2 \eta_3) - F(0)  + 2 [ (1 - \sum_j \eta_j) \ln (1 - \sum_j \eta_j) - (1 - \sum_j \eta_j) ] + 2 \pi i (n + 1/2) (1 - \sum_j \eta_j)] }
\end{aligned}
\end{align}
for $\text{Re} \ b^{-2} < 0$ and
\begin{align}
\begin{aligned}
&C(\alpha_1 , \alpha_2 , \alpha_3)
\sim \frac{1}{ e^{- \frac{\pi i }{b^2}(1 - \sum_j \eta_j)} - e^{\frac{\pi i }{b^2}(1 - \sum_j \eta_j)} } \\
& \quad \times
e^{  - \frac{1}{b^2} [ (\sum_j \eta_j -1) \ln \lambda + F(\eta_1 + \eta_2 - \eta_3) +  F(\eta_1 + \eta_3 - \eta_2) +  F(\eta_2 + \eta_3 - \eta_1) + F(\sum_j \eta_j) ] } \\
& \quad \times e^{ - \frac{1}{b^2} [ - F(2 \eta_1) - F(2 \eta_2) - F(2 \eta_3) - F(0)  + 2 [ (1 - \sum_j \eta_j) \ln (1 - \sum_j \eta_j) - (1 - \sum_j \eta_j) ] + 2 \pi i (n + 1/2) (1 - \sum_j \eta_j)] }
\end{aligned}
\end{align}
for $\text{Re} \ b^{-2} > 0$.

As in the case of two-point function, we can rewrite the three-point function in the language of dual gravity theory. We begin with the case of positive cosmological constant, which implies $\text{Re} \, b^{-2} < 0$ as in \eqref{eq:reb2}.
The three-point function is related to the wave functional of universe with the three conical defects with angle $2 \pi (1 - 2 \eta_j)$ as explained above.
The wave functional can be decomposed into
\begin{align}
 \Psi_\text{dS} = \sum_{n=-1,0}  \Psi_n
\end{align} 
with
\begin{align}  \label{eq:Psin3pt}
 & \Psi_n \sim e^{\frac{\ell_\text{dS}}{4G}(2 n + 1) \pi (1 -  \sum_j \eta_j)} |z_{12}|^{ i\frac{ \ell_\text{dS}}{2 G} (\delta_1 + \delta_2 - \delta_3)} |z_{13}|^{i\frac{\ell_\text{dS}}{2 G}(\delta_1 + \delta_3 - \delta_2)} |z_{23}|^{i\frac{\ell_\text{dS}}{2 G} (\delta_2 + \delta_3 - \delta_1)} \nonumber \\
 & \quad \times
e^{i\frac{\ell_\text{dS}}{4 G} \{ (\sum_j \eta_j -1) \ln \lambda + F(\eta_1 + \eta_2 - \eta_3) +  F(\eta_1 + \eta_3 - \eta_2) +  F(\eta_2 + \eta_3 - \eta_1) + F(\sum_j \eta_j ) \} } \\
& \quad \times e^{i \frac{\ell_\text{dS}}{4 G} \{ - F(2 \eta_1) - F(2 \eta_2) - F(2 \eta_3) - F(0)  + 2 [ (1 - \sum_j \eta_j) \ln (1 - \sum_j \eta_j) - (1 - \sum_j \eta_j) ] + 2 \pi i (n + 1/2) (1 - \sum_j \eta_j) \} } \, . \nonumber
\end{align}
The wave functional of universe consists of two saddles with $n = -1,0$ and the Gibbons-Hawking entropy from each saddle is
\begin{align} \label{eq:GH3}
 G_\text{GH}^{(n)} =  \frac{(2 n +1) \pi \ell_\text{dS} (1 - \sum_j \eta_j)}{2G} \, .
\end{align}
We then consider the case of negative cosmological constant, i.e., $\text{Re}\, b^{-2} > 0$.
The three-point function is dual to a geometry with three conical defects with angle $2 \pi (1 - 2 \eta_j)$.
The gravity partition function can be decomposed as
\begin{align}
 \mathcal{Z}_\text{AdS} = \sum_{n=0}^\infty \mathcal{Z}_n
\end{align} 
with
\begin{align} \label{eq:Zn3pt}
 &\mathcal{Z}_n \sim e^{\frac{\ell_\text{AdS}}{4G}(2 n + 1) \pi i (1 -  \sum_j \eta_j)} |z_{12}|^{- \frac{\ell_\text{AdS}}{2 G} (\delta_1 + \delta_2 - \delta_3)} |z_{13}|^{ - \frac{\ell_\text{AdS}}{2 G}(\delta_1 + \delta_3 - \delta_2)} |z_{23}|^{ -\frac{\ell_\text{AdS}}{2 G} (\delta_2 + \delta_3 - \delta_1)} \nonumber \\
 & \quad \times
e^{  - \frac{\ell_\text{AdS}}{4 G} \{ (\sum_j \eta_j -1) \ln \lambda + F(\eta_1 + \eta_2 - \eta_3) +  F(\eta_1 + \eta_3 - \eta_2) +  F(\eta_2 + \eta_3 - \eta_1) + F(\sum_j \eta_j ) \} } \\
& \quad \times e^{ - \frac{\ell_\text{AdS}}{4 G} \{ - F(2 \eta_1) - F(2 \eta_2) - F(2 \eta_3) - F(0)  + 2 [ (1 - \sum_j \eta_j) \ln (1 - \sum_j \eta_j) - (1 - \sum_j \eta_j) ] + 2 \pi i (n + 1/2) (1 - \sum_j \eta_j)\} } \, . \nonumber
\end{align}
As in the case of two-point function, that there are infinitely many saddles labeled by $n=0,1,2,\ldots$ and the dependence on $n$ appears in the pure phase factor as
\begin{align} \label{eq:phase3}
\Theta_n  = e^{\frac{\ell_\text{AdS}} {2G} n   \pi i (1 - \sum_j \eta_j)} \, .
\end{align}

\section{Revisiting the series expansion of the phase factor}\label{app:regularization}
Here we consider the CFT dual of AdS geometries with conical deficits, i.e. $b,\eta_i\in\mathbb{R}$. Following \cite{Harlow:2011ny}, the semi-classical approximation of the two-point function with $\alpha=b/2+\eta/b$ is given by 
\begin{align}\begin{aligned}
    \left\langle V_\alpha(z_1)V_\alpha(z_2)\right\rangle&\sim \frac{1}{e^{i\pi (1-2\eta)/b^2}-e^{-i\pi (1-2\eta)/b^2}}\\
    & \times e^{-\frac{2}{b^2}\left[(1-2\eta)\log(1-2\eta)-(1-2\eta)\right]}\lambda^{(1-2\eta)/b^2}|z_{12}|^{-4\eta(1-\eta)/b^2}\delta(0) \ .
\end{aligned}\end{align}

Let us carefully discuss the expansion of the phase factor.
If $\eta$ were not real-valued, the expansion can be easily obtained as 
\begin{align}\label{expand}
    \frac{1}{e^{i\pi (1-2\eta)/b^2}-e^{-i\pi (1-2\eta)/b^2}}=\pm \sum_{n=0}^\infty e^{\mp 2\pi i(n+ 1/2)(1-2\eta)/b^2},
\end{align}
where the two signatures take the upper one if $\Im \eta>0$ and the lower one if $\Im \eta<0$ since the radius of convergence of the series expansion
\begin{align}
    \frac{1}{1-z}=\sum_{n=0}^\infty z^n   
\end{align} 
is $1$. However, if $\eta$ has no imaginary part, we cannot expand as \eqref{expand} in the usual sense. Nevertheless, we may be able to define the expansion of the phase factor by using the notion of hyperfunction. 

To define a hyperfunction $f(x)$ for a real variable $x$, we prepare a holomorphic function $F(z)$ defined on the complex plane except for the real axis and define $f(x)$ as 
\begin{align}
    f(x):=F(x+i0)-F(x-i0) \ .
\end{align}
For example, the defining function $F$ for the delta function $\delta(a)$ is given by 
\begin{align}
    F(z)=-\frac{1}{2\pi i}\frac{1}{z-a} \ , \quad z\in \mathbb{C}\setminus \mathbb{R} \ .
\end{align}
For our case, the appropriate $F(\eta)$ for $\eta\in \mathbb{C}\setminus \mathbb{R}$ may be
\begin{align}\label{Fdef}
    F(\eta)=
    \begin{cases}
        \sum_{n=0}^\infty e^{- 2\pi i(n+ 1/2)(1-2\eta)/b^2} \ , & \Im \eta >0 \ ,\\
        \sum_{n=0}^\infty e^{ 2\pi i(n+ 1/2)(1-2\eta)/b^2} \ , & \Im \eta<0 \ .
    \end{cases}
\end{align}
We define the series that can be thought of as the series expansion of the phase factor by using this $F(\eta)$.
Therefore the precise expression of the expansion could be
\begin{align}
    \frac{1}{e^{i\pi (1-2\eta)/b^2}-e^{-i\pi (1-2\eta)/b^2}}=F(\eta+i0)-F(\eta-i0)
\end{align}
for real $\eta$. Roughly speaking, one may understand this result as follows: The two-point function in the current case has the contributions from infinite saddles (i.e. labeled by integers $\mathbb{Z}$), not semi-infinite saddles. 
Naively treating the hyperfunction, we write the series expansion as 
\begin{align}
    \frac{1}{e^{i\pi (1-2\eta)/b^2}-e^{-i\pi (1-2\eta)/b^2}}=2i\sum_{n=0}^\infty \sin\left[\pi\left(2n+1\right)(1-2\eta)\frac{1}{b^2}\right],
\end{align}
by pairing the terms with same $n$ in $F(\eta+i0)$ and $F(\eta-i0)$,
although of course the right-hand side is literally divergent. A justification to use this naive notation is given by computing the sum inversely with adopting a simple regularization
\begin{align}\begin{aligned}
    \sum_{n=0}^\infty \sin\left[\left(2n+1\right)x\right]&\to\lim_{\epsilon\to 0}\sum_{n=0}^\infty e^{-\epsilon(2n+1)}\sin\left[\left(2n+1\right)x\right]\\
    &=\lim_{\epsilon\to 0}\frac{\sin x+\mathcal{O}(\epsilon^2)}{2\sin^2x+\mathcal{O}(\epsilon^2)}\\
    &=\frac{1}{2\sin x} \ .
\end{aligned}\end{align}
Finally we obtain the two-point function 
\begin{align}\begin{aligned}\label{expandinf}
    \left\langle V_\alpha(z_1)V_\alpha(z_2)\right\rangle&\sim \sum_{n=0}^\infty \sin\left[\pi\left(2n+1\right)(1-2\eta)\frac{1}{b^2}\right] \\
    & \times e^{-\frac{2}{b^2}\left[(1-2\eta)\log(1-2\eta)-(1-2\eta)\right]}\lambda^{(1-2\eta)/b^2}|z_{12}|^{-4\eta(1-\eta)/b^2}\delta(0) \ .
\end{aligned}\end{align}

Here let us comment on an ambiguity for the definition of $F(\eta)$ \eqref{Fdef}. Actually, there would exist other choices of $F(\eta)$ that seems to give the series expansion \eqref{expandinf}. For example, the regularization adopted in \cite{Chang:2016ftb}, which simply makes an infinitesimal shift $\eta\to\eta+i\epsilon$ in the imaginary direction, is realized as the hyperfunction defined by using
\begin{align}
    F(\eta)=
    \begin{cases}
        \sum_{n=0}^\infty e^{- 2\pi i(n+ 1/2)(1-2\eta)/b^2}, & \Im \eta >0 \ ,\\
        0 \, , & \Im \eta<0 \ .
    \end{cases}
\end{align} 
This ambiguity is analogous to the variety of definitions of the Green's function, where Feynman's $i\epsilon$ prescription corresponds to our regularization \eqref{Fdef}. One advantage of the current definition is that the expansion looks purely imaginary as expected from the original form. Furthermore, the result seems to respect the reflection symmetry under $\eta\to 1-\eta$.

What do the saddles in \eqref{expandinf} correspond to in the gravity side? The two-point solution to the Liouville equation of motion approximates as
\begin{align}
    \varphi(z,\bar{z})&\to -4\eta\log|z-z_i|+C_i \ ,\quad \text{as}\ \ z\to z_i \ ,\\
    \varphi(z,\bar{z})&\to -4\log |z|+C_\infty \ , \quad \text{as}\ \ z\to \infty \  ,
\end{align}
where the constants are 
\begin{align}
    C_1&=2\pi i \left(N+\frac{1}{2}\right)-\log\lambda -2\log\Tilde{\kappa}+(4\eta-4)\log |z_{12}| \ ,\\
    C_2&=2\pi i \left(N+\frac{1}{2}\right)-\log\lambda +2\log\Tilde{\kappa}+4 \log (1-2\eta) \ ,\\
    C_\infty&=2\pi i \left(N+\frac{1}{2}\right)-\log\lambda -2\log\Tilde{\kappa}-2\log\left(1+\frac{1}{\Tilde{\kappa}^2(1-2\eta)^2|z_{12}|^2}\right) \ .
\end{align}
Under the infinitesimal shift $\eta \to\eta+ i0$, the Liouville field shifts as $\varphi\to\varphi-i0$ at least the regimes near $z_1,z_2$. If we adopt the following ansatz 
\begin{align}
    ds^2=d\rho^2+\cosh^2\rho\, e^{\varphi(z,\bar{z})}dz d\bar{z} \ ,
\end{align}
the shift is equivalent to $\rho\to\rho +i0$ \footnote{From $\cosh^2(\rho\pm i\epsilon)=(\cosh\rho\cos\epsilon \pm i\sinh\rho\sin\epsilon)^2 \simeq \cosh^2\rho \pm 2i\epsilon \sinh\rho\cosh\rho$}. On the other hand, the regularization of $\eta\to\eta-i0$ is equivalent to $\rho\to\rho-i0$.

\section{Interpretation of saddles as complex metrics}
\label{app:geometry}

In subsection \ref{sec:complexsaddles}, we argued that the geometries labeled by an integer $n$ are characterized by the winding numbers $\pi_3(\mathbb{S}^3)$. Here we calculate the winding numbers for the geometries we constructed in subsection \ref{sec:dSmini} for the dS$_3$ case and check they actually realize the proposed geometries in section \ref{sec:complexsaddles}.
The vielbein $e^a$ and the spin connection $\omega^a$ for the ansatz \eqref{minidS} are%
\footnote{Here we use a slightly different notation adopted in section \ref{sec:claads}.}
\begin{alignat}{3}
    e^0&=\ell_{\text{dS}}Nd\tau\,, &\qquad e^1&=\ell_{\text{dS}}a(\tau)d\theta\,, &\qquad e^2&=\ell_{\text{dS}}a(\tau)\sin\theta d\phi\,,\\
    \omega^0&=\cos\theta d\phi\,,& \omega^1&=-\frac{a'(\tau)}{N}\sin\theta d\phi\,, & \omega^2&=\frac{a'(\tau)}{N}d\theta\,.
\end{alignat}
Here $(\theta,\phi)$ are the standard coordinates of $\mathbb{S}^2$.  
Since we are adopting the Euclidean notation, the connection one-forms $A,\tilde{A}$ in the Chern-Simons formalism with positive cosmological constant are expressed as 
\begin{align}
    A=\left(\omega^a+\frac{1}{\ell_{\text{dS}}}e^a\right)J_a\,,\qquad \tilde{A}=\left(\omega^a-\frac{1}{\ell_{\text{dS}}}e^a\right)J_a\,,
\end{align}
where $J_a\ (a=0,1,2)$ are $\mathfrak{su}(2)$ generators 
\begin{align}\label{su2}
    J_0=\frac{1}{2}
    \begin{pmatrix}
        i & 0\\
        0 & -i
    \end{pmatrix},\qquad
    J_1=\frac{1}{2}
    \begin{pmatrix}
        0 & i\\
        i & 0
    \end{pmatrix},\qquad
    J_2=\frac{1}{2}
    \begin{pmatrix}
        0 & -1\\
        1 & 0
    \end{pmatrix}.
\end{align}
Inserting the solution \eqref{dSabar}, the connections are expressed as 
\begin{align}
    A=b(\tau)^{-1}db(\tau)+b(\tau)^{-1}\alpha(\theta,\phi)b(\tau)\,,\qquad \tilde{A}=b(\tau)db(\tau)^{-1}+b(\tau)\alpha(\theta,\phi)b(\tau)^{-1}\,,
\end{align}
where
\begin{align}
    b(\tau)=e^{N\tau J_0},\qquad \alpha(\theta,\phi)=\frac{1}{2}
    \begin{pmatrix}
        i\cos\theta d\phi & -\frac{a_1}{\sin N}(d\theta+i\sin\theta d\phi)\\
        \frac{a_1}{\sin N}(d\theta-i\sin\theta d\phi) & -i\cos\theta d\phi
    \end{pmatrix}\,.
\end{align}
When $N=N^+_m$, for example, 
\begin{align}
    b(\tau)=e^{N^+_m\tau J_0},\qquad \alpha(\theta,\phi)=\frac{1}{2}
    \begin{pmatrix}
        i\cos\theta d\phi & -(-1)^m(d\theta+i\sin\theta d\phi)\\
        (-1)^m(d\theta-i\sin\theta d\phi) & -i\cos\theta d\phi
    \end{pmatrix}\,.
\end{align}
In this case, we can rewrite $\alpha(\theta,\phi)$ in the form of the pure gauge $\alpha(\theta,\phi)=h(\theta,\phi)^{-1}dh(\theta,\phi)$ with 
\begin{align}
    h(\theta,\phi)=e^{\phi J_0}e^{(-1)^m\theta J_2}\,,
\end{align}
and then $A$ and $\tilde{A}$ can also be expressed in the pure gauge
\begin{align}
    A=g^{-1}dg\,,\qquad \tilde{A}=\tilde{g}^{-1}d\tilde{g}\,,
\end{align}
with $g=b(\tau)h(\theta,\phi)$ and $\tilde{g}=b(\tau)^{-1}h(\theta,\phi)$.

The winding number 
\begin{align}
    W_3[g]=\frac{1}{24\pi^2}\int \tr[g^{-1}dg\wedge g^{-1}dg\wedge g^{-1}dg]
\end{align}
for the solution with $N^+_m$ is 
\begin{align}
    W_3[g^{(N^+_m)}]=-W_3[\tilde{g}^{(N^+_m)}]=\frac{N^+_m}{4\pi}=\frac{1}{4}\left(m+\frac{1}{2}\right)+\frac{i}{4\pi}\log\left(a_1+\sqrt{a_1^2-1}\right)\,.
\end{align}
In total
\begin{align}
    W_3[g^{(N^+_m)}]-W_3[\tilde{g}^{(N^+_m)}]=\frac{1}{2}\left(m+\frac{1}{2}\right)+\frac{i}{2\pi}\log\left(a_1+\sqrt{a_1^2-1}\right)\,.
\end{align}

Let us move to the AdS$_3$ case.
A saddle point solution $\bar{a}^{(N_m^+)}(r)$ for a saddle $N^+_m$ takes the form 
\begin{align}
    \bar{a}^{(N_m^+)}(r)=\frac{a_1}{\sinh\left( N^+_m\right)}\sinh\left(N^+_mr\right)\,.
\end{align}
When $m=0$, for example, 
\begin{align}
    \bar{a}^{(N^+_0)}(r)=\sinh\left(r\arcsinh a_1\right)=\sinh\rho\,,
\end{align}
where we have defined $\rho\equiv r\arcsinh a_1$. We then obtain the metric of the ordinary Euclidean AdS$_3$ space
\begin{align}
    ds^2=\ell_{\text{AdS}}^2\left(d\rho^2+\sinh^2\rho\, d\Omega_2^2\right)\,.
\end{align}
For $m\neq 0$, the metric \eqref{minisuperspace} has complex-valued elements. The plots of $a^{(N^+_m)}(r)$ for $m=1,2,3,4$ are shown in figure \ref{fig:DDsaddleplot}. Let us first consider $m=1$. At the crossing point of the blue curve with the imaginary axis, the spatial slice shrinks to zero volume, so at that point the outer geometry is connected to another geometry with spherical topology (the left part of the blue curve to the imaginary axis). For higher $m$, we can see that the curve has more crossing points than $m=1$. This can be regarded as a geometry connected with $m$ additional geometries each with spherical topology. 
\begin{figure}
    \centering
    \includegraphics[width=0.45\linewidth]{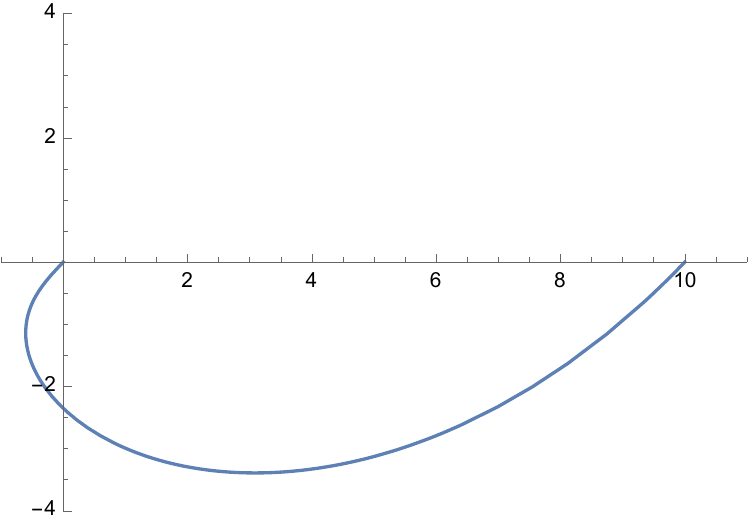}
    \includegraphics[width=0.45\linewidth]{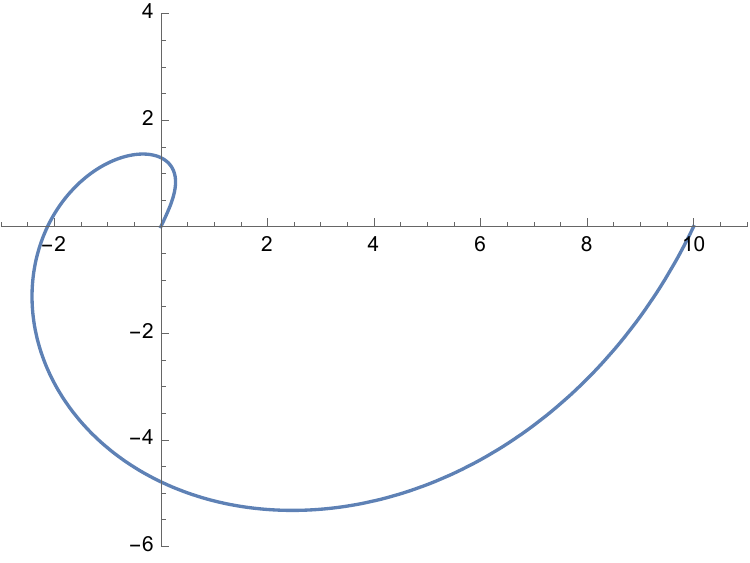}
    \includegraphics[width=0.45\linewidth]{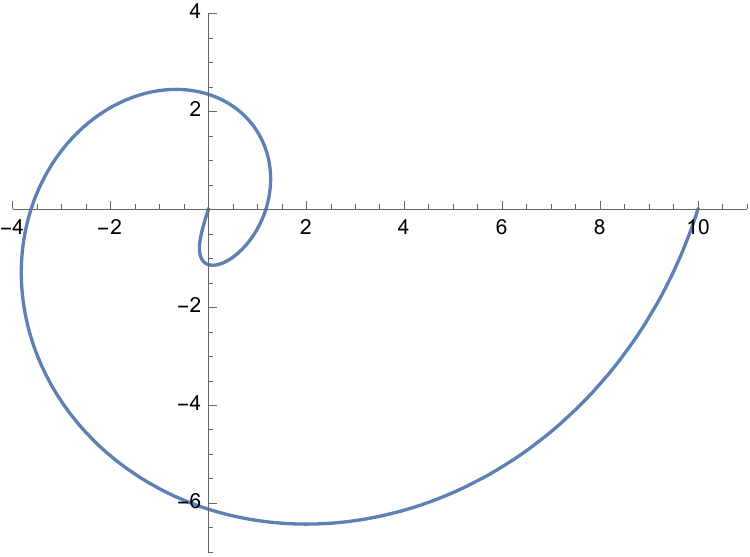}
    \includegraphics[width=0.45\linewidth]{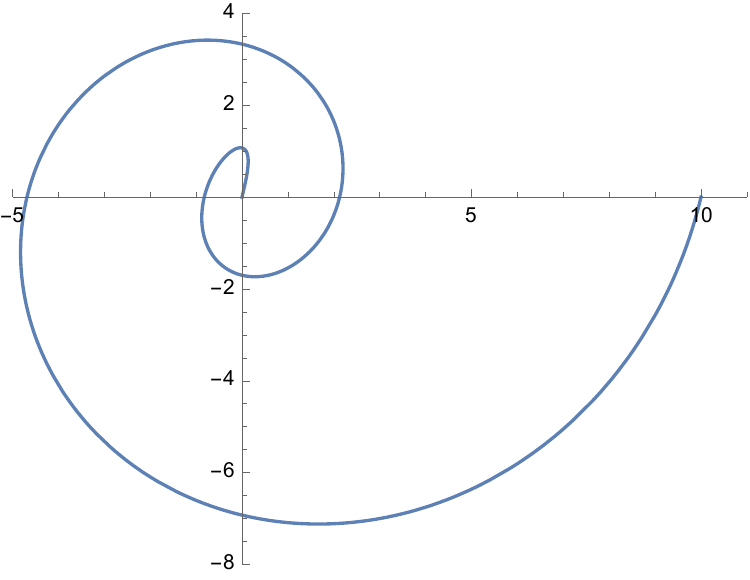}
    \caption{The plots of $\bar{a}^{(N^+_m)}(r)$ for $m=1,2,3,4$ in the range $0<r<1$. The parameters are fixed as $a_1=10$ and $\ell=1$. }
    \label{fig:DDsaddleplot}
\end{figure}

Another good way to understand the saddle points is to express a solution as that of the Chern-Simons formulation of 3d gravity as above. Since we are now considering the Euclidean AdS$_3$, The gauge fields are 
\begin{align}
    A=\left(\omega^a+\frac{i}{\ell_{\text{AdS}}}e^a\right)J_a \,,\quad \bar{A}=\left(\omega^a-\frac{i}{\ell_{\text{AdS}}}e^a\right)J_a\,.
\end{align}
Here $J_a\ (a=0,1,2)$ are the $\mathfrak{su}(2)$ generators with expressions \eqref{su2}.
The expressions of the gauge fields that correspond to our ansatz \eqref{minisuperspace} with \eqref{DDsolution} are given by 
\begin{align}
    A=b(r)^{-1} db(r)+b(r)^{-1}\alpha(\theta,\phi)b(r)\,,\quad \bar{A}=b(r)db(r)^{-1}+b(r)\alpha(\theta,\phi)b(r)^{-1}
\end{align}
where 
\begin{align}
    b(r)&=e^{iNrJ_0}\,,\\
    \alpha(\theta,\phi)&= \frac{1}{2}
    \begin{pmatrix}
        i\cos\theta d\phi & -\frac{a_1}{\sinh N}(d\theta+i\sin\theta d\phi)\\
        \frac{a_1}{\sinh N}(d\theta-i\sin \theta d\phi) & -i\cos\theta d\phi
    \end{pmatrix}\,.
\end{align}
We can then repeat similar analysis as in the previous dS$_3$ case.

\bibliographystyle{JHEP}
\bibliography{Resurgence}

\end{document}